\documentclass[12pt]{article}
\usepackage{url} 

\newcommand{\blind}{0}

\usepackage[utf8]{inputenc}
\usepackage{comment}

\usepackage{lscape} 
\usepackage{varioref}
\usepackage[linktocpage=true]{hyperref}
\usepackage{appendix}
\usepackage[]{natbib}
\usepackage{bookmark}
\usepackage{multirow}
\usepackage{booktabs}
\usepackage[flushleft]{threeparttable}
\usepackage{amsmath}
\usepackage[capitalise,noabbrev]{cleveref}
\usepackage{amssymb}
\usepackage{amsthm}
\usepackage{graphicx}
\usepackage{float}
\usepackage{soul}
\usepackage{algorithm}
\usepackage{algpseudocode}
\usepackage{xcolor}

\newcommand{\bfX}{\mathbf{X}}
\newcommand{\bfx}{\mathbf{x}}
\newcommand{\bfY}{\mathbf{Y}}
\newcommand{\bfy}{\mathbf{y}}
\newcommand{\bfW}{\mathbf{W}}
\newcommand{\bfw}{\mathbf{w}}

\newcommand{\vect}[1]{{\boldsymbol{#1}}} 
\def\one{\mathbf{1}} 

  {
      \theoremstyle{plain} 
      \newtheorem{assumption}{Assumption}
      \newtheorem{theorem}{Theorem}

  }
\begin{document}

\def\spacingset#1{\renewcommand{\baselinestretch}%
{#1}\small\normalsize} \spacingset{1}


\if0\blind
{
  \title{\bf Conformal Prediction with Time-Series Data via Sequential Conformalized Density Regions }
  \author{Max Sampson \\
    Department of Statistics and Actuarial Science, University of Iowa\\
    and \\
    Kung-Sik Chan \\
    Department of Statistics and Actuarial Science, University of Iowa}
  \maketitle
} \fi

\if1\blind
{
  \bigskip
  \bigskip
  \bigskip
  \begin{center}
    {\LARGE\bf Title}
\end{center}
  \medskip
} \fi

\bigskip
\begin{abstract}

We propose a new conformal prediction method for time-series data with a guaranteed asymptotic conditional coverage rate, Sequential Conformalized Density Regions (SCDR), which is flexible enough to produce both prediction intervals and disconnected prediction sets, signifying the emergence of bifurcations. Our approach uses existing estimated conditional highest density predictive regions to form initial predictive regions. We then use a quantile random forest conformal adjustment to provide guaranteed coverage while adaptively changing to take the non-exchangeable nature of time-series data into account. 

We show that the proposed method achieves the guaranteed coverage rate asymptotically under certain regularity conditions. In particular, the method is doubly robust -- it works if the predictive density model is correctly specified and/or if the  scores follow a nonlinear autoregressive model with the correct order specified. 

Simulations reveal that the proposed method outperforms existing methods in terms of empirical coverage rates and set sizes. We illustrate the method using two real datasets, the Old Faithful geyser dataset and the Australian electricity usage dataset. Prediction sets formed using SCDR for the geyser eruption durations include both single intervals and unions of two intervals, whereas existing methods produce wider, less informative, single-interval prediction sets.

\end{abstract}

\noindent%
\vfill

\newpage
\spacingset{1.8} 

\section{Introduction}\label{sec:introduction}

Consider a univariate time-series where the first $T$ observations, $Y_1, Y_2, \ldots, Y_T$, are observed. The subsequent value, $Y_{T+1}$, is unobserved and represents the outcome of interest.  Additionally, there are $T + 1$ observed features, $\vect{X}_1, \vect{X}_2, \ldots, \vect{X}_T, \vect{X}_{T + 1}$, which are used to help predict $Y_{T+1}$. These features can include past values of the time series, as well as predictors. Our focus in this paper is not on a point prediction for $Y_{T+1}$, but on prediction sets which are valid regardless of the quality of the model.

Conformal prediction concerns the development of methods for constructing prediction sets with guaranteed coverage rates for parametric or nonparametric machine learning algorithms, and it has found wide application. However, the literature on conformal prediction heavily focuses on exchangeable data. This is because the exchangeability assumption facilitates the construction of conformal prediction sets using suitable (non)conformity scores and the quantiles of the observed (non)conformity scores \citep{conformal_book}.

Conformal prediction for exchangeable data have valid marginal coverage,
\[
\mathbb{P}(Y_{t} \in \hat{C}(\vect{X}_t)) = 1 - \alpha.
\]
With additional assumptions on the form of the data or the model, some conformal prediction sets also have some form of conditional coverage.
One form of conditional validity is conditional on the observed features at time $t$,
\[
\mathbb{P}(Y_{t} \in \hat{C}(\vect{X}_t) \mid \vect{X}_t) = 1 - \alpha,
\]
In this paper we aim to create sharp prediction sets that achieve asymptotic conditional validity with minimal assumptions,
\begin{equation}\label{eq:conditional_asymp_cov}
|\mathbb{P}(Y_{t} \in \hat{C}(\vect{X}_{t}) \mid \vect{X}_t) - (1 - \alpha)| \to 0 \quad \forall t \quad \text{as}\quad T \to \infty.
\end{equation}

\subsection{Motivation and Preview for SCDR}

Because time-series data are generally time-irreversible, they tend to be  non-exchangeable. Conformal prediction for time-series data is relatively underexplored, although several methods have recently been proposed (for example, \cite{gibbs2021adaptive, SPCI_Xu_xie_2023, feldman2022achieving, zaffran2022adaptive, bellman_conformal_yan_candes_2024, PID_angelopoulos}). The problem is challenging because the one-step predictive distribution is generally specific to the past realization of the time series. 

\citet{SPCI_Xu_xie_2023} addressed this by estimating two conditional quantiles of the signed residual given several past lags of the residuals, via the quantile random forest. Meanwhile, \citet{PID_angelopoulos} and  \citet{bellman_conformal_yan_candes_2024} proposed methods that iteratively adjust the nominal coverage rate or the conformal adjustment to ensure the long-run coverage rate reaches the desired level. These approaches also attempt to optimize the size of the prediction intervals to some extent. However, existing methods generally require large sample sizes for the guaranteed coverage rate to hold, they are designed to output only prediction intervals, and they fail to have strong guarantees on the conditional coverage.

In practice, the time-series generating process may be nonlinear, yielding possibly multimodal predictive distributions, limit cycles, bifurcations, and/or conditional heteroscedasticity (volatility), see \citet{tong1990non} for a more thorough discussion. Thus, it is desirable to develop conformal prediction methods for time-series data that are capable of producing disconnected prediction sets.

Inspired by the approach of \cite{SPCI_Xu_xie_2023}, we propose to model past non-conformity scores with a non-parametric conditional quantile model to form conformal prediction sets. 
This approach allows us to take past realizations of the time-series into account when computing a conformal adjustment, enabling us to form model agnostic prediction sets without exchangeable data. We extend their approach from the signed regression residuals to general non-conformity scores, accommodating a wider range of models, including quantile and density-based approaches. Empirically, starting with an existing prediction set and adjusting it with a model based (non)conformity score provides much better finite sample coverage than forming a prediction interval based entirely on the (non)conformity scores. 

In this paper we focus on forming prediction sets using the conditional probability density. This allows us to capture the geometry of the conditional distribution, which can help inform practitioners of the possibly \emph{nonlinear} data generating mechanisms -- a task that is increasingly common with large datasets \citep{shrestha2021algorithm, elragal2017theory}. We demonstrate this in section~\cref{sec:geyser} with data on eruption duration and time between eruptions of the Old Faithful Geyser, which is known to erupt due to two different mechanisms, causing both the eruption duration and waiting time between eruptions to be bimodal \citep{kieffer1984seismicity, hutchinson1997situ}. This descriptive property is unique to our approach, as other conformal prediction approaches for time-series focus strictly on prediction intervals. 

Our approach, Sequential Conformalized Density Regions (SCDR), uses a multiplicative, model-based, conformal adjustment to adjust unconformalized conditional highest predictive density sets. The model-based conformal adjustment, which utilizes a conditional quantile random forest, can be viewed as a local (conditional) adjustment based on the data characteristics, while the conditional density describes the data geometry. This combination of models results in the first, to our knowledge, doubly robust conformal prediction method for time-series data, guaranteeing asymptotic conditional coverage when the conditional density model or the conformal adjustment model is correctly specified. Not only is SCDR doubly robust, but because it is an adjusted highest predictive density set, the prediction sets it forms are asymptotically approach the smallest possible prediction sets under some regularity conditions. 

\subsection{Related Work}

We briefly outline three current methods for producing conformal prediction intervals for time-series data. These methods focus on prediction intervals that either apply a conformal like update to the regression error for point predictors, $Y_t - \hat{f}(\vect{X}_t)$, or adaptively adjust the coverage rate for prediction intervals produced by the forecasting algorithm, $1 - \alpha_t^*$ \citep{SPCI_Xu_xie_2023, bellman_conformal_yan_candes_2024, PID_angelopoulos, gibbs2021adaptive}. 

Conformal PID control (PID), proposed by \cite{PID_angelopoulos}, treats the system for producing prediction sets as a proportional-integral-derivative (PID) controller. Assuming a model has been trained that outputs a point prediction, $\hat{Y}_t$, PID consists of three steps: quantile tracking, error integration, and scorecasting. Quantile tracking consists of tracking the $1 - \alpha$ quantile of the non-conformity scores, for example the signed conformal score $s_t = Y_t - \hat{Y}_t$ \citep{Linusson_2014_signed_conformal}. When one score is received at a time, this can be updated with the following approach,
\[
q_{t+1} = q_t + \eta\times (\text{err}_t - \alpha),
\]
where $\text{err}_t$ is 0 if $s_t \leq q_t$ and 1 otherwise.  This can be viewed as decreasing the quantile when the response was covered in the last step, and increasing the quantile when the interval miscovered in the last step. The second part is error integration,which is a generalized version quantile tracking replacing the constant $\eta$ with $r_t$, a saturation function,
\[
q_{t+1} = r_t   \Big(\sum_{i=1}^t(\text{err}_t - \alpha)\Big).
\]
Lastly, the score caster is a model that attempts to capture any remaining dependency in the scores. It takes past scores in and outputs an estimate of the $1 - \alpha$ quantile error, $\hat{q}_{t-1}$. This allows any leftover noise or signal that was not captured by the original base forecaster to be included in the prediction intervals, helping to reduce systematic errors.

Combining all three results in
\[
q_{t+1} = \eta g_t + r_t  \Big(\sum_{i=1}^t g_t\Big) + g'_t,
\]
where $\eta > 0$ is a learning rate, $g_t = (\text{err}_t - \alpha)$, and $g'_t$ is the output from the scorecaster. This is then used as the conformal adjustment  \citep{PID_angelopoulos}.

Bellman Conformal Inference (BCI), introduced by \cite{bellman_conformal_yan_candes_2024},  attempts to model and control the coverage level for a model, $\alpha_t$ by using the past data and coverage levels. Define $L_{t}(\beta) = |C_{t}(1 - \beta)|$, where $|\cdot|$ denotes the length (Lebesgue measure) of the enclosed set,  as the function that maps the miscoverage rate, $\beta$, to the length of the prediction interval for $Y_{t}$. Denote $F_{t}$, as the estimated marginal distribution of $\beta_{t}$ which is estimated using the past observations. For example, the empirical cdf, $F_{t}$, of $\{\beta_{t-1}, \ldots, \beta_{t - B} \}$ for a ``large" lag $B$. Note, BCI can be generalized to more than one step ahead prediction intervals, which is not further discussed here for ease of understanding. At time $t$ the following optimization problem is solved to determine the adjusted $\alpha_t^*$, 
\[
\alpha_t^* =\arg\min_{\alpha_t}\mathbb{E}_{\beta_{t} \sim F_{t}}\Bigg[ L_{t}(\alpha_t) + 
\lambda_t \times \max (\text{err}_t - {\alpha}, 0)
\Bigg],
\]
where $\text{err}_t = \one{(\alpha_t > \beta_t)} $ denotes the coverage error indicator at time $t$, $\lambda_t$ denotes the relative weight on the miscoverage level that is used to guarantee coverage, and ${\alpha}$ is the nominal miscoverage rate. $ L_{t}(\alpha_t)$ controls the length of the prediction interval, while  $\lambda_t \times \max (\text{err}_t - {\alpha}, 0)$ controls the coverage level, attempting to ensure a sharp interval with the nominal $1-\alpha$ coverage. Note the possible value $\alpha_t$ is used in place of $\alpha$ at time $t$. The solution $\alpha_t^*$ acts as an adjustment for misspecified models or incorrect assumptions to ensure the formed prediction intervals have asymptotic coverage of at least $1 - \alpha$ while minimizing the size of the prediction set \citep{bellman_conformal_yan_candes_2024}. This is similar to Adaptive Conformal Inference (ACI), but ACI fails to optimize the prediction interval lengths, which can lead to infinite prediction intervals \citep{bellman_conformal_yan_candes_2024, gibbs2021adaptive}. 

Sequential Predictive Conformal Inference for Time Series (SPCI), proposed by \cite{SPCI_Xu_xie_2023}, starts with a point estimate, $\hat{f}(\cdot)$. Signed residuals are then computed using either a leave-one-out approach, or a bootstrap and aggregation approach. The bootstrap approach trains $B$ models, and computes residuals on the out-of-sample data for each of the $B$ models. In standard signed conformal regression the resulting prediction intervals would then be,
\[
[\hat{f}(\vect{X}_{t}) - \hat{q}_1, \hat{f}(\vect{X}_t) + \hat{q}_2],
\]
where $\hat{q}_1$ and $\hat{q}_2$ are the empirical quantiles found using a grid search to minimize the prediction interval length while maintaining $1 - \alpha$ coverage under exchangeability. In SPCI, a quantile random forest model (QRF) is instead fit autoregressively on the residuals, using the past $w \geq 1$ residuals to predict the future (unobserved) residual. This allows the dependence from the past $w$ residuals to be incorporated into the adjustment, leading to the correct asymptotic nominal coverage when the QRF is correctly specified, which can be assessed by examining the autocorrelation and autocorrelation plots. The QRF for the residuals is retrained at each prediction index using a sliding window of the most recent $k$ residuals. A grid search is then done to find the smallest $1 - \alpha$ interval using the predicted quantiles of the future residuals \citep{SPCI_Xu_xie_2023}. 

The preceding methods provide asymptotic guarantees on nominal coverage, as well as some finite sample coverage bounds, but fail to have guarantees on the optimal prediction set size or conditional coverage. Some of the methods attempt to solve an optimization problem at each step that can produce smaller prediction intervals, but these approaches don't provide any theoretical guarantees on optimal prediction set length, and can result in needlessly large intervals in practice. We attempt to solve these problems by introducing a method that combines conditional highest density prediction sets with a minor adjustment that guarantees asymptotic conditional coverage if the conditional density model is misspecified. 

We extend the ideas introduced in SPCI to a larger class of scores by replacing the empirical quantile used in standard conformal prediction with a conditional quantile regression estimate, where the conditioning variables are the previous non-conformity scores \citep{SPCI_Xu_xie_2023}.  Sequential Conformalized Density Regions (SCDR), starts with a conditional density estimator. It then uses the division non-conformity score first introduced in \cite{sampson_CHCDS_2025}, to adjust the highest level density set.

We derive a doubly robust property, that the nominal conditional coverage is guaranteed when either the conditional density estimator or the conditional quantile model is correctly specified. This doubly robust property is, to our knowledge, the first of its kind for conformal prediction with time-series data. We also derive sufficient conditions under which SCDR converges to the optimal prediction set, that is the set that satisfies the nominal conditional coverage level with the smallest Lebesgue measure. We then demonstrate these properties via numerical results.

\section{Methodology}\label{sec:methodology}

Throughout we let $\vect{X}$ represent features, which may contain past outcomes, $Y$ the response, and $(\vect{X}_1, Y_1), \ldots, (\vect{X}_T, Y_T)$ the observed data up to time $T$. We also note that this method also holds for multivariate responses, $\vect{Y}$, though our focus in this paper remains on univariate responses. The extension to multivariate responses is briefly discussed in~\cref{sec:conclusion}. 

We now describe our method, sequential conformalized density regions (SCDR), which applies a conformal adjustment to existing conditional highest density prediction regions to provide a conditional coverage guarantee for time-series data. Given any conditional density estimating function, $\mathcal{B}$, we iterate through the observed data to fit leave-one-out conditional density estimators:
\[
\hat{f}_{-j} = \mathcal{B}( \{ (X_i, Y_i) : i \neq j \} ).
\]

Using these leave-one-out (LOO) models, unadjusted $1-\alpha$ upper level density set cutoffs are computed for the left-out data, $\hat{c}(\vect{X}_j)$. Next, the estimated $1- \alpha$ level set cutoff is compared with the estimated height of the density, and summarized into a conformity score:
\[
V_j = \hat{f}_{-j}(Y_j \mid \vect{X}_j) / \hat{c}(\vect{X}_j), \> j = 1, 2, \ldots, T.
\]
We make a note here that one does not need to compute LOO scores for all $T$ of the series, but only the number of scores, say $k$, that will be used to train the initial QRF. 

After LOO scores have been computed, the conditional density model is refit on the full data:
\[
\hat{f}_{T+1} = \mathcal{B}( \{ (X_i, Y_i) : i =1, \ldots, T\} ).
\]

At the same time, a second model is fit to predict the $\alpha$ quantile of the scores. This modeled quantile replaces the empirical quantile used with conformal prediction when the data are exchangeable. Any condtional quantile model can be used here, but the results in~\cref{sec: coverage results} assume it to be a conditional quantile random forest. The past $w$ scores are included as covariates in the model to account for any remaining dependence: 
\[
\hat{Q} = \mathcal{Q}( \{(V_i, [V_{i-1}, \ldots, V_{i - w}], \alpha):i= t-1,\ldots, t - k - 1\})
\]
We describe why only past scores are used as covariates, and not the scores and observed data, in~\cref{sec: QRF covariate justification}. 

This QRF model is then used to predict the $\alpha$ quantile of the new score,
\[
\hat{q}_{T+1} = \hat{Q}(V_{t-1}, \ldots, V_{t-w}).
\]
Next, $\hat{c}(\vect{X}_{T+1})$ is estimated using $\hat{f}_{T+1}$. The resulting SCDR prediction set is then
\[
\hat{C}(\vect{X}_{T+1}) = \{y: \hat{f}_{T+1}(y \mid \vect{X}_{T+1}) > \hat{c}(\vect{X}_{T+1}) \times \hat{q}_{T+1} \},
\]
where multiplying by $\hat{q}_{T+1}$ can be viewed as an adjustment to the unconformalized level set cutoff, $\hat{c}(\vect{X}_{T+1})$, to provide a guarantee on the coverage level in the presence of model misspecification. 

As new observations are observed, the scores and models are sequentially updated in the same way. The algorithm for SCDR with the LOO score approach is given in~\cref{alg:SCDR_loo}. To avoid computing LOO scores a bootstrap approach with SCDR can also be used. 

In the bootstrap approach, $B$ bootstrap samples are taken with replacement, which are used to fit $B$ conditional density models,
\[
\hat{f}^b = \mathcal{B}( \{ (\vect{X}_i, Y_i) : i \in \mathcal{I}_b \}) \text{ for }b=1, 2,\ldots, B,
\]
where $\mathcal{I}_b$ indexes the indices included in the $b$th bootstrap sample. 

Both the estimated conditional density and the density level set cutoffs used to compute the scores are aggregates of the out of bootstrap sample density level set cutoffs:
\[
\hat{f}(Y_j \mid \vect{X}_j) = \phi( \{\hat{f}^b(Y_j \mid \vect{X}_j) : j \notin \mathcal{I}_b\}),
\]
and
\[
\hat{c}(\vect{X}_j) = \phi ( \{\hat{c}(\vect{X}_j) : j \notin \mathcal{I}_b\}),
\]
where $\hat{c}^b(\vect{x})$ is the estimated density level set cutoff for $Y \mid \vect{X} = \vect{x}$ found using $\hat{f}^b$ and $\phi(\cdot)$ is an aggregator function, for example the mean. 

The scores are then computed using the aggregated estimates,
\[
V_j = \hat{f}(Y_j \mid \vect{X}_j) / \hat{c}(\vect{X}_j).
\]

With the scores computed, the approach for forming the prediction sets is nearly identical to the LOO SCDR approach. First, a conditional quantile model is fit on the past $k$ scores using the previous $w$ lagged scores as covariates in the model:
\[
\hat{Q} = \mathcal{Q}( \{(V_i, [V_{i-1}, \ldots, V_{i - w}], \alpha):i= T,\ldots, T - k\}).
\]
This is then used to predict the $\alpha$ quantile of the new score,
\[
\hat{q}_{T+1} = \hat{Q}(V_{T}, \ldots, V_{T-w}). 
\]

An aggregated density level cutoff is then computed, and multiplied by the model misspecification adjustment $\hat{q}_{T+1}$ to form the final prediction sets,
\[
\hat{c}(\vect{X}_{T+1}) =  \phi(\{\hat{c}^b(\vect{X}_t)\}_{b=1}^B),
\]
leading to
\[
\hat{C}(\vect{X}_{T+1}) = \{y: \hat{f}(y|\vect{X}_{T+1}) >\hat{c}(\vect{X}_{T+1}) \times \hat{q}_{T+1} \}, 
\]
where $\hat{f}(y|\vect{X}_{T+1}) = \phi(\{\hat{f}^b(y \mid \vect{X}_{T+1})\}_{b=1}^B)$.

As new observations are observed, the new scres are computed in the following way: 
\[
 \hat{f}(Y_{T+1}\mid\vect{X}_{T+1}) = \phi(\{\hat{f}^b(Y_{T+1} \mid \vect{X}_{T+1})\}_{b=1}^B),
\]
and 
\[
V_{T+1} =  \hat{f}(Y_{T+1}\mid\vect{X}_{T+1}) / \hat{c}(\vect{X}_{T+1}).
\]

The algorithm for one step ahead prediction with bootstrap aggregation is given in~\cref{alg:SCDR_boot}. One possible extension of SCDR to multi-step ahead prediction sets is discussed in Section A of the supplementary materials. 
Three key differences exist between SCDR and SPCI. The first is that SCDR starts with a prediction set that is later adjusted, instead of a conditional point estimate. Thus, the SCDR adjustment is not solely responsible for forming valid prediction intervals, allowing for better finite sample coverage than SPCI. The second difference is that SCDR attempts to model the dependency in both the conditional density model and the QRF, whereas SPCI does not include lagged responses as features. However, in our numerical results, we attempt to model dependency in both the conditional point estimate and the QRF for SPCI by including lagged responses and lagged residuals into the respective models. The third is that SCDR only requires estimating one quantile, not two. As with the first difference, this leads to SCDR having substantially better finite sample coverage than SPCI.

\begin{algorithm}[H]
    \caption{SCDR LOO}\label{alg:SCDR_loo}
    \textbf{Input:} miscoverage level $\alpha$, data =$(Y_i, \vect{X}_i)_{t=1}^{T}$,  conditional density algorithm $\mathcal{B}$, quantile regression algorithm $\mathcal{Q}$, conditional density sliding window $k_1$, score sliding window $k_2$, and nonlinear autoregressive order of the scores $w$. \newline
    \textbf{Procedure:}
    \begin{algorithmic}[1]

    \State Initialize $\vect{V} =\{\}$
    \For{($j = 1, \ldots T$)}
        \State Fit $\hat{f}_{-j} = \mathcal{B}(\{(\vect{X}_i, Y_i):i = j-1, , \ldots, j-k_2 - 1\})$
        \State Calculate $\hat{c}(\vect{X}_j)$, where $\hat{c}(\vect{x})$is the estimated $1-\alpha$ upper level density set cutoff for $Y\mid \vect{X} = \vect{x}$ found using $\hat{f}_{-j}$
        \State $V_j = \hat{f}_{-j}(Y_j \mid \vect{X}_j) / \hat{c}(\vect{X}_j)$
        \State $\vect{V} = \vect{V} \cup V_j$
    \EndFor
    \For{($t = T+1, \ldots T + m$)}
        \State Fit $\hat{f}_{t} = \mathcal{B}(\{(\vect{X}_i, Y_i):i = t - 1,\ldots, t- k_2- 1\})$
        \State Fit $\hat{Q} = \mathcal{Q}( \{(V_i, [V_{i-1}, \ldots, V_{i - w}], \alpha):i= t-1,\ldots, t - k_1 - 1\})$
        \State $\hat{q}_t = \hat{Q}(V_{t-1}, \ldots, V_{t-w})$
        \State Calculate $\hat{c}(\vect{X}_t) $ using $\hat{f}_t$
        \State     \textbf{Output:} $\hat{C}(\vect{X_t}) = \{y: \hat{f}_t(y|\vect{X}_t) >\hat{c}(\vect{X}_t) \times \hat{q}_t \}$

        \State $V_t = \hat{f}_t(Y_t \mid \vect{X}_t) / \hat{c}(\vect{X}_t)$
        \State $\vect{V} = \vect{V}\cup V_t$
    \EndFor
    \end{algorithmic}
    
\end{algorithm}

\begin{algorithm}[H]
    \caption{SCDR Bootstrap}\label{alg:SCDR_boot}
    \textbf{Input:} miscoverage level $\alpha$, data =$(Y_i, \vect{X}_i)_{t=1}^{T}$,  conditional density algorithm $\mathcal{B}$, quantile regression algorithm $\mathcal{Q}$, bootstrap number $B$, aggregator function $\phi$, score sliding window size $k$, and nonlinear autoregressive order of the scores $w$. \newline
    \textbf{Procedure:}
    \begin{algorithmic}[1]

    \For{($b = 1,\ldots, B$)}
        \State Sample with replacement $T$ indices from $\{1, 2, \ldots, T \}$, $\mathcal{I}_b = \{b_1, \ldots, b_T\}$
        \State Fit and store $\hat{f}^b = \mathcal{B}(\{(\vect{X}_i, Y_i):i \in \mathcal{I}_b\})$
    \EndFor
    \State Initialize $\vect{V} =\{\}$
    \For{($j = 1, \ldots T$)}
        \State $\hat{c}(\vect{X}_j) = \phi(\{\hat{c}^b(\vect{X}_j): j\notin \mathcal{I}_b\})$, where $\hat{c}^b(\vect{x})$ is the estimated $1-\alpha$ upper level density set cutoff for $Y\mid \vect{X} = \vect{x}$ using $\hat{f}^b$
        \State $\hat{f}(Y_j \mid \vect{X}_j) = \phi(\{ \hat{f}^b(Y_j \mid \vect{X}_j): j \notin \mathcal{I}_b\})$
        \State $V_j = \hat{f}(Y_j \mid \vect{X}_j) / \hat{c}(\vect{X}_j)$
        \State $\vect{V} = \vect{V} \cup V_j$
    \EndFor
    \For{($t = T+1, \ldots T + m$)}
        \State $\hat{Q} = \mathcal{Q}( \{(V_i, [V_{i-1}, \ldots, V_{i - w}], \alpha):i= t-1,\ldots, t - k\})$
        \State $\hat{q}_t = \hat{Q}(V_{t-1}, \ldots, V_{t-w})$
        \State $\hat{c}(\vect{X}_t) = \phi(\{\hat{c}^b(\vect{X}_t)\}_{b=1}^B)$
        \State  \textbf{Output:} $\hat{C}(\vect{X_t}) = \{y: \hat{f}(y|\vect{X}_t) >\hat{c}(\vect{X}_t) \times \hat{q}_t \}$, where $\hat{f}(y|\vect{X}_t) = \phi(\{\hat{f}^b(y \mid \vect{X}_t)\}_{b=1}^B)$
        \State $\hat{f}(Y_t\mid\vect{X}_t) = \phi(\{\hat{f}^b(Y_t \mid \vect{X}_t)\}_{b=1}^B)$
        \State $V_t = \hat{f}(Y_t \mid \vect{X}_t) / \hat{c}(\vect{X}_t)$
        \State $\vect{V} = \vect{V}\cup V_t$
    \EndFor
    \end{algorithmic}

\end{algorithm}

One note on~\cref{alg:SCDR_loo} is that one could use $j \neq i$ in step 3 and $i = 1, \ldots, t - 1$ in step 9 to get true leave-one-out scores. Though, we find that unless the model is perfectly specified, training on the most recent $k_2$ data points is more effective than training on all the data points, as those in the far past do not provide meaningful information. 

We use Normal mixtures to jointly model $(Y, \vect{X})$ with SCDR because, in independent and identically distributed (i.i.d.\@) settings, they are known to be universal approximators capable of approximating any smooth density to an arbitrary precision using sufficiently many components \citep[Chapter 3]{roeder1994graphical, bengio2017deep}. We believe this property holds in some settings outside the i.i.d.\@ case, including time-series, which we explore theoretically in~\cref{sec: density consistency}. Additionally, modeling the joint density as a mixture of Normals ensures the conditional density estimator has a closed form, which is also a mixture of Normals. We provide a proof of this in the supplementary material for completeness, building on the fact that the conditional distributions resulting from a joint Normal distribution are themselves Normal \citep{zimmerman2020linear}.

\subsection{Covariates in the QRF}\label{sec: QRF covariate justification}

The numerator of the scores, $\hat{f}(Y_t \mid \vect{X}_t)$, is a measure of how typical an observation is, but it fails to account for heteroskedasticity. This makes the direct comparison of $\hat{f}(Y_t \mid \vect{X}_t)$ and $\hat{f}(Y_{t-1} \mid \vect{X}_{t-1})$ difficult. The denominator of the scores, $\hat{c}(\vect{X}_t)$, serves as a measure of the volatility at time $t$. In a volatile state the conditional density is more spread out; a ``typical" response will have a small conditional density, but the corresponding density level-set cutoff will also be proportionally small. Thus, the scores discard where $Y_t$ is and instead capture if $Y_t$ is typical given $\vect{X}_t$.  

Because the model used is imperfect, residual temporal dependence remains, which is captured by the scores. To see this, consider that if an observation at time $t$ is typical ($V_t \geq 1$), the next observation is also likely to be typical ($V_{t+1} \geq 1$). The converse is also true: atypical observations tend to cluster. Consequently, the scores can be viewed as an indicator, $W_t = \one(V_t < 1)$, which indicates if the prediction set needs to expand. 

When training the quantile random forest, we strictly use past scores, $V_{t-1}$, as covariates instead of the observed data, $(Y_{t-1}, \vect{X}_{t-1})$. Because this temporal dependence is efficiently summarized by the scores, very little predictive power for $V_t$ is lost by excluding $Y_{t-1}$ and $\vect{X}_{t-1}$. While one could theoretically include the observed data, the curse of dimensionality will cause the resulting estimates to be noisier than necessary. That is, conditioning on the raw observed data results in a worse estimate for the quantile adjustment. 

The justification for strictly using past scores holds under the following two cases:
\begin{itemize}
    \item Under suitable regularity conditions (see~\cref{thm:doubly robust}) when the conditional density model is well-specified, the scores are both nearly i.i.d.\@ and independent of the observed data. Hence, the scores efficiently capture the information required to form prediction sets with valid coverage. 
    \item Regardless of the conditional density model, under stationarity, the joint distribution of the data remains constant over time. Consequently, the one-dimensional summary of recent observations describes whether or not the future score will need to grow or shrink the unadjusted prediction set. Introducing the potentially high-dimensional observed data is unnecessary because it does not provide any additional information about the adjustment required for the prediction set to have valid coverage. It merely results in a noisier estimate. 
\end{itemize}

\section{Theoretical Results}\label{sec:theory}
All proofs can be found in the supplementary materials Section B. As with the previous section, we note that these results can easily be extended to a multivariate response, $\vect{Y}$, which we discuss further in~\cref{sec:conclusion}. 

\subsection{Coverage Results}\label{sec: coverage results}

Denote the scores as $\{e_t\}_{t=1}^T$. Assume the $pth$ quantile of the scores is unique. 

The following 4 assumptions are identical to those from \cite{SPCI_Xu_xie_2023}. We discuss when these assumptions hold for Gaussian mixture density models after their statement.

\begin{assumption}\label{as: QRF1}
Define $U_t := F_{e}(e_t \mid \vect{X} = \vect{X}_t)$ as the quantile of the scores $e_t$ conditioning on $w$ past scores, $\vect{X}_t$, where $U_t \sim \text{Unif }[0,1]$. For $\vect{x} \in \mathbb{B} := \text{Supp}(\{\vect{X}_t\}_{t \geq 1})$, define the scalar $z[\vect{x}] := F_{e}(z \mid \vect{X} = \vect{x})$. Given
\[
g(i,j,\vect{x}_1,\vect{x}_2) := \text{Cov}(\one(U_i \leq z[\vect{x}_1]), \one(U_j \leq z[\vect{x}_2])),
\]
we require that for any pair of $\vect{x}_1, \vect{x}_2 \in \mathbb{B}$,
\[
g(i,j,\vect{x}_1,\vect{x}_2) = g(|i-j|,\vect{x}_1,\vect{x}_2) \text{ for } i \neq j. 
\]

In addition, there exists $\tilde{g}$ such that
\[
g(k,\vect{x}_1,\vect{x}_2) \leq \tilde{g}(k) \ \forall \vect{x}_1,\vect{x}_2 \in \mathbb{B}, \ k \geq 1 
\]
and
\[
\lim_{{T} \to \infty} \left[ \int_1^{{T}} \int_1^s \tilde{g}(u) \, du \, ds \right] / {T}^2 = 0. 
\]
\end{assumption}

The QRF can be viewed as a weighted quantile of the previously observed scores (as in \cite{SPCI_Xu_xie_2023} and \cite{meinshausen2006quantile}),
\begin{equation}\label{eq: QRF weighted}
\hat{F}_e(z \mid \vect{X}=\vect{x}) = \sum_{t=1}^{T} w_t(\vect{x}) \mathbf{1}\{e_t \leq z\}.
\end{equation}
\begin{assumption}\label{as: QRF2}
The weights in~\cref{eq: QRF weighted} satisfy that for all $\vect{x} \in \mathbb{B}$, $w_t(\vect{x}) = O(1/T)$.
\end{assumption}

\begin{assumption}\label{as: QRF3}
The true conditional distribution function of the scores is Lipschitz continuous,
\[
\sup_{z} |F_e(z \mid \vect{X} = \vect{x}) - F_e(z \mid \vect{X}=\vect{x}')| \leq L \lVert \vect{x} - \vect{x}' \rVert_{1}.
\]
\end{assumption}

\begin{assumption}\label{as: QRF4}
For every $\vect{x}$ in the support of $\vect{X}$, the conditional distribution function of the score $F_e(z \mid \vect{X} = \vect{x})$ is continuous and strictly monotonically increasing in $z$. 
\end{assumption}

Assumptions \ref{as: QRF1}, \ref{as: QRF3}, and \ref{as: QRF4} hold with the CHCDS score and a Normal mixture model assuming that the mixing weights, component means, and component variances are Lipschitz continuous in $\vect{X}$, and that the component variances are bounded away from zero. Assumption \ref{as: QRF2} depends on the forest construction and can be verified empirically to ensure that no weights concentrate excessively.

\begin{theorem}\label{thm:coverage}
    When assumptions \ref{as: QRF1} - \ref{as: QRF4} hold, the prediction sets output by SCDR, both the leave-one-out approach and the bootstrap approach, achieve asymptotic $1 - \alpha$ conditional coverage as defined in~\cref{eq:conditional_asymp_cov}. 
\end{theorem}

We omit the proof of the above theorem as it is similar to that of \citealt[Theorem 2]{SPCI_Xu_xie_2023}. 

The following 5 assumptions are required for the doubly robustness property to hold.  
\begin{assumption}\label{as: location scale}
$Y \mid \vect{X}$ follows a location-scale family of the following form:
\[
Y_t = \mu(\vect{X}_t) + \sigma(\vect{X}_t) \epsilon_t,
\]
where the innovations are independent and identically distributed, $\epsilon_t \overset{i.i.d.}{\sim} P$, with common probability density function $g(\cdot)$. 
\end{assumption}
\begin{assumption}\label{as: dr1}
The sample space of $(Y, \vect{X})$ is compact and of the form $\mathcal{K} = \mathcal{Y}\times \mathcal{X} \subseteq \mathbb{R} \times \mathbb{R}^p$.
\end{assumption}
\begin{assumption}\label{as: dr2}
The conditional density estimator is uniformly consistent for the population conditional density, 
\[
\sup_{(y, \vect{x}) \in \mathcal{K}} \left| \hat{f}(y \mid \vect{x}) - f(y\mid \vect{x}) \right| = o_p(1)
\]
\end{assumption}
\begin{assumption}\label{as: dr3}
The conditional density, $f(y \mid \vect{x})$ is continuous in $(y, \vect{x})$.
\end{assumption}
\begin{assumption}\label{as: dr4}
Define $H(c\mid \vect{x})$ as the probability mass of the density level set at threshold $c$. 
\[
H(c \mid \vect{x}) = \int_{ \{ y: f(y \mid \vect{x}) \geq c \} }f(y \mid \vect{x})dy
\]
For every $\vect{x} \in \mathcal{X}$ and for all $c > 0$, the Lebesgue measure of the set $\{ y: f(y \mid \vect{x}) = c \}$ is zero. Consequently, $H(c \mid \vect{x})$ is continuous and strictly monotonic in $c$ for all $c$ such that $0< H(c\mid \vect{x})<1$.
\end{assumption}

We note that Assumption~\ref{as: dr1} and Assumption~\ref{as: dr3} are mild regularity conditions. Assumption~\ref{as: dr4} is required to ensure that there are no flat spots of the density near the $1 - \alpha$ density level cutoff.

\begin{theorem}\label{thm:density set convergence}
    Under assumptions ~\ref{as: dr1} - \ref{as: dr4}, when the conditional density estimator converges to the population conditional density uniformly, the density level-set cutoff estimate, $\hat{c}(\vect{x})$, converges uniformly to the population density level-set cutoff, $c(\vect{x})$. 
\end{theorem}

\begin{theorem}\label{thm:doubly robust}
    Under assumptions ~\ref{as: QRF1} - \ref{as: dr4} we have doubly robust asymptotic conditional coverage. When either the conditional density estimator converges to the conditional density uniformly or the QRF is correctly specified (or both), SCDR using  the leave-one-out approach achieves asymptotic $1 - \alpha$ conditional coverage as defined in~\cref{eq:conditional_asymp_cov}. 
\end{theorem}

A formal proof of this is given in the supplementary materials Section B. The result of this doubly robust property can be seen through the following high level argument. First, if the QRF is correctly specified, then the asymptotic conditional coverage holds due to~\cref{thm:coverage}. 

Second, if the conditional density estimator is correctly specified, then the density level-set cutoffs are correct due to~\cref{thm:density set convergence}. When the data generating process is a location-scale family ($Y_t = g(\vect{X}_t) + \sigma(\vect{X}_t)\epsilon_t$), the ratio of the conditional density divided by the density level-set cutoffs removes the conditional heteroskedasticity while the conditional density itself removes any dependence on the conditional mean, resulting in an i.i.d.\@ sequence. Since there is no dependence in an i.i.d.\@ sequence, the QRF can only be correctly specified or overspecified, and~\cref{thm:coverage} again applies. 

\begin{theorem}\label{thm:smallest set}
Under the same assumptions as~\cref{thm:doubly robust}, the prediction sets output by SCDR asymptotically achieve the minimum Lebesgue measure among all prediction sets that achieve $1-\alpha$ conditional coverage.
\end{theorem}



Our approach differs from that in \cite{SPCI_Xu_xie_2023} because we don't require the scores to be regression residuals. We note that as long as assumptions~\ref{as: QRF1}-~\ref{as: QRF4} hold, the scores can be from any conformal approach. For example, this approach can also be used with quantile regression, including unconformalized prediction intervals, and the conformalized quantile regression score \citep{romanocqr}. In this paper we use conditional densities instead of a regression model to ensure that we have small (and hence informative) prediction sets. 

No other method, to our knowledge, has a doubly robust property for conformal prediction with time-series data. SPCI is the only other method that has a conditional coverage guarantee, though in the numerical results we show that SCDR achieves coverage close to the nominal $(1-\alpha)\times 100\%$ in much smaller sample sizes. In the supplementary materials we prove that the doubly robust property holds under slightly different conditions for other classes of scores, including both signed residuals and conformalized quantile regression scores. 

SCDR is also unique because it is the only, to our knowledge, method that has a formal guarantee on the prediction set size. Both BCI and SPCI attempt to minimize the length of the prediction intervals they output but they: 1. lack any theoretical guarantee on the prediction interval size and 2. only output prediction intervals while SCDR can output both prediction intervals and prediction sets, depending on the model used as well as the underlying data structure. 

BCI has the following marginal guarantee, as long as the prediction intervals are monotonic in the coverage rate, 
\[
\Bigg|\frac{1}{K}\sum_{t=m}^{K+m}\text{err}_t - \alpha\Bigg| \leq \frac{c+1}{cK},
\]
where $\gamma = c\lambda_{max}$. We note that this bound can be very large, for example when $\gamma = 0.6$, $\lambda_{max} = 20$, and $K = 500$, $c = 0.03$ and the bound is $0.07$, or a difference of 7\% between the desired and observed coverage. In many of the scenarios in \cite{bellman_conformal_yan_candes_2024}, as well as ours, $\lambda_{max}$ is much larger than 20, and the bound is larger than 7\%. 

PID guarantees asymptotic marginal coverage as long as the scores are bounded,
\[
\Bigg|\frac{1}{T}\sum_{t=1}^{T}\text{err}_t - \alpha\Bigg| \leq \frac{ch(T) + 1}{T},
\]
where $x \geq ch(t) \implies r_t(x) \geq b$ and $x \leq -ch(t) \implies r_t(x) \leq -b$ for a saturation function $r_t(x)$ and constants $c, b > 0$. This coverage bound is much stronger than that of BCI, which is corroborated by our numerical results. Though, PID without further assumptions cannot guarantee conditional coverage because of its reactive response to a miscoverage. 

\subsection{Density Consistency Results}\label{sec: density consistency}

The proofs of the following results can be found in the supplementary materials. 

Let $\bfW^\top=(\bfY, \bfX^\top)$. Let $f_{\bfW}(\cdot), f_{\bfX}(\cdot)$ denote the joint pdf of $\bfW$ and that of $\bfX$, respectively. Then the conditional pdf of $\bfY$ given $\bfX$ is given by $f_{\bfY|\bfX}(\bfy|\bfx)=f_{\bfW}(\bfw)/f_\bfX(\bfx)$. We consider approximating the unknown true $f_{\bfW}$ by a  multivariate normal mixture of the following form: 
\begin{eqnarray*}
    f(\bfw|F)&=& \int_S \phi(\bfw|\mu_{\bfw}, \Sigma_{\bfw, \bfw})  dF(\mu_W, \Sigma_{\bfW, \bfW})  \\
    &=&\int_S \frac{1}{(2\pi)^{(p+1)/2}|\Sigma_{\bfw,\bfw}|^{1/2}} \exp(-(\bfw-\mu_\bfw)^\top\Sigma_{\bfw, \bfw}^{-1} (\bfw-\mu_\bfw)/2) dF(\mu_\bfW, \Sigma_{\bfW, \bfW}),
\end{eqnarray*}
where $F$ is a probability distribution on the parameter space $S=S_a=\{(\mu_\bfW, \Sigma_{\bfW, \bfW}): \|\mu_\bfW\|\le a, \|\Sigma_{\bfW, \bfW}\|_F \le a,  \Sigma_{\bfW, \bfW} \mbox{ is positive definite, whose eigenvalues lie between } 0< \underline{\lambda}< \overline{\lambda}< \infty\}$, where $\|\cdot\|_F$ denotes the Frobenius norm of the matrix and $\phi(\cdot|\mu, \Sigma)$ is the pdf of the multivariate normal distribution with mean vector $\mu$ and covariance  matrix $\Sigma$. The normal mixture pdf for $\bfW$ induces a mixture pdf estimate for $\bfX$, specifically, $f_{\bfX}(\bfx|F)=\int f_{\bfW}(\bfw)d\bfy$ and hence an estimate for the conditional pdf of $\bfY$ given $\bfX=\bfx$: $f(\bfy|\bfx; F)=f(\bfw|F)/f(\bfx|F)$ where we write $f(\bfw|F)$ and $f(\bfx|F)$ for $f_{\bfW}(\bfw|F)$ and $f_{\bfX}(\bfx|F)$, respectively, i.e., the argument signifies the function.   It is assumed that the bound $a\le L \left( \log\frac{1}{\epsilon}\right)^\gamma$ and $\gamma\ge 1/2$ and $L>0$ are constants. The eigenvalue bounds $0<\underline{\lambda} <\overline{\lambda}<\infty$ are fixed constants.  This eigenvalue condition is assumed henceforth. Our goal is to quantify the proximity of the approximate conditional density to its true counterpart.
\begin{theorem}~\label{Thm: convergence rate for the marginal pdf estimator}
    Let $\hat{f}_n$ be the MLE with the support of its mixing distribution inside $ S_a$ with $a\lesssim (\log n)^\gamma$, where $\gamma\ge 1/2$, any eigenvalue of the covariance matrices admitted by $S_a$ falls within  $[\underline{\lambda},\overline{\lambda}]$ where  $0<\underline{\lambda}<\overline{\lambda}<\infty$, and $f_0$ be the true pdf.
Suppose that  the process $\{\bfW_t\}$ is a stationary $\beta$-mixing sequence with the mixing rate $\beta(j)\le \beta_0 j^{-\zeta}$ for some $\beta_0>0, \zeta>2$, and that the stationary marginal distribution of $\bfW_t$ is a normal mixture with the support of the true mixing distribution $F_0$ being a compact subset of $S_a$ for some $a>0$. Let $f_0$ denote the true pdf. Then, for any compact set $\mathcal{K}\subset\mathbb{R}^p$,
 \begin{equation}
    \int \int |\hat{f}_n(\bfy|\bfx)- f_0(\bfy|\bfx)|^2 I(\bfx\in \mathcal{K}) d\bfy d\bfx=O_p(\epsilon_n^2)
    \label{Eq: convergence rate}
    \end{equation}
    where $\epsilon_n= (\log n)^{(p+1)\max(\gamma, 1/2)+1}/\sqrt{n}$.
    
    Furthermore, $\hat{f}_n(\bfy|\bfx) \to f_0(\bfy|\bfx)$ uniformly for $(\bfy, \bfx^\top)^\top$ in a compact set, as $n\to\infty$, in probability. Thus, for all $\bfx\in \mathcal{K}$ and bounded subset $B$ of the real line, $\int_B\hat{f}_n(\bfy|\bfx)d\bfy\to \int_B f_0(\bfy|\bfx)d\bfy$ as $n\to \infty$, in probability.  
\end{theorem}

The following result shows that in practice we can approximate a conditional distribution by that of a finite normal mixture with the number of mixing components of the order $O_p((\log n)^{p+1})$ where $p$ is the dimension of the features, $\bfX_t$. The result quantifies the curse of dimensionality as the number of components increases exponentially with the feature dimension. 
\begin{theorem}
Suppose that all the assumptions of   Theorem~\ref{Thm: convergence rate for the marginal pdf estimator} hold, with $\gamma=1/2$. Then the conclusions of Theorem~\ref{Thm: convergence rate for the marginal pdf estimator}  still hold for the sieve MLE whose mixing distribution has  at most $O_p((\log n)^{p+1})$ atoms in $S_a$.  
\end{theorem}

\section{Numerical Results}\label{sec:numerical_results}
\subsection{Non-stationary Simulation Studies}\label{sec:sim_studies}
In the first two simulation studies, we follow \citet{SPCI_Xu_xie_2023} by generating a non-stationary time-series of the following form,
\[
Y_t = f({X}_t) + \epsilon_t,
\]
where 
\begin{align*} 
    f({X}_t) &= g(t)h({X}_t), \>
    g(t) = \log(t') \sin(2\pi t' / 12), \> t' =\text{mod}(t, 12), \> {X}_t = Y_{t-1}, \\ h(t) &= ({X}_t + {X}_t^2 + {X}_t^3)^{1/4}, \>
    \epsilon_t = 0.6 \epsilon_{t-1} + e_t, \text{ where } e_t \overset{i.i.d.}{\sim} \mathcal{N}(0, 1).
\end{align*} 

The simulation size was $1,000$, with $T = 2,000$ initially observed data points, $\alpha = 0.10$, a bootstrap number of $B = 30$, the mean as the bootstrap aggregator function, $\phi$, and 5 new data points in each simulation with one step ahead prediction. That is, 2,000 data points were initially observed and prediction sets were formed for the next data point. That data point was then observed, coverage and size were recorded for each method, and new prediction sets were then formed for the 2,002nd data point. This was repeated for 5 ``new" data points. The bootstrap approach was used for both SCDR and SPCI. For all initial conditional point or conditional density estimators, the predictors were $X_t$, and $t'$. For SPCI, the point estimator was computed with a random forest model. For SCDR, the joint density was estimated with a Normal mixture with up to 4 components. The quantile model used for both approaches was a random forest quantile regression trained on the previous 100 residuals, with the past 5 residuals as predictors. The initial prediction intervals computed with PID were formed with an ARMA model, where $p$ and $q$ were chosen using the auto.arima function the \textit{forecast} R package \citep{hyndman_forecast_pkg}. Following \cite{PID_angelopoulos, wang_hyndman_multistepTS_2024} the signed residual score was used with a Theta model as the scorecaster with a learning rate of $0.01 \hat{\beta}_t$, where $\hat{\beta}_t = \max\{V_{t - \Delta + 1}, \ldots, V_{t - 1}\}$ is the highest score over a trailing window of length $\Delta = 100$. We also used a saturation function of $r_t(x) = 30\tan(x\log(t) / (0.53t))$. An AR(5) model was used with BCI, selected via the auto.arima function in the \textit{forecast} R package assuming Normal innovations for the initial prediction intervals \citep{hyndman_forecast_pkg}. The maximum $\lambda$ value was 200, with an intial value of $\lambda$ set to be 4, $\gamma = 0.6$, and $B = 100$. The simulation results are given in~\cref{tab:tt_1000_bootstrap}. Conditional coverage for the values of $t' = 8, 9, 10, 11, 0$ can be found in~\cref{tab:tt_boot_conditional}.

A second simulation was performed with the same simulation size. This time the observed sample size was decreased to $T = 500$, and the leave-one-out approach, instead of the bootstrap approach, was used for both SCDR and SPCI. One new data point was generated to check coverage in each simulation with one step ahead prediction. For the initial conditional point or conditional density estimator, the predictors were $X_t$, and $t'$. For SPCI, the point estimator was computed with a random forest model. For SCDR, a joint density estimator was fit using a Normal mixture model with up to 3 components. The quantile model used for both approaches was a random forest quantile regression trained on the previous 100 residuals, with the past 5 scores as predictors. The setup for PID was used in this simulation was the same as the previous simulation, except the saturation function was  $r_t(x) = 30\tan(x\log(t) / (0.55t))$. The setup for BCI was the same as in the previous simulation scenario. The results are given in~\cref{tab:tt_500_loo}.

\begin{table}
\begin{center}
\caption{2,000 Observed Data Points, Bootstrap Approach nout = 5}

\begin{tabular}{|c|l l|}
\hline
Method & Coverage & Size \\
\hline
SPCI & 0.866 (0.005) & 4.909 (0.038) \\
PID &  0.931 (0.003) & 7.716 (0.076)\\
BCI & 0.844 (0.005) & 3.899 (0.029) \\
SCDR & 0.883 (0.005) & 3.565 (0.015) \\
\hline
\end{tabular}\label{tab:tt_1000_bootstrap}
\end{center}
\end{table}

\begin{table}
\begin{center}
\caption{2,000 Observed Data Points, Conditional Coverage for Varying Values of t'}

\begin{tabular}{|c|l l l l|}
\hline
t' & SPCI & PID & BCI & SCDR \\
\hline
8 & 0.825 (0.012) & 0.967 (0.006) & 0.953 (0.007) & 0.905 (0.009) \\
9 &  0.829 (0.012) & 0.951 (0.007) & 0.904 (0.009) & 0.877 (0.010) \\
10 & 0.877 (0.010) & 0.902 (0.009) & 0.829 (0.012) & 0.879 (0.010) \\
11 & 0.910 (0.009) & 0.897 (0.010) & 0.595 (0.016) & 0.884 (0.010) \\
0 & 0.890 (0.010) & 0.938 (0.008) & 0.930 (0.008) & 0.868 (0.011) \\
\hline
\end{tabular}\label{tab:tt_boot_conditional}
\end{center}
\end{table}

\begin{table}
\begin{center}
\caption{500 Observed Data Points, Leave One Out Approach nout = 1}

\begin{tabular}{|c|l l|}
\hline
Method & Coverage & Size \\
\hline
SPCI & 0.833 (0.012) & 3.872 (0.024) \\
PID & 0.873 (0.011) & 6.721 (0.057) \\
BCI & 0.951 (0.007) & 3.923 (0.034) \\
SCDR & 0.894 (0.010) & 4.828 (0.042) \\
\hline
\end{tabular}\label{tab:tt_500_loo}
\end{center}
\end{table}

None of the models in the non-stationary simulation (outside of the Random Forest point estimator used with SPCI) are able to handle the non-linearity of the data very well. Though, through the results we can see that SCDR has the closest coverage to the nominal 90\% in both scenarios while also having the smallest prediction region size. The coverage is also less variable over differing values of $t'$, while the other methods have conditional coverage that differs by between 7\% to over 30\%. 

\subsection{AR Simulation Study}

A third simulation study was conducted using an AR(2) data generating process,
\[
Y_t = 0.5 Y_{t-1} - 0.3 Y_{t-2} + \epsilon_t,
\]
where $\epsilon_t \overset{i.i.d.}{\sim}t_{3}$.

The initial observed sample size was 300 data points, with 10 out of sample data points generated for one-step ahead prediction, similar to the first simulation scenario, but with 10 ``new" and sequentially observed data points instead of 5. The simulation size was 500. All simulations used the same parameters as in the second simulation scenario, except the initial model for SPCI, PID, and BCI, which used an AR(2) model with estimated AR coefficients using the ordinary least squares method, and initial prediction intervals formed assuming Normal innovations. The simulation results are given in~\cref{tab:AR2}. 

\begin{table}
\begin{center}
\caption{AR(2) Simulation Results}

\begin{tabular}{|c|l l|}
\hline
Method & Coverage & Size \\
\hline
SPCI & 0.861 (0.005) & 4.987 (0.059) \\
PID & 0.891 (0.004) & 5.553 (0.057) \\
BCI & 0.803 (0.007) & 2.707 (0.035) \\
SCDR & 0.882 (0.005) & 5.514 (0.087) \\
\hline
\end{tabular}\label{tab:AR2}
\end{center}
\end{table}

The results of this simulation demonstrate that SCDR is just as good as competing methods when the competing methods begin with the oracle point estimates. It also demonstrates that the objective function BCI optimizes can be very sensitive to the scale of the response. 

\subsection{Real Data Analyses}
\subsubsection{Geyser}
\label{sec:geyser}

In our first real data analysis we look at Old Faithful Geyser eruption durations and waiting times collected continuously from August 1st until August 15th, 1985, which contains 298 observations \citep{geyser_azzalini_1990}. The response was the eruption duration, the features used are the previous eruption duration and the previous waiting time. All models and tuning parameters used are the same as in the second leave-one-out simulation, except for PID where the saturation function was $r_t(x) = 30\tan(x\log(t) / (0.52t))$. The past 100 residuals were used to fit the quantile model for both methods, with the previous 3 residuals as predictors.

\begin{figure}[ht]
    \centering
\includegraphics[scale = 0.6]{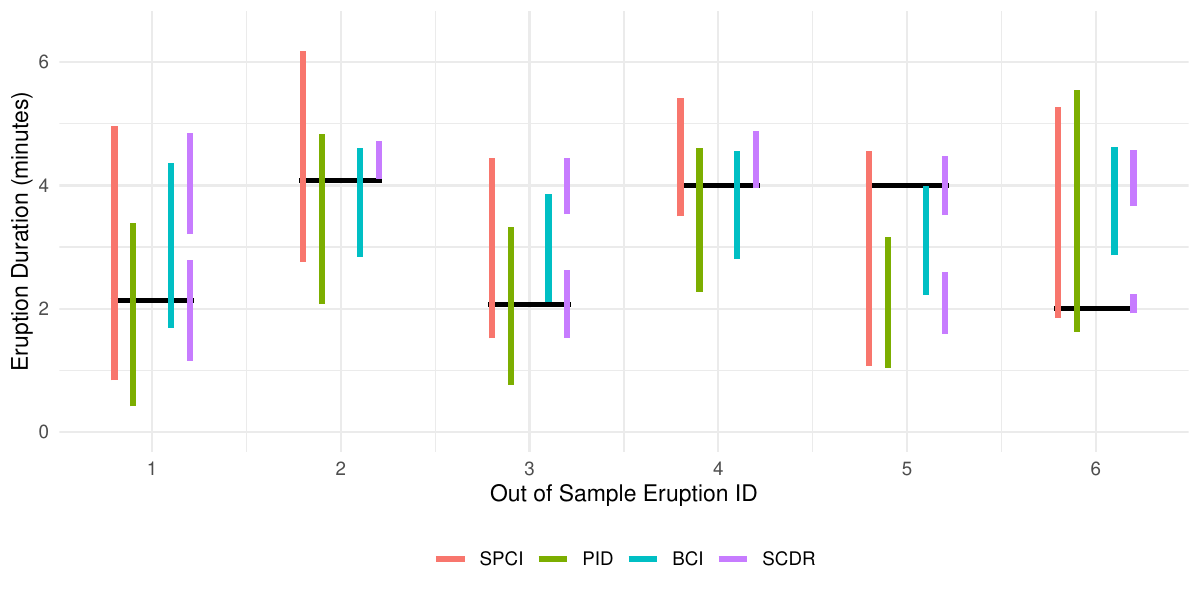}
\caption{Prediction Regions for 6 Randomly Selected Eruption Durations. They are from left to right: SPCI, PID, BCI, and SCDR.}\label{fig:geyser_coverage}
\end{figure}

We treated 200 of the data points as observed, and made sequential one step ahead predictions for the remaining 98 data points.  The nominal coverage level was set to be 90\%. The results can be found in~\cref{tab:geyser_dat}. Examples of the prediction sets for 6 of the predictions can be found in~\cref{fig:geyser_coverage}. 64 of the out of sample points had eruption durations larger than 3.5 minutes, 34 had eruption durations less than 3.5 minutes. 

\begin{table}
\begin{center}
\caption{Old Faithful Eruption Duration Predictions}

\begin{tabular}{|c|l l l l|}
\hline
Method & Coverage & Size & Cov $|\, \text{Duration}> 3.5$ & Cov $|\, \text{Duration} \leq 3.5$ \\
\hline
SPCI & 0.837 (0.038) & 2.909 (0.079) & 0.844 (0.066) & 0.824 (0.046)\\
PID & 0.898 (0.031) & 4.069 (0.099) & 0.922 (0.034) & 0.853 (0.062) \\
BCI & 0.837 (0.038) & 2.199 (0.056) & 0.938 (0.030) & 0.647 (0.083)\\
SCDR  & 0.908 (0.029) & 1.837 (0.078) & 0.922 (0.034) & 0.882 (0.056)\\

\hline
\end{tabular}\label{tab:geyser_dat}
\end{center}
\end{table}

\subsubsection{Electricity}

The second real dataset we analyzed was the quantity of electricity transferred between New South Wales, Australia and Queensland, Australia updated every 30 minutes in a 2.5 year period between 1996-1999 \citep{electricity_data}. For all four methods, the previous 4 transfer values as well as the price and demand in both states were used as features. The models and tuning parameters used were the same as in the geyser data analysis for PID, BCI, and SPCI. All features as well as the response were standardized to be between $0$ and $1$ with a mean of $0.5$. 

For SCDR, the previous 1,000 data points were used to train a Normal Mixture density model with up to 8 mixtures. For both SPCI and SCDR, the previous 250 residuals were used to train the QRF with 10 lag residuals used as the features.
The nominal coverage level was set to be 90\%. Marginal coverage and length, and the coverage when the demand in Victoria was above and below the third quartile can be seen in~\cref{tab:elect_dat}. 

It is clear from both of the real data analyses that SCDR consistently has the smallest prediction set sizes while maintaining near marginal and conditional coverage. PID also maintains marginal coverage, but has much larger prediction sets (2-3 times) without providing a guarantee on the conditional coverage. Both SPCI and BCI fail to have marginal coverage that is near the nominal 90\%. 

\begin{table}
\begin{center}
\caption{Electricity Predictions}

\begin{tabular}{|c|l l l l|}
\hline
Method & Coverage & Size & Cov $|\, \text{Vic Demand}> 0.554$ & Cov $|\, \text{Vic Demand} \leq 0.554$ \\
\hline
SPCI & 0.859 (0.010) & 0.133 (0.001) & 0.890 (0.017) & 0.848 (0.011)\\
PID & 0.900 (0.008) & 0.632 (0.015) & 0.905 (0.016) & 0.899 (0.010) \\
BCI & 0.832 (0.010) & 0.306 (0.002) & 0.829 (0.011) & 0.833 (0.006)\\
SCDR & 0.892 (0.009) & 0.121 (0.001) & 0.872 (0.019) & 0.899 (0.010)\\

\hline
\end{tabular}\label{tab:elect_dat}
\end{center}
\end{table}

\subsection{Doubly Robust Demonstration}

In this section, we demonstrate the doubly robust properties of SCDR with a simple AR(1) model,
\[
Y_t = 0.5Y_{t-1} + \epsilon_t,
\]
where $\epsilon_t \sim \mathcal{N}(0, 1)$. 

First, we use a Normal density that uses the correct mean and variance to form prediction regions for $Y_{t+1}$, that is we use the density of a $\mathcal{N}(0.5 Y_{t}, 1)$ to create prediction regions for $Y_{t+1}$. For the conformal adjustment, we use the empirical quantiles of the non-conformity scores, with no conditional quantile random forest adjustment. We refer to this approach as the correct conditional density with unadjusted scores.

Second, we use a Normal density with an incorrectly specified mean and variance to form prediction regions for $Y_{t+1}$, $\mathcal{N}(0.6 Y_{t}, 0.8^2)$. For the conformal adjustment, we use a conditional quantile random forest with the past 5 scores as covariates, trained on the previous 130 scores. We refer to this approach as the incorrect conditional density with quantile adjusted scores approach.

Third, we use a Normal density with the correctly specified mean and variance with the conformal adjustment coming from a conditional quantile random forest with the past 5 scores as covariates, trained on the previous 130 scores. We refer to this approach as the correct conditional density and quantile adjusted scores.

Finally, we use the incorrectly specified model with no quantile adjustment; the prediction regions were the unadjusted 90\% highest density sets from a Normal distribution,  $\mathcal{N}(0.6 Y_{t}, 0.8^2)$.

For all scenarios, the simulation size was 1,000. Within each simulation, the initial observed sample size was 150 data points, with 10 out of sample data points generated for one-step ahead prediction. The results can be found in~\cref{tab:DR_demo}.

These results clearly demonstrate that when either the conditional density estimator or the QRF are correctly specified, SCDR achieves coverage close to $100 \times (1 - \alpha)\%$. We can also see that the prediction set sizes are smaller when the correct conditional density is used.

\begin{table}
\begin{center}
\caption{Doubly Robust Demonstration}

\begin{tabular}{|c|ll|}
\hline
Method & Coverage & Size \\
\hline
Correct Conditional Density \& Unadjusted Scores & 0.904 (0.003) & 3.356 (0.008) \\
Incorrect Conditional Density \& Quantile Adjusted Scores & 0.882 (0.003) & 3.851 (0.012) \\
Correct Conditional Density \& Quantile Adjusted Scores & 0.881 (0.003) & 3.231 (0.009)\\
Incorrect Conditional Density \& No Adjustment & 0.808 (0.004) & 3.006 \\
\hline
\end{tabular}\label{tab:DR_demo}
\end{center}
\end{table}

\section{Conclusion}\label{sec:conclusion} SCDR proposes a new approach to building conformal prediction sets for univariate time-series data. Theoretical results state that the proposed approach should achieve both nominal marginal and conditional coverage while maintaining small prediction sets. The numerical results demonstrate that SCDR achieves coverage (both marginal and conditional) near the nominal target with significantly smaller set sizes than existing methods. The real data analyses also demonstrates that SCDR is able to capture multimodality and output prediction sets that are smaller and more descriptive than prediction intervals output by other methods. It is also capable of outputting sharp prediction intervals when the target is unimodal. 

There are two future areas of exploration related to SCDR. The first involves proving consistency results for other conditional distributions to broaden the range of models that can be used with SCDR. For example, justifying the use of conditional normalizing flows would allow the modeling of highly complex datasets where Gaussian mixtures are insufficient. A second area for future research is the development of joint $H$-step ahead prediction regions. While Section A in the supplementary materials introduces an extension of the SCDR method that provides an asymptotic guarantee on the individual prediction sets, we believe certain problems benefit from a coverage guarantee on the joint prediction sets over $H$ steps. Although our framework can accommodate multivariate responses, visualizing and updating these sets as data are sequentially observed remains a non-trivial challenge.

\newpage

\bibliography{ref}

\end{document}


\def\spacingset#1{\renewcommand{\baselinestretch}%
{#1}\small\normalsize} \spacingset{1}


\if0\blind
{
  \title{\bf Appendix for SCDR }
  \author{Max Sampson \\
    Department of Statistics and Actuarial Science, University of Iowa\\
    and \\
    Kung-Sik Chan \\
    Department of Statistics and Actuarial Science, University of Iowa}
  \maketitle
} \fi

\if1\blind
{
  \bigskip
  \bigskip
  \bigskip
  \begin{center}
    {\LARGE\bf Title}
\end{center}
  \medskip
} \fi

\bigskip
\begin{abstract}
Section A concerns the $H$-step ahead prediction with SCDR. Section B contains all proofs of the results given in Section 3 of the main paper, including the doubly robust property of SCDR as well as the consistency of the sieve MLE density estimator. Section C contains ACF \& PACF plots of the scores from the main paper. 
\end{abstract}

\noindent%
{\it Keywords:} Conformal Predictive Inference; Doubly Robustness; Highest Predictive Density Sets;
  Time Series Prediction;
  Uncertainty Quantification. \vfill

\newpage
\spacingset{1.8} 

\appendix

\appendix

\section{Multi-step Ahead Prediction}

Below we outline one approach to extending SCDR from being a 1-step ahead sequential prediction framework for time-series data to an $H$-step ahead sequential prediction framework.

As with SCDR for sequential 1-step ahead prediction, the $H$-step ahead prediction begins by fitting leave-one-out conditional density models by iterating through the observed data:
\[
\hat{f}_{-j} = \mathcal{B}( \{ (\vect{X}_i, Y_i) : i \neq j \} ).
\]
Next, we construct $H$ score vectors. The scores we calculate on the training data reflect the scores we wish to predict, so when finding the $j$-th scores, we must replace the unobserved lags in the covariate vector, $\vect{X}_j$, with a point estimate $\hat{y}_{j, h}$ for $h = 2, \ldots, H$. Denote the recursive covariate vector for the $h$-step prediction at time $j$ as $\tilde{\vect{X}}_{j, h}$. We estimate $\hat{y}_{j, h}$ using the estimated conditional density: 
\[
\hat{y}_{j, h} = \int y \hat{f}_{-j}(y \mid \tilde{\vect{X}}_{j, h}) dy \quad h = 1,\ldots, H \quad \text{and} \quad j = 1,\ldots, T.
\]
Next, with our recursive covariate vector and LOO conditional density estimate, we find the unadjusted $1 - \alpha$ upper density level sets, $\hat{c}(\tilde{\vect{X}}_{j, h})$. 
We then compute the scores. Let $V_{j, h}$ denote the $j$-th score for the $h$-step ahead prediction sets:
\[
V_{j, h} = \frac{\hat{f}_{-j}(Y_j \mid \tilde{\vect{X}}_{j, h})}{\hat{c}(\tilde{\vect{X}}_{j, h})} \quad h = 1, \ldots, H \quad \text{ and } \quad j = 1,\ldots, T.
\]

With all of the LOO scores computed, a conditional density estimator is fit with all of the observed training data:
\[
\hat{f}_{T+1} = \mathcal{B}( \{ (\vect{X}_i, Y_i) : i =1, \ldots, T\} ).
\]
Next, to predict the $\alpha$ quantile of each of the $H$ score vectors, $H$ separate QRF models are fit using a training window of size $k$:
\[
\hat{Q}_h = \mathcal{Q}( \{(V_{i, h}, [V_{i-1, h}, \ldots, V_{i - w, h}], \alpha):i= T,\ldots, T - k \}) \text{ for } h = 1, \ldots, H.
\]
These models are then used to estimate the $\alpha$ quantile of the new scores for each of the $H$ steps:
\[
\hat{q}_{T+h} = \hat{Q}_h(V_{T, h}, \ldots, V_{T-w+1, h}) \quad h = 1,\ldots, H.
\]
Next, $\hat{c}( \tilde{\vect{X}}_{T+h, h})$ is estimated for $h = 1,\ldots, H$ using $\hat{f}_{T+1}$. The resulting $h$-step ahead SCDR prediction set is then:
\[
\hat{C}(\tilde{\vect{X}}_{T+h, h}) = \{y: \hat{f}_{T+1}(y \mid \tilde{\vect{X}}_{T+h, h}) > \hat{c}(\tilde{\vect{X}}_{T+h, h}) \times \hat{q}_{T+h} \}
\]
As new data is observed, the scores and models are sequentially updated in the same way. An algorithm for $H$-step ahead prediction is given in~\cref{alg:SCDR_Hstep}. 

Two simulations were done to demonstrate the method for $5$-step ahead prediction using the following AR(2) data generating process:
\[
Y_t = 0.5 Y_{t-1} - 0.3 Y_{t-2} + \epsilon_t,
\]
where $\epsilon_t \overset{i.i.d.}{\sim}t_{10}$ in the first simulation and $\epsilon_t \overset{iid}{\sim}\mathcal{N}(0, 1)$ in the second simulation. 

The initial observed sample size was 300 data points, with 100 out of sample data points generated for $5$-step ahead prediction. That is, we initially observed 300 data points. Data were observed in batches of 5, so prediction sets were formed for 1, 2, 3, 4, and 5 step ahead prediction at each time point. The simulation size was 500. A joint density estimator was fit using a Normal mixture model with up to 3 components. The quantile model used for both approaches was a random forest quantile regression trained on the previous 100 residuals, with the past $w = 5$ scores as predictors.

The simulation results are given in~\cref{tab:ARHstep} and~\cref{tab:ARHstep2}. 

\begin{table}[ht]
\begin{center}
\caption{H-step Ahead Simulation Results $t_{10}$}
\begin{tabular}{|c|l l|}
\hline
Method & Coverage & Size \\
\hline
SCDR (1-step) & 0.871 (0.003) & 3.703 (0.016) \\
SCDR (2-step)& 0.876 (0.003) & 4.061 (0.016) \\
SCDR (3-step)& 0.873 (0.003) & 4.044 (0.015) \\
SCDR (4-step) & 0.874 (0.003) & 4.087 (0.015) \\
SCDR (5-step) & 0.873 (0.003) & 4.102 (0.015) \\
\hline
\end{tabular}\label{tab:ARHstep}
\end{center}
\end{table}

\begin{table}[ht]
\begin{center}
\caption{H-step Ahead Simulation Results $\mathcal{N}(0, 1)$}
\begin{tabular}{|c|l l|}
\hline
Method & Coverage & Size \\
\hline
SCDR (1-step) & 0.879 (0.003) & 3.268 (0.011) \\
SCDR (2-step)& 0.877 (0.003) & 3.634 (0.012) \\
SCDR (3-step)& 0.876 (0.003) & 3.637 (0.013) \\
SCDR (4-step) & 0.882 (0.003) & 3.679 (0.012) \\
SCDR (5-step) & 0.873 (0.003) & 3.682 (0.012) \\
\hline
\end{tabular}\label{tab:ARHstep2}
\end{center}
\end{table}

From the results, we can see that the asymptotic coverage guarantees are still valid for this $H$-step ahead SCDR method. The prediction set length does not vary highly for the 2nd through 5th steps because the AR(2) process is simple enough to be well estimated by a Gaussian mixture. In fact, the results are very similar to the length of symmetric prediction intervals when derived using the Yule-Walker equations with Normal innovations, which are 3.290 for a 1-step ahead prediction interval and 3.678 for a 2-step ahead prediction intervals. The prediction set sizes found using the $H$-step ahead SCDR method with t-distributed innovations are larger than those suggested from the Yule-Walker equations (about 3.241 and 3.597, respectively) because a 3-component Gaussian mixture is an underspecified model for $(Y_t, Y_{t-1}, Y_{t-2})$ when the innovations are not Gaussian. 

We note that the $H$-step ahead SCDR approach is more computationally expensive because it requires fitting $H$ QRF models. Though, these models are not very computationally expensive and fitting/predicting with them is quite fast. 

\newpage

\begin{algorithm}
    \caption{SCDR H-Step Ahead LOO}\label{alg:SCDR_Hstep}
{\renewcommand{\baselinestretch}{0.8}\selectfont
\textbf{Input:} miscoverage level $\alpha$, initial data $(Y_t, \vect{X}_t)_{t=1}^{T}$, conditional density algorithm $\mathcal{B}$, quantile regression algorithm $\mathcal{Q}$, autoregressive order $w$, and block update size $H$. \par}
\textbf{Procedure:}
\begin{algorithmic}[1]
    \scriptsize
    \State Initialize score matrix $\mathbf{V} \in \mathbb{R}^{T \times H}$
    \For{($j = 1, \ldots, T$)} \Comment{LOO Training for Initial Matrix}
        \State Fit $\hat{f}_{-j} = \mathcal{B}(\{(\vect{X}_i, Y_i):i \neq j\})$ 
        \State Initialize recursive estimates $\hat{y}_{est} = \{\}$ 
        \Comment{to ensure that we aren't training the QRF on scores that used the true response when we only have an estimate}
        \For{($h = 1, \ldots, H$)}
        \State Construct input $\tilde{\vect{X}}_{j,h}$ using observed history (if $h=1$) or $\hat{y}_{est}$ (if $h>1$)
        \State Compute $\hat{y}_{j,h} = \int y \cdot \hat{f}_{-j}(y \mid \tilde{\vect{X}}_{j,h}) dy$ and append to $\hat{y}_{est}$ \Comment{Calculate point estimate (mean) from estimated density}
             \State Compute and update scores $V_{j,h} = \hat{f}_{-j}(Y_j \mid \tilde{\vect{X}}_{j,h}) / \hat{c}(\tilde{\vect{X}}_{j,h})$ and $\mathbf{V}[j, h] \leftarrow V_{j,h}$
        \EndFor
    \EndFor
    
    \State Let current time $t_{now} = T$
    \While{($t_{now} < T_{total}$)}
        \State Fit $\hat{f}_{block} = \mathcal{B}(\{(\vect{X}_i, Y_i): i \le t_{now}\})$ 
        
        \For{($h = 1, \ldots, H$)} \Comment{Train specific QRF for each horizon}
            \State Construct dataset $\mathcal{D}_h$ using lags of column $\mathbf{V}_{\cdot, h}$
            \State Fit $\hat{Q}_h = \mathcal{Q}(\mathcal{D}_h, \alpha)$
        \EndFor

        \State Initialize block estimates $\hat{y}_{block} = \{\}$
        \For{($h = 1, \ldots, H$)} \Comment{Predict step-by-step}
            \State Set target time $t = t_{now} + h$
            \State Construct $\tilde{\vect{X}}_{t, h}$ using observed history (for $h=1$) or $\hat{y}_{block}$ (for $h>1$)
            \State Predict density $\hat{f}_{block}(\cdot \mid \tilde{\vect{X}}_{t, h})$ and threshold $\hat{c}(\vect{\tilde{X}}_{t, h})$
            \State Compute and store point estimate $\hat{y}_{t}$ into $\hat{y}_{block}$ 
            
            \State Predict quantile $\hat{q}_t = \hat{Q}_h(\text{lags of } \mathbf{V}_{\cdot, h})$ \Comment{Uses history of horizon $h$}
            \State \textbf{Output:} $\hat{C}(\tilde{\vect{X}}_{t, h}) = \{y: \hat{f}_t(y) > \hat{c}(\tilde{\vect{X}}_{t, h}) \times \hat{q}_t \}$
        \EndFor
        
        \State Observe true block, $\mathbf{Y}_{new} = (Y_{t_{now}+1}, \ldots, Y_{t_{now}+H})$, and update the observed responses
        
        \For{($k = 1, \ldots, H$)} \Comment{Retrospective Score Update}
            \State Let index $i = t_{now} + k$
            \State Compute scores $V_{i, 1}, \ldots, V_{i, H}$ using $\hat{f}_{block}$ and newly observed data
        \EndFor
        
        \State $t_{now} = t_{now} + H$
    \EndWhile
\end{algorithmic}    
\end{algorithm}

\newpage

\section{Proofs for Section 3}

\subsection{Coverage Results}

We begin by proving Theorem 2: Under Assumptions B.1-B.4, when the conditional density estimator converges to the population conditional density uniformly, the density level-set cutoff estimate, $\hat{c}(\vect{x})$, converges uniformly to the population density level-set cutoff, $c(\vect{x})$. 

\noindent We will use the following assumptions.
\begin{enumerate}
\item[B.1] The sample space of $(Y, \vect{X})$ is compact and of the form $\mathcal{K} = \mathcal{Y}\times \mathcal{X} \subseteq \mathbb{R} \times \mathbb{R}^p$.
\item[B.2] The conditional density estimator is uniformly consistent for the population conditional density, 
\[
\sup_{(y, \vect{x}) \in \mathcal{K}} \left| \hat{f}(y \mid \vect{x}) - f(y\mid \vect{x}) \right| = o_p(1)
\]
\item[B.3] The conditional density, $f(y \mid \vect{x})$ is continuous in $(y, \vect{x})$.
\item[B.4] Define $H(c\mid \vect{x})$ as the probability mass of the density level set at threshold $c$. 
\[
H(c \mid \vect{x}) = \int_{ \{ y: f(y \mid \vect{x}) \geq c \} }f(y \mid \vect{x})dy
\]
For every $\vect{x} \in \mathcal{X}$ and for all $c > 0$, the Lebesgue measure of the set $\{ y: f(y \mid \vect{x}) = c \}$ is zero. Consequently, $H(c \mid \vect{x})$ is continuous and strictly monotonic in $c$ for all $c$ such that $0< H(c\mid \vect{x})<1$.
 

\end{enumerate}

The first and third assumptions are mild regularity conditions. Assumption 4 is required to ensure that the density has no flat spots near the $1-\alpha$ density-level cutoff, so that $c(\vect{x})$ is unique and well defined. It is equivalent to assuming that the random variable $f(Y|\vect{x})$  does not have any atoms, for all $\vect{x} \in \mathcal{X}$.

\begin{proof}
First, we show that $c(\vect{x})$ is a continuous function of $\vect{x}$ by letting $\{\vect{x}_m\}$ be a sequence such that $\vect{x}_m \to \vect{x}$ as $m \to \infty$. Denote $c(\vect{x}_m)$ as $c_m$ and $H(\cdot \mid \cdot)$ be the volume function defined in  Assumption B.4. \\
By the assumption that $f(y\mid \vect{x})$ is continuous on a compact set, by Scheffé's lemma
\[
\left| H(c_m \mid \vect{x}_m) - H(c_m \mid \vect{x})\right| \to 0 \text{ as } m\to \infty.
\]
This implies that $H(c_m \mid \vect{x}) \to 1- \alpha$ as $m\to \infty$.\\
Then, by assumption B.4, for any $\tau > 0$
\[
H(c(\vect{x}) + \tau\mid \vect{x}) < 1-\alpha < H(c(\vect{x}) - \tau\mid \vect{x}).
\]
Because $H(c_m \mid \vect{x}) \to 1-\alpha$ and $H(\cdot \mid \vect{x})$ is strictly monotic, for a large enough $m$ $c_m \in (c(\vect{x}) - \tau, c(\vect{x}) + \tau)$. Because $\tau > 0$ is arbitrary, $c_m \to c(\vect{x})$ and $c(\vect{x})$ is continuous. 

Now, we note that the response space being compact by the Heine-Borel Theorem implies that the Lebesgue measure (volume) is finite. Denote the volume of the response space as $V_{\max}$. Define $\epsilon_n = \sup_{(y, \vect{x}) \in \mathcal{K}} \left| \hat{f}(y\mid \vect{x}) - f(y \mid \vect{x})\right|$. By assumption B.2 we know that $\epsilon_n = o_p(1)$.\\
Next, we bound the difference in probability mass between the level sets induced by the estimated and true conditional densities at an arbitrary threshold $c$: 
\[
\Delta=\left|\hat{H}(c \mid \vect{x}) - H(c \mid \vect{x}) \right| = \left| \int_{\hat{A}_c} \hat{f}(y \mid \vect{x})dy - \int_{{A}_c} {f}(y \mid \vect{x})dy \right|,
\]
where $A_c = \{y: f(y\mid \vect{x}) \geq c \}$ and $\hat{A}_c = \{y: \hat{f}(y\mid \vect{x}) \geq c \}$. 
By the triangle inequality,
\[
\Delta \leq \left| \int_{\hat{A}_c}(\hat{f}(y\mid\vect{x}) - f(y \mid \vect{x}))dy \right| + \left| \int_{\hat{A}_c}{f}(y\mid \vect{x})dy - \int_{A_c}f(y\mid \vect{x})dy\right|
\]
\[
\leq  \int_{\hat{A}_c}\left|(\hat{f}(y\mid\vect{x}) - f(y \mid \vect{x}))\right|dy + \int_{\hat{A}_c\triangle A_c}f(y\mid \vect{x})dy,
\]
where $\triangle$ denotes the symmetric set difference.\\
Applying the uniform bound, $\epsilon_n$, to the first term leads to:
\[
\leq \epsilon_n V_{\max} + \int_{\hat{A}_c\triangle A_c}f(y\mid \vect{x})dy.
\]
To bound the second term, note that by the uniform convergence of $\hat{f}(y \mid \vect{x})$ we have $\hat{A}_c\triangle A_c \subseteq B\equiv  \{ y: c-\epsilon_n \leq f(y \mid \vect{x}) \leq c+\epsilon_n\}$. 
\[
\int_{\hat{A}_c\triangle A_c}f(y\mid \vect{x})dy \leq \int_{B}f(y\mid \vect{x})dy = H(c-\epsilon_n \mid \vect{x}) - H(c + \epsilon_n \mid \vect{x}).
\]
By assumption B.4, $H(\cdot \mid \vect{x})$ is a continuous function. Therefore, as $n \to \infty$, $\epsilon_n \to 0$, and
\[
H(c-\epsilon_n \mid \vect{x}) - H(c + \epsilon_n \mid \vect{x}) \to 0.
\]
Because $\epsilon_n = o_p(1)$, 
\[
\sup_{c, \vect{x}\in \mathcal{X}}\left|\hat{H}(c \mid \vect{x}) - H(c \mid \vect{x}) \right| = o_p(1). 
\]

The last step is to use the above two results to demonstrate that $\hat{c}(\vect{x})$ converges uniformly to $c(\vect{x})$. 
We note that because $c(\vect{x})$ is continuous over a compact set, $\mathcal{X}$, by the extreme value theorem there are minimum and maximum density thresholds, $0 < c_{\min} < c_{\max} < \infty$.\\
Also, by the strict monotonicity of $H$ there exists $K_1$ such that for any $\eta$, $0 <\eta < \eta_0$, and every $\vect{x} \in \mathcal{X}$:
\[
H(c(\vect{x}) + \eta \mid \vect{x}) \leq 1 - \alpha - K_1 \eta,
\]
and
\[
H(c(\vect{x}) - \eta \mid \vect{x}) \geq 1 - \alpha + K_1 \eta,
\]
where $\eta_0$ is chosen to be sufficiently small such that $c(\vect{x}) \pm \eta_0 $ remains strictly within the positive part of the density for all $\vect{x} \in \mathcal{X}$. Let $\delta_n =\sup_{c, \vect{x} \in \mathcal{X}} \left| \hat{H}(c \mid \vect{x}) - H(c \mid \vect{x})\right|.$ By the uniform convergence of $\hat{H}$ we know that $\delta_n = o_p(1)$. 
Note that by construction, $H(c(\vect{x}) \mid \vect{x}) = 1-\alpha$. We then have that with probability approaching 1
\[
\hat{H}(c(\vect{x}) + \eta\mid \vect{x}) \leq 1 - \alpha -  K_1\eta + \delta_n
\]
and
\[
\hat{H}(c(\vect{x}) - \eta \mid \vect{x}) \geq 1 - \alpha +  K_1\eta - \delta_n.
\]
Let $\eta_n = \frac{2}{K_1}\delta_n$. Because $\delta_n = o_p(1)$, $\eta_n \to 0$ in probability as $n\to \infty$. So, with probability approaching 1, $\eta_n < \eta_0$. We then have that
\[
\hat{H}(c(\vect{x}) + \eta_n\mid \vect{x}) \leq 1 - \alpha - \delta_n < 1 - \alpha 
\]
and
\[
\hat{H}(c(\vect{x}) - \eta_n \mid \vect{x}) \geq 1 - \alpha + \delta_n > 1 - \alpha.
\]
Also note that because $\hat{f}(y \mid \vect{x}) \geq 0$, $\hat{H}(c \mid \vect{x})$ is monotonically non-increasing in $c$. 
By construction, $\hat{H}(\hat{c}(\vect{x}) \mid \vect{x}) = 1-\alpha$. Because $\hat{H}(\cdot \mid \vect{x})$ is monotonically decreasing, $\hat{c}(\vect{x}) \in (c(\vect{x}) - \eta_n, c(\vect{x}) + \eta_n)$, which implies that
\[
\sup_{\vect{x} \in \mathcal{X}} \left| \hat{c}(\vect{x}) - c(\vect{x}) \right| < \eta_n = \frac{2}{K_1}\delta_n = o_p(1),
\]
as desired. 

\end{proof}

Below we relist the first four assumptions from the main paper, which are required for the pointwise convergence of the QRF estimates.

\begin{enumerate}
    \item[Q.1]Define $U_t := F_{e}(e_t \mid \vect{X} = \vect{X}_t)$ as the quantile of the scores $e_t$ conditioning on $w$ past scores, $\vect{X}_t$, where $U_t \sim \text{Unif }[0,1]$. For $\vect{x} \in \mathbb{B} := \text{Supp}(\{\vect{X}_t\}_{t \geq 1})$, define the scalar $z[\vect{x}] := F_{e}(z \mid \vect{X} = \vect{x})$. Given
\[
g(i,j,\vect{x}_1,\vect{x}_2) := \text{Cov}(\one(U_i \leq z[\vect{x}_1]), \one(U_j \leq z[\vect{x}_2])),
\]
we require that for any pair of $\vect{x}_1, \vect{x}_2 \in \mathbb{B}$,
\[
g(i,j,\vect{x}_1,\vect{x}_2) = g(|i-j|,\vect{x}_1,\vect{x}_2) \text{ for } i \neq j. 
\]

In addition, there exists $\tilde{g}$ such that
\[
g(k,\vect{x}_1,\vect{x}_2) \leq \tilde{g}(k) \ \forall \vect{x}_1,\vect{x}_2 \in \mathbb{B}, \ k \geq 1 
\]
and
\[
\lim_{{T} \to \infty} \left[ \int_1^{{T}} \int_1^s \tilde{g}(u) \, du \, ds \right] / {T}^2 = 0. 
\]
    \item[Q.2]The QRF can be viewed as a weighted quantile of the previously observed scores,
\begin{equation}\label{eq: QRF weighted}
\hat{F}_e(z \mid \vect{X}=\vect{x}) = \sum_{t=1}^{T} w_t(\vect{x}) \mathbf{1}\{e_t \leq z\}.
\end{equation}
The weights in~\cref{eq: QRF weighted} satisfy that for all $\vect{x} \in \mathbb{B}$, $w_t(\vect{x}) = O(1/T)$.
    \item[Q.3]The true conditional distribution function of the scores is Lipschitz continuous,
\[
\sup_{z} |F_e(z \mid \vect{X} = \vect{x}) - F_e(z \mid \vect{X}=\vect{x}')| \leq L \lVert \vect{x} - \vect{x}' \rVert_{1}.
\]
    \item[Q.4]For every $\vect{x}$ in the support of $\vect{X}$, the conditional distribution function of the score $F_e(z \mid \vect{X} = \vect{x})$ is continuous and strictly monotonically increasing in $z$. 

\end{enumerate}

Here we list a lemma, which comes directly from \citealt[Proposition 2]{SPCI_Xu_xie_2023}. 
\begin{lemma}
Under assumptions Q.1-Q.4, the QRF estimate converges pointwise to the true quantile.
\end{lemma}

Next we prove Theorem 3:
Under assumptions B.1-B.5 and Q.1-Q.4, we have doubly robust asymptotic conditional coverage. When either the conditional density estimator converges to the conditional density uniformly or the QRF is correctly specified (or both), SCDR using  the leave-one-out approach achieves asymptotic $1 - \alpha$ conditional coverage as defined in Equation 1 of the main text. 

\begin{enumerate}
    \item[B.5] $Y \mid \vect{X}$ follows a location-scale family of the following form:
\[
Y_t = \mu(\vect{X}_t) + \sigma(\vect{X}_t) \epsilon_t,
\]
where the innovations are independent and identically distributed (i.i.d.), $\epsilon_t \overset{iid}{\sim} P$, with common probability density function $g(\cdot)$. 
\end{enumerate}

\begin{proof}
We note that if the QRF is correctly specified, we have asymptotic $1-\alpha$ conditional coverage by Lemma 1. 

Next, we show that if the true density is used, then the ideal scores are i.i.d.
Under the location-scale family assumption, the conditional density is given by:
\[
f(y \mid \vect{x}) = \frac{1}{\sigma(\vect{x})} g\Big(\frac{y - \mu(\vect{x})}{\sigma(\vect{x})}\Big).
\]
The unique density level cutoff, $c(\vect{x})$, satisfies:
\[
\int_{\{y: f(y \mid \vect{x}) \geq c(\vect{x}) \}} f(y\mid \vect{x})dy = 1-\alpha.
\]
Let $z = \frac{y - \mu(\vect{x})}{\sigma(\vect{x})}$. Performing the change of variable, with $dy = \sigma(\vect{x})dz$, the above integral can be written in terms of the error density $g(\cdot)$:
\[
1 - \alpha = \int_{\{z: \frac{1}{\sigma(\vect{x})} g(z) \geq c(\vect{x}) \}} g(z)dz.
\]
Note that the condition in the integral bound becomes $\frac{1}{\sigma(\vect{x})} g(z) \geq c(\vect{x})$, or equivalently $ g(z) \geq c(\vect{x}) \sigma(\vect{x})$. \\
Let $c_g$ be the unique $1-\alpha$ level set threshold for $g(\cdot)$, defined by $\int_{\{z: g(z) \geq c_g \}} g(z)dz = 1-\alpha$. By comparison of the previous two integrals, $c_g = \sigma(\vect{x})c(\vect{x})$. \\
We can then rewrite the ideal scores:
\[
\tilde{V}_t = \frac{f(Y_t \mid \vect{X}_t)}{c(\vect{X}_t)} = \frac{ \frac{1}{\sigma(\vect{X}_t)}g\Big( \frac{Y_t - \mu(\vect{X}_t)}{\sigma(\vect{X}_t)}\Big)}{\frac{1}{\sigma(\vect{X}_t)} c_g} = \frac{g(\epsilon_t)}{c_g}.
\]
Because $c_g$ is a constant and $\epsilon_t$ are iid by assumption, $\tilde{V}_t = \frac{g(\epsilon_t)}{c_g}$ are iid. 

Now, consider the scores that are used in practice:
\[
V_t = \frac{\hat{f}(Y_t \mid \vect{X}_t)}{\hat{c}(\vect{X}_t)} = \frac{f(Y_t \mid \vect{X}_t) + \Delta f(Y_t \mid \vect{X}_t)}{c(\vect{X}_t) + \Delta c(\vect{X}_t)},
\]
where $\Delta f(y \mid \vect{x}) = \hat{f}(y \mid \vect{x}) - f(y \mid \vect{x})$ and $\Delta c(\vect{x}) = \hat{c}(\vect{x}) - c(\vect{x})$. \\
By assumption B.2 and Theorem 2, we know that $\sup \left|\Delta c(\vect{x})\right|$ and $\sup\left|\Delta f(y \mid \vect{x})\right|$ are $o_p(1)$. We also note that because $c(\vect{x})$ is continuous over the compact set $\mathcal{X}$, by the extreme value theorem there is a minimum density threshold $c_{\min} > 0$. Using the same logic, because $f(y\mid \vect{x})$ is continuous on the compact set $\mathcal{K}$ there is a maximum density value $f_{\max} < \infty$.\\
Now, look at the difference between the ideal and the actual scores:
\[
\sup_{(y, \vect{x}) \in \mathcal{K}}\left| \tilde{V} - V \right| = \sup_{(y, \vect{x}) \in \mathcal{K}} \left|\frac{f(y \mid \vect{x}) }{c(\vect{x}) } - \frac{{f}(y \mid \vect{x}) + \Delta f(y \mid \vect{x})}{{c}(\vect{x}) + \Delta c(\vect{x})} \right| = \sup_{(y, \vect{x}) \in \mathcal{K}}\left| \frac{f(y \mid \vect{x}) \Delta c(\vect{x}) - c(\vect{x}) \Delta f(y \mid \vect{x})}{c(\vect{x})(c(\vect{x}) + \Delta c(\vect{x}))} \right|
\]
Note that $c(\vect{x}) \geq c_{\min} > 0$, $f(y \mid \vect{x}) \leq f_{\max} < \infty$, $\sup_{(y, \vect{x}) \in \mathcal{K}} \left|\Delta f(y \mid \vect{x})\right| = o_p(1)$, and $\sup_{\vect{x} \in \mathcal{X}}\left|\Delta c(\vect{x})\right| = o_p(1)$. Thus, the numerator converges to zero while the denominator remains bounded away from zero. 

\noindent Therefore, $\sup_{(y, \vect{x}) \in \mathcal{K}}\left| \tilde{V} - V\right| = o_p(1)$. This implies that
\[
\max_{1 \leq t \leq T} \left| \tilde{V}_t - V_t \right| \leq \sup_{(y, \vect{x}) \in \mathcal{K}}\left| \tilde{V} - V\right| = o_p(1). 
\]
Because the maximum difference between any scores vanishes in probability, the sequence $\{  V_t\}$ is asymptotically equivalent to $\{ {{V_t}}\}$, implying they are asymptotically i.i.d.

Because the scores are asymptotically i.i.d, the QRF cannot be underspecifed, only correctly specified or overspecified, so the asymptotic conditional coverage from Theorem 1 again applies. Therefore, when either the QRF or the conditional density estimator are correctly specifed, the prediction sets output by SCDR have asymptotic $1-\alpha$ conditional coverage. 
\end{proof}

Next is our proof of Theorem 4, that under the same assumptions used in Theorem 3, the prediction sets output by SCDR asymptotically achieve the minimum Lebesgue measure among all prediction sets that achieve $1-\alpha$ conditional coverage.

\begin{proof}
By the definition of c($\vect{X}_t$) we have that:
\[
P\left[f(Y_t\mid \vect{X}_t) \geq c(\vect{X}_t) \mid \vect{X}_t\right] = 1-\alpha.
\]
Rearranging this in terms the oracle scores:
\[
P\left(\frac{f(Y_t\mid \vect{X}_t)}{c(\vect{X}_t)} \geq 1 \mid \vect{X}_t\right) = P(\tilde{V}_t \geq 1 \mid \vect{X}_t) = 1-\alpha.
\]
Because the scores are unconditionally i.i.d., the $1-\alpha$ quantile is $q_t = 1$. Therefore, by the consistency of the QRF,
\[
\hat{q}_t \to q_t = 1.
\]

Now, looking at the prediction sets output by SCDR:
\[
\hat{C}(\vect{X}_t) = \{y: \frac{\hat{f}(y \mid \vect{X}_t)}{\hat{c}(\vect{X}_t)} \geq \hat{q}_t \} = \{ y: \hat{f}(y\mid\vect{X}_t) \geq \hat{q}_t \hat{c}(\vect{X}_t) \}.
\]
Define the oracle set, which is known to have the smallest Lebesgue measure of all sets with conditional $1-\alpha$ coverage, to be:
\[
C(\vect{X}_t) = \{y: {f}(y\mid\vect{X}_t) \geq {c}(\vect{X}_t) \}.
\]

Next, the symmetric set difference between the SCDR prediction set and the oracle set is
\[
\hat{C}(\vect{X}_t)\triangle C(\vect{X}_t)= 
\]
\[
\{y: ([\hat{f}(y\mid \vect{X}_t) \geq \hat{q}_t\hat{c}(\vect{X}_t)] \cap [f(y \mid \vect{X}_t) < c(\vect{X}_t)]) \cup ([\hat{f}(y\mid \vect{X}_t) < \hat{q}_t\hat{c}(\vect{X}_t)] \cap [f(y \mid \vect{X}_t) \geq c(\vect{X}_t)]) \}.
\]

Define \[
\Delta(y) = \hat{f}(y \mid \vect{X}_t) - f(y \mid \vect{X}_t)
\]
and
\[
\delta = \hat{q}_t \hat{c}(\vect{X}_t) - c(\vect{X}_t).
\]

The disagreement between the prediction sets occurs during two cases. The first is when $\hat{f}(y \mid \vect{X}_t) \geq \hat{q}_t\hat{c}(\vect{X}_t)$ and $f(y \mid \vect{X}_t) < c(\vect{X}_t)$. The second is when $\hat{f}(y \mid \vect{X}_t) < \hat{q}_t\hat{c}(\vect{X}_t)$ and $f(y \mid \vect{X}_t) \geq c(\vect{X}_t)$.

Looking at the first case we have:
\[
\hat{f}(y \mid \vect{X}_t) \geq \hat{q}_t\hat{c}(\vect{X}_t),
\]
which can be rewritten as
\[
f(y \mid \vect{X}_t) + \Delta(y) \geq c(\vect{X}_t) + \delta
\]
\[
f(y \mid \vect{X}_t) - c(\vect{X}_t) \geq \delta - \Delta(y).
\]
Because $f(y\mid \vect{X}_t) < c(\vect{X}_t)$:
\[
\left| f(y \mid \vect{X}_t) - c(\vect{X}_t) \right| = c(\vect{X}_t) -f(y \mid \vect{X}_t) \leq \Delta(y) - \delta.
\]
Through an application of the triangle inequality and taking the supremum on the right hand side we have:
\[
\left| f(y \mid \vect{X}_t) - c(\vect{X}_t) \right| \leq \sup_{y\in\mathcal{Y}}|\Delta(y)| + |\delta|.
\]

Similarly, for the second case we have:
\[
\hat{f}(y \mid \vect{X}_t) < \hat{q}_t\hat{c}(\vect{X}_t),
\]
which can be rewritten as
\[
f(y\mid \vect{X}_t) + \Delta(y) < c(\vect{X}_t) + \delta
\]
\[
f(y\mid \vect{X}_t) - c(\vect{X}_t) < \delta - \Delta(y).
\]
Because $f(y\mid \vect{X}_t) \geq c(\vect{X}_t)$:
\[
\left| f(y\mid \vect{X}_t) - c(\vect{X}_t) \right| = f(y\mid \vect{X}_t) - c(\vect{X}_t) < \delta - \Delta(y).
\]
Through an application of the triangle inequality and taking the supremum on the right hand side we have:
\[
\left| f(y \mid \vect{X}_t) - c(\vect{X}_t) \right| \leq \sup_{y\in\mathcal{Y}}|\Delta(y)| + |\delta|.
\]
Therefore, to show the prediction sets output by SCDR converge to the oracle sets, we can show that
\[
\sup_{y\in\mathcal{Y}}|\Delta(y)| + |\delta| \to 0.
\]

By our assumption of the conditional density being uniformly consistent, $\sup_{y\in\mathcal{Y}}|\Delta(y)| =o_p(1)$. Furthermore, because $\hat{c}(\vect{X}_t) \to  c(\vect{X}_t)$ and $\hat{q}_t \to 1$, by Slutsky's Theorem, $\hat{q}_t\hat{c}(\vect{X}_t) \to c(\vect{X}_t)$, meaning that $\delta = o_p(1)$. Therefore,
\[
\sup_{y\in\mathcal{Y}}|\Delta(y)| + |\delta| =o_p(1).
\]

This, along with assumption B.4 which guarantees that $\{y: f(y \mid \vect{X}_t) = c(\vect{X}_t) \}$ has a measure of zero, proves the sets output by SCDR converge to the oracle sets, which have the smallest Lebesgue measure for a given conditional coverage. 
\end{proof}

\subsection{Doubly Robustness of other scores}

We've shown that the density ratio scores used with SCDR are doubly robust when the family is a location-scale family. In this section we prove what data generating process assumptions are required for other scores to be doubly robust. 

\subsubsection{Survival Score}

Under the same assumptions as those used in Theorem 3, when either the conditional density estimator converges uniformly to the conditional density or the QRF is correctly specified (or both), the sequential approach that utilizes the survival score achieves asymptotic $1-\alpha$ conditional coverage, where the survival score is defined as,
\[
V_t = \hat{H}(\hat{f}(Y_t \mid \vect{X}_t) \mid \vect{X}_t) = \int_{\{y: \hat{f}(y \mid \vect{X}_t) \geq \hat{f}(Y_t\mid \vect{X}_t)\}} \hat{f}(y\mid \vect{X}_t)dy.
\]

\begin{proof}
    We note and will use the fact that in the proof of Theorem 3, it was shown that the existing conditions imply that the estimated survival function converges uniformly to the population survival function.

    We begin by defining the oracle scores which use the population conditional density function as,
    \[
    \tilde{V}_t = H(f(Y_t \mid \vect{X}_t) \mid \vect{X}_t). 
    \]
    By the probability integral transformation, $\tilde{V}_t \mid \vect{X}_t \sim \text{Unif}(0, 1)$. Because the $\text{Unif}(0, 1)$ distribution is free of $\vect{X}_t$, the sequence of oracle scores, $\{ \tilde{V}_t\}$, is unconditionally i.i.d., and thus exchangeable.

    Now, look at the difference between the ideal and the actual scores:
\[
\sup_{(y, \vect{x}) \in \mathcal{K}}\left| \tilde{V} - V \right| = \sup_{(y, \vect{x}) \in \mathcal{K}} \left|H(f(y \mid \vect{x}) \mid \vect{x}) -\hat{H}(\hat{f}(y \mid \vect{x}) \mid \vect{x})\right|
\]
\[
\leq \sup_{(y, \vect{x}) \in \mathcal{K}} \left|H(f(y \mid \vect{x}) \mid \vect{x}) -H(\hat{f}(y \mid \vect{x}) \mid \vect{x})\right| + \sup_{(y, \vect{x}) \in \mathcal{K}} \left|{H}(\hat{f}(y \mid \vect{x}) \mid \vect{x}) -\hat{H}(\hat{f}(y \mid \vect{x}) \mid \vect{x})\right|,
\]
where the last inequality follows from an application of the triangle inequality. 

By Assumption B.4, $H(c \mid \vect{x})$ is strictly continuous with respect to $c$ and $\vect{x}$. By Assumption B.3 and the Extreme Value Theorem, there exists a finite maximum of the conditional density over $\mathcal{K}$, $f_{\max}$. Therefore, the threshold $c$ is bounded in $[0, f_{\max}]$. $H$ is a continuous function over a compact space, so by the Heine-Cantor Theorem $H$ is uniformly continuous. Therefore, because a uniformly continuous function maps uniformly converging sequences to uniformly converging sequences, and $\sup_{\mathcal{K}} |\hat{f}(y\mid \vect{x}) - f(y\mid \vect{x})| = o_p(1)$, the first term is $o_p(1)$. 

The second term is also $o_p(1)$ by the uniform convergence of the conditional survival function, 
\[
\sup_{(y, \vect{x}) \in \mathcal{K}}\left| H(\hat{f}(y \mid \vect{x})\mid \vect{x}) - \hat{H}(\hat{f}(y \mid \vect{x}) \mid \vect{x})\right| \leq \sup_{c \in [0, f_{\max}], \vect{x} \in \mathcal{X}}\left| H(c\mid \vect{x}) - \hat{H}(c \mid \vect{x})\right| = o_p(1).
\]

Therefore, $\sup_{(y, \vect{x}) \in \mathcal{K}}\left| \tilde{V} - V \right| = o_p(1)$, which implies that $\{{V_t}\}$ is asymptotically equivalent to $\{\tilde{{V_t}} \}$, implying asymptotic exchangeability.  
\end{proof}

\subsubsection{Residual Score}

We make the following assumptions, which are similar to those used in Theorem 3, but modified to use the regression residual as the score.
\begin{enumerate}
    \item The covariate space, $\mathcal{X} \subseteq \mathbb{R}^p$, is compact.
    \item The conditional point estimator $\hat{g}(\vect{x})$ is uniformly consistent for the true regression function $g(\vect{x})$:
    \[
    \sup_{\vect{x} \in \mathcal{X}}\left|g(\vect{x}) - \hat{g}(\vect{x}) \right| = o_p(1).
    \]
    \item The true data generating process is a location family:
    \[
    Y_t = g(\vect{X}_t) + \epsilon_t,
    \]
    where $\epsilon_t \overset{i.i.d.}{\sim} P$ and are independent of $\vect{X}_t$. 
\end{enumerate}

\begin{proof}
Define the oracle scores as 
\[
\tilde{V}_t = Y_t - g(\vect{X}_t) = g(\vect{X}_t) + \epsilon_t - g(\vect{X}_t) = \epsilon_t,
\]
which are i.i.d. by Assumption 3. 

Next, define the scores used in practice as 
\[
V_t = Y_t - \hat{g}(\vect{X}_t). 
\]

Finally, looking at the difference between the scores we have:
\[
\sup_{\vect{x}  \in \mathcal{X}} \left|\tilde{V} - V \right| = \sup_{\vect{x} \in \mathcal{X}} \left| g(\vect{x}) - \hat{g}(\vect{x})\right| = o_p(1),
\]
where the last equality follows from Assumption 2. Therefore, $\{{V_t}\}$ is asymptotically equivalent to $\{\tilde{{V_t}} \}$, implying asymptotic exchangeability.  
\end{proof}

\subsubsection{Quantile Score}

We make the following assumptions to establish when the conformalized quantile regression (CQR) scores are asymptotically exchangeable. 
\begin{enumerate}
    \item The covariate space, $\mathcal{X} \subseteq \mathbb{R}^p$, is compact.
    \item The conditional quantile estimators $\hat{q}_{\tau}(\vect{x})$ are uniformly consistent for the true conditional quantiles $q_{\tau}(\vect{x})$, for $\tau \in \{\alpha_{\lo}, \alpha_{\hi} \}$:
    \[
    \sup_{\vect{x} \in \mathcal{X}}\left|q_{\tau}(\vect{x}) - \hat{q}_{\tau}(\vect{x}) \right| = o_p(1), \quad \text{for } \tau \in \{\alpha_{\lo}, \alpha_{\hi} \}.
    \]
    \item The true data generating process is a location family:
    \[
    Y_t = g(\vect{X}_t) + \epsilon_t,
    \]
    where $\epsilon_t \overset{i.i.d.}{\sim} P$ and are independent of $\vect{X}_t$. 
\end{enumerate}

\begin{proof}
By Assumption 3, the population quantiles of $Y_t$ are $q_{\tau}(\vect{X}_t) = g(\vect{X}_t) + z_{\tau}$, where $z_{\tau}$ is the $\tau$-th quantile of the error distribution. 

Define the oracle scores as:
\[
\tilde{V}_t = \max\{q_{\alpha_{\lo}}(\vect{X}_t) - Y_t, Y_t - q_{\alpha_{\hi}}(\vect{X}_t)\} 
\]
\[
= \max\{g(\vect{X}_t) +  z_{\alpha_{\lo}}  - g(\vect{X}_t) -\epsilon_t, g(\vect{X}_t) +\epsilon_t -g(\vect{X}_t) - z_{\alpha_{\hi}} \} 
\]
\[
= \max\{z_{\alpha_{\lo}} - \epsilon_t, \epsilon_t - z_{\alpha_{\hi}}  \}.
\]

Because $z_{\alpha_{\hi}}$ and $z_{\alpha_{\lo}}$ are constants and $\epsilon_t$ are i.i.d. by Assumption 3, the oracle scores are exchangeable. 

The CQR scores used in practice are
\[
V_t= \max\{\hat{q}_{\alpha_{\lo}}(\vect{X}_t) - Y_t, Y_t - \hat{q}_{\alpha_{\hi}}(\vect{X}_t)\}.
\]

Looking at the difference between the actual and oracle scores:
\[
\sup_{\vect{x} \in \mathcal{X}} \left|\tilde{V} - V\right| \leq \sup_{\vect{x} \in \mathcal{X}} \max \{ \left| {q}_{\alpha_{\lo}}(\vect{x}) - \hat{q}_{\alpha_{\lo}}(\vect{x})\right|, \left| \hat{q}_{\alpha_{\hi}}(\vect{x}) - {q}_{\alpha_{\hi}}(\vect{x})\right| \},
\]
where the inequality follows from the fact that:
\[\left| \max(A, B) - \max(C, D) \right|\leq \max(\left|A-C \right|, \left| B - D\right|). \]

The maximum of two sequences which are $o_p(1)$ is also $o_p(1)$, so we have that
\[
\sup_{\vect{x} \in \mathcal{X}} \left|\tilde{V} - V\right| = o_p(1).
\]
Therefore, $\{V_t \}$ is asymptotically equivalent to $\{\tilde{V}_t \}$, implying asymptotic exchangeability of the CQR scores.
\end{proof}

\subsubsection{A Comparison of Doubly Robust Scores}

The result of the preceding three subsections brings up the question: Why use the density ratio scores, which have a doubly robust property when the data generating process (DGP) is location-scale, instead of the survival function scores, which have a doubly robust property without needing to place restrictions on the DGP? The answer can be seen in the simulation studies, where the residual score is used with SPCI when the DGP is a location family. 

When using the survival or residual score, conformal prediction does ``all" the work forming prediction sets. When using the density ratio score or the CQR score, conformal prediction is only an adjustment to get the coverage closer to the nominal level. This allows the density ratio score to have good finite sample coverage. 

We demonstrate this using the same simulation setup as the doubly robust simulation in the main text. The data generating model is a simple AR(1) model,
\[
Y_t = 0.5Y_{t-1} + \epsilon_t,
\]
where $\epsilon_t \sim \mathcal{N}(0, 1)$. 

We correctly specify all parameters of the models: the mean and variance of the conditional density for the density ratio and survival scores, the conditional quantiles for the CQR score, and the conditional mean for the residual score. For all approaches, we use a conditional quantile random forest with the past 5 scores as covariates, trained on the previous 130 scores. The simulation size was 1,000. Results can be found in~\cref{tab:DR_comp}.

\begin{table}
\begin{center}
\caption{Doubly Robust Comparison}

\begin{tabular}{|c|ll|}
\hline
Method & Coverage & Size \\
\hline
Density Ratio Score & 0.881 (0.003) & 3.231 (0.009)\\
Survival Score & 0.874 (0.003) & 3.159 (0.009) \\
Residual Score  & 0.856 (0.003) & 3.109 (0.008) \\
CQR Score  & 0.872 (0.003) & 3.157 (0.009) \\
\hline
\end{tabular}\label{tab:DR_comp}
\end{center}
\end{table}

From this simulation we can see that the density ratio score used with SCDR has coverage closest to the nominal $1-\alpha$ quantile. All methods in this scenario would achieve perfect nominal conditional coverage asymptotically, but the density ratio score has the best finite sample coverage while having a doubly robust property that holds under the very general location-scale family data generating process. Through this, we can also see that the signed residual score approach has the worst nominal coverage.

\subsection{Consistency of the sieve MLE density estimator}

The class of multivariate normal mixtures  with a general mixing distribution for the mean and covariance matrix parameters is infinite dimensional. It renders the numerical computation and theoretical analysis of the maximum likelihood estimator (MLE) challenging. Instead, we consider sieve estimation which approximates the MLE by the MLEs constrained to lie within a sequence of finite-dimensional parameter spaces indexed by the sample size that grow into the original infinite-dimensional space asymptotically. Here,  the  finite normal mixture models with the number of mixing components increasing with the sample size provide an effective sieve. For estimating the conditional pdf, we show below that the number of mixing components need only be of the order $\log n$ raised to the $p+1$-th power, where $n$ is the sample size and $p$ the feature dimension.

\citet{ghosal2001entropies} studied the theoretical properties of the estimator of the pdf of a real random variable, via the normal mixture with a  general joint mixing distribution for the mean and variance parameter of the normal distribution. They derived the convergence rate of the normal-mixture MLE using the asymptotics for the sieve MLE developed by \citet{wong1995probability}. Their results hold for independent and identically distributed (iid) data and  require the assumption that the variance parameter is bounded away from zero and infinity, with known bounds, and the true pdf is a normal mixture subject to some regularity conditions.   The key to their derivation of the convergence rate is the  bounds for the metric entropy for the class of  normal mixtures of the form: 
$$
p(y)=p_F(y)=\int \phi_\sigma(y-z)dF(z,\sigma)
$$
where $\phi(\cdot)$ is the standard normal pdf, $F(\cdot, \cdot)$ is a probability distribution on $\mathscr{R}\times (0, \infty)$.
Let $\mathscr{F}_a=\{p_F: F \mbox{ is a probability distribution with support inside } [-a,a]\times [\underline{\sigma}, \overline{\sigma}])\}$, where $a\le L \left( \log \frac{1}{\epsilon}\right)^\gamma$ and $\gamma\ge 1/2$. 
\citet{ghosal2001entropies}  showed that for $0<\epsilon<1/2$ the  entropy of the preceding class of normal mixtures with $F\in \mathscr{F}_a$ can be bounded as follows:

\begin{eqnarray}
   & \log N(\epsilon, \mathscr{F}_a, \|\cdot\|_\infty)\lesssim \left(\log\frac{1}{\epsilon}\right)^{4\gamma+1}\\
   & \log N_{[]}(\epsilon, \mathscr{F}_a, \|\cdot\|_1)\lesssim \left(\log\frac{1}{\epsilon}\right)^{4\gamma+1}\\
   & \log N_{[]}(\epsilon, \mathscr{F}_a, d)\lesssim \left(\log\frac{1}{\epsilon}\right)^{4\gamma+1}
\end{eqnarray}
where $N(\epsilon, \mathscr{F}_a, \|\cdot\|_\infty)$ is the minimal number of balls of radius $\epsilon$ needed to cover $\mathscr{F}_a$, with the max norm as the metric, and  $N_{[]}(\epsilon, \mathscr{F}_a, d)$ ($N_{[]}(\epsilon, \mathscr{F}_a, \|\cdot\|_1)$ ) is the minimal number of brackets needed to cover the class of densities, with $d$  being the Helllinger distance  and $ \|\cdot\|_1$ the $L^1$-norm. For any pair of functions $g_1\le g_2$, the bracket $[g_1, g_2]$ is the collection of functions bounded below by $g_1$ and above by $g_2$.

We derive similar bounds for the estimation of the conditional density of $\bfY$ given the p-dimensional $\bfX$ via the pdf of a multivariate normal mixture. Let $\bfW^\top=(\bfY, \bfX^\top)$. Let $f_{\bfW}(\cdot), f_{\bfX}(\cdot)$ denote the joint pdf of $\bfW$ and that of $\bfX$, respectively. Then the conditional pdf of $\bfY$ given $\bfX$ is given by $f_{\bfY|\bfX}(\bfy|\bfx)=f_{\bfW}(\bfw)/f_\bfX(\bfx)$. We consider approximating the unknown true $f_{\bfW}$ by a  multivariate normal mixture of the following form: 
\begin{eqnarray*}
    f(\bfw|F)&=& \int_S \phi(\bfw|\mu_{\bfw}, \Sigma_{\bfw, \bfw})  dF(\mu_W, \Sigma_{\bfW, \bfW})  \\
    &=&\int_S \frac{1}{(2\pi)^{(p+1)/2}|\Sigma_{\bfw,\bfw}|^{1/2}} \exp(-(\bfw-\mu_\bfw)^\top\Sigma_{\bfw, \bfw}^{-1} (\bfw-\mu_\bfw)/2) dF(\mu_\bfW, \Sigma_{\bfW, \bfW}),
\end{eqnarray*}
where $F$ is a probability distribution on the parameter space $S=S_a=\{(\mu_\bfW, \Sigma_{\bfW, \bfW}): \|\mu_\bfW\|\le a, \|\Sigma_{\bfW, \bfW}\|_F \le a,  \Sigma_{\bfW, \bfW} \mbox{ is positive definite, whose eigenvalues lie between } 0< \underline{\lambda}< \overline{\lambda}< \infty\}$, where $\|\cdot\|_F$ denotes the Frobenius norm of the matrix,  $\phi(\cdot|\mu, \Sigma)$ is the pdf of the multivariate normal distribution with mean vector $\mu$ and covariance  matrix $\Sigma$. The normal mixture pdf for $\bfW$ induces a mixture pdf estimate for $\bfX$, specifically, $f_{\bfX}(\bfx|F)=\int f_{\bfW}(\bfw)d\bfy$ and hence an estimate for the conditional pdf of $\bfY$ given $\bfX=\bfx$: $f(\bfy|\bfx; F)=f(\bfw|F)/f(\bfx|F)$ where we write $f(\bfw|F)$ and $f(\bfx|F)$ for $f_{\bfW}(\bfw|F)$ and $f_{\bfX}(\bfx|F)$, respectively, i.e., the argument signifies the function.   It is assumed that the bound $a\le L \left( \log\frac{1}{\epsilon}\right)^\gamma$ and $\gamma\ge 1/2$ and $L>0$ are constants. The eigenvalue bounds $0<\underline{\lambda} <\overline{\lambda}<\infty$ are fixed constants.  This eigenvalue condition is assumed henceforth. Our goal is to quantify the proximity of the approximate conditional density to its true counterpart.

As we are interested in estimating the conditional pdf with time series data, we shall make use of the theories developed by \citet{chen1998sieve} for sieve MLE with time series data. According to their Theorem 1, the MLE enjoys the convergence rate of $O_p(\max(\tilde{d}(\theta_0, \pi_n(\theta_0)),\delta_n))$ where $\delta_n$ depends on the complexity of the sieve parameter space, $\tilde{d}$ is a psuedo-metric equivalent to the square root of the Kullback-Leibler divergence (i.e., the psuedo-metric is bounded below and above by some multiples of the Kullback-Leibler divergence), $\pi_n(\theta_0)$ is the parameter in  the $n$-th sieve closest to the true parameter where  $n$ is the sample size.  Theorem~1 of \citet{chen1998sieve} requires the following assumptions:
\begin{enumerate}
    \item[A.1.]  The data are generated by some $\beta$-mixing process with polynomial decaying mixing rate.
    \item[A.2.] The variance of the Kullback-Liebler divergence for a single datum is maximally of the order $\epsilon^2$ for all parameters within a ball centered at the true parameter $\theta_0$ and of radius $\epsilon$, in terms of the psuedo-metric.
    \item[A.3.] The convergence rate bound $\delta_n$ alluded to above is given by
$$
\delta_n=\sup\left \{\delta>0: \delta^{-2} \int_{b\delta^2}^{a\delta} H^{1/2}(w, \mathcal{F}_n)dw\le C n^{1/2}\right\},
$$
where $a,b, C$ are positive constants, $n$ is the sample size,  $\mathcal{F}_n$  the approximating finite-dimensional sieve,  $H(\cdot,\mathcal{F}_n)$ is the bracket entropy, w.r.t. the $L^2$-norm, which measures the complexity of the approximating sieve.
\item[A.4.] The supremum of the  likelihood ratio relative to the true model for a single datum, over a ball centered at the true parameter $\theta_0$ and of radius $\epsilon$ is upper-bounded by the product of $\epsilon^s$, where $s\in (0,2)$, and some measurable function of the datum which has finite $\gamma$-moment for $\gamma>2$.  
\end{enumerate}
See \citet{chen1998sieve} for details and helpful remarks on these regularity conditions. 
 

 The following theorem derives a convergence rate for the sieve maximum likelihood normal mixture density estimator.
 \setcounter{theorem}{4}
\begin{theorem}~\label{Thm: convergence rate for the marginal pdf estimator}
Let $\hat{f}_n$ be the MLE with the support of its mixing distribution inside $ S_a$ with $a\lesssim (\log n)^\gamma$, where $\gamma\ge 1/2$, any eigenvalue of the covariance matrices admitted by $S_a$ falls within  $[\underline{\lambda},\overline{\lambda}]$ where  $0<\underline{\lambda}<\overline{\lambda}<\infty$, and $f_0$ be the true pdf.
Suppose that  the process $\{\bfW_t\}$ is a stationary $\beta$-mixing sequence with the mixing rate $\beta(j)\le \beta_0 j^{-\zeta}$ for some $\beta_0>0, \zeta>2$, and that the stationary marginal distribution of $\bfW_t$ is a normal mixture with the support of the true mixing distribution $F_0$ being a compact subset of $S_a$ for some $a>0$. Let $f_0$ denote the true pdf. Then, for any compact set $\mathcal{K}\subset\mathbb{R}^p$,
 \begin{equation}
    \int \int |\hat{f}_n(\bfy|\bfx)- f_0(\bfy|\bfx)|^2 I(\bfx\in \mathcal{K}) d\bfy d\bfx=O_p(\epsilon_n^2)
    \label{Eq: convergence rate}
    \end{equation}
    where $\epsilon_n= (\log n)^{(p+1)\max(\gamma, 1/2)+1}/\sqrt{n}$.
    
    Furthermore, $\hat{f}_n(\bfy|\bfx) \to f_0(\bfy|\bfx)$ uniformly for $(\bfy, \bfx^\top)^\top$ in a compact set, as $n\to\infty$, in probability. Thus, for all $\bfx\in \mathcal{K}$ and bounded subset $B$ of the real line, $\int_B\hat{f}_n(\bfy|\bfx)d\bfy\to \int_B f_0(\bfy|\bfx)d\bfy$ as $n\to \infty$, in probability.  
\end{theorem}
\begin{proof}
    We first derive the convergence rate of $\hat{f}_n(\bfw)$, by leveraging  Theorem 1 of \citet{chen1998sieve}. We now verify A.1--A.4. This requires constructing a psuedo-metric $\tilde{d}(\cdot, \cdot)$ for the mixing distributions such that  whenever  $ \tilde{d}(f(\cdot|F_0), f(\cdot|F))\le \epsilon$, then the variance of $\log f(\bfw|F_0)/f(\bfw|F)
    \lesssim \epsilon^2$. The pseudo-metric is, furthermore, required to be equivalent to the square root of Kullback-Leibler divergence, i.e.,  there exist two positive constants $c_1, c_2$ such that 
    $c_1 K(f(\cdot|F_0), f(\cdot|F) )\le \tilde{d}(f(\cdot|F_0), f(\cdot|F)) \le c_2 K(f(\cdot|F_0), f(\cdot|F))$. Lemma~\ref{lemma: K and V} below shows that the Hellinger metric almost satisfies this requirement except that the variance is bounded by a multiple of $\epsilon^2 (\log 1/\epsilon)^2$  and not  $\epsilon^2$. However, it inspires the following  metric obtained as a increasing function of the Hellinger metric. Specifically,  for any two pdfs $p, q$, define  $\tilde{d}(p,q)=h(d(p,q))$ where $h(x)= \min(x\log(1/x), e^{-1})$, for $x\ge 0$. By continuity, $h(0)=0$. Since the function $y=x\log(1/x)$ is an increasing function for $x\in [0,e^{-1}]$ and attains its maximum value $e^{-1}$ at $x=e^{-1}$, $h(x)=x\log(1/x)$ over  $x\in [0,e^{-1}]$, and stays at $e^{-1}$ for all larger $x$. It can be  checked that $\tilde{d}$ is a metric. Note that  $d(\cdot, \cdot) \le \tilde{d}(\cdot,\cdot)$ whenever the latter is $\le e^{-1}$, hence any convergence rate in terms of $\tilde{d}$ applies equally for the Hellinger distance $d$.  Moreover, thanks to Lemma 7, assumption A.2 holds. Assumption A.3 also holds by setting  $\delta_n= 
    \frac{1}{\sqrt{n}}(\log n)^{2(p+1)\max(\tau, 1/2)+1/2}$. Theorem ~\ref{Thm: generalization of Thm 5 of Wong and Shen} below implies that Assumption A.4 holds with $\gamma=3$. It then follows from Theorem 1 of \citet{chen1998sieve} that the Hellinger distance between the MLE of the density of $\bfW=(\bfY, \bfX^\top)^\top$ and its true counterpart is $O_p(\frac{1}{\sqrt{n}}(\log n)^{(p+1)\max(\tau, 1/2)+1/2})$. The convergence rate also holds for the $L^2$-norm because the eigenvalue condition implies that the densities are uniformly bounded. Similarly, the rate of convergence of the MLE of the density of $\bfX$ is $O_p(\frac{1}{\sqrt{n}}(\log n)^{p\max(\tau, 1/2)+1/2})$. 

It follows from Lemma~\ref{Result 1} that the class of pdfs $f(\cdot; F)$'s, for either $\bfW$ or $\bfX$,  is equicontinuous, thanks to 
the eigenvalue condition $0<\underline{\lambda}<  \overline{\lambda}<\infty$. Thus, $\hat{f}_n$ converges to $f_0$ uniformly over compact sets, in probability as $n\to\infty$, either as the pdf of $\bfW$ or $\bfX$.  Consequently, $\hat{f}_n(\bfy|\bfx)$ converges uniformly to $f_0(\bfy|\bfx)$ over compact sets in probability. 
     Then \eqref{Eq: convergence rate} follows from the following bound:
    $$
    |\hat{f}_n(\bfy|\bfx)- f_0(\bfy|\bfx)|^2 \le 2\left| \frac{\hat{f}_n(\bfy, \bfx)-f_0(\bfy, \bfx)}{\hat{f}_n(\bfx)}\right|^2+2 \left| \frac{f_0(\bfy, \bfx) (\hat{f}_n(\bfx)-f_0(\bfx))}{\hat{f}_n(\bfx)f_0(\bfx) } \right|^2,
    $$
    where we write $\hat{f}_n(\bfw)$ as 
    $\hat{f}_n(\bfy, \bfx)$, etc.
    \end{proof}

The following result shows that in practice we can approximate a conditional distribution by that of a finite normal mixture with the number of mixing components of the order $O_p((\log n)^{p+1})$ where $p$ is the dimension of the features, $\bfX_t$. The result quantifies the curse of dimensionality as the number of components increases exponentially with the feature dimension. 
\begin{theorem}
Suppose that all the assumptions of   Theorem~\ref{Thm: convergence rate for the marginal pdf estimator} hold, with $\gamma=1/2$. Then the conclusions of Theorem~\ref{Thm: convergence rate for the marginal pdf estimator}  still hold for the sieve MLE whose mixing distribution has  at most $O_p((\log n)^{p+1})$ atoms in $S_a$.  
\end{theorem}

\begin{proof}
Our proof is similar to that of  \citet[Theorem 4.2]{ghosal2001entropies}. Given $\epsilon>0$ and $\eta>0 $  to be determined below, Lemma~\ref{Result 4} ensures that there exists a discrete distribution $F_0'$ supported on a compact set with $N\lesssim \left( \log \frac{1}{\eta}\right)^{p+1}$ atoms such that $\|f(\cdot|F_0)- f(\cdot|F_0')\|_\infty \lesssim \eta$, where  $F_0$ is the true mixing distribution. It follows from Lemmas~\ref{Result: e-net} and \ref{lemma: K and V} that 
\begin{eqnarray}
        K(f_{\bfW}(\cdot|F_0), f_{\bfW}(\cdot| F_0') )&\lesssim & \eta\left(\log\frac{1}{\eta} \right)^{(p+3)/2}\label{eq: K1-inequality} \\
        V(f_{\bfW}(\cdot|F_0), f_{\bfW}(\cdot| F_0')  ) &\lesssim & \eta\left(\log\frac{1}{\eta} \right)^{(p+5)/2}, 
        \label{eq: V1-inequality}
    \end{eqnarray}
    where  for any two pdfs $p,q$, $K(p, q)=\int \log (p/q)p$ is the Kullback-Leibler divergence and $V(p,q)=\int (\log p/q)^2 p$ is an upper bound of the variance of the Kullback Leibler divergence.
    Let us choose $\eta$ to be the solution of the equation $\epsilon=\eta^{1/2}\left(\log\frac{1}{\eta} \right)^{(p+5)/4}$, which admits a unique positive solution for all sufficiently small, positive $\epsilon$.   This is because  the function $y=\eta^{1/2}\left(\log\frac{1}{\eta} \right)^{(p+5)/4}$ is an increasing function for $0\le \eta\le e^{-(p+5)/2}$. 
    Note that $\log\frac{1}{\eta}\sim \log \frac{1}{\epsilon}$.
    For $n$ sufficiently large, set $\epsilon=\epsilon_n= (\log n)^{(p+1)/2+1}/\sqrt{n}$ and the corresponding $\eta=\eta_n$.  Then, there exists a discrete distribution $F_n'$ with $N\lesssim \left( \log \frac{1}{\eta_n}\right)^{p+1}$ atoms such that $\tilde{d}(f(\cdot|F_0)-f(\cdot|F_n'))\le \epsilon_n$.  It then follows from Theorem 1 of \citet{chen1998sieve} that the finite normal-mixture MLE with the number of components of the order $(\log n)^{p+1}$ enjoys the same convergence rate stated in Theorem~\ref{Thm: convergence rate for the marginal pdf estimator}, and the rest of the proof is the same as that of Theorem~\ref{Thm: convergence rate for the marginal pdf estimator}.  
\end{proof}

The following results are instrumental for proving the preceding main results. 
The size of the collection of the multivariate normal mixtures with the mixing distribution having its support inside $S_a$ is characterized by the following theorem. 
\begin{theorem}
Let $\epsilon>0$ be given. It holds that 
\begin{eqnarray}
   & \log N(\epsilon, \mathscr{F}_a, \|\cdot\|_\infty)\lesssim \left(\log\frac{1}{\epsilon}\right)^{2(p+1)\gamma+1}\\
   & \log N_{[]}(\epsilon, \mathscr{F}_a, \|\cdot\|_1)\lesssim \left(\log\frac{1}{\epsilon}\right)^{2(p+1)\gamma+1}\\
   & \log N_{[]}(\epsilon, \mathscr{F}_a, \|\cdot\|_2)\lesssim \left(\log\frac{1}{\epsilon}\right)^{2(p+1)\gamma+1}\\
   & \log N_{[]}(\epsilon, \mathscr{F}_a, d)\lesssim \left(\log\frac{1}{\epsilon}\right)^{2(p+1)\gamma+1}
\end{eqnarray}
where $\mathscr{F}_a$ is the collection of pdfs $\{f(\bfw|F), F\in S_a\}$ with $\bfw=(\bfy,\bfx^\top)^\top$ and  $\|f(\cdot| F)-f(\cdot| F')\|_\infty = \sup_{\bfw} |f(\bfw| F)-f(\bfw| F')|$, 
$\|f(\cdot| F)-f(\cdot| F')\|_k = (\int|f(\bfw| F)-f(\bfw| F')|^kdy)^{1/k}$, for $k\ge 1$ and 
$d(f(\cdot| F), f(\cdot| F'))=  (\int|f^{1/2}(\bfw| F)-f^{1/2}(\bfw| F')|^2dy)^{1/2}$ is the Hellinger metric. 
\end{theorem}

\begin{proof}
    The proof is similar to that of \citet[Theorem 3.2]{ghosal2001entropies}, by lifting the technical results there for the 1-dimensional case to the multivariate setting, as elaborated in Lemmas~\ref{Result 1}--\ref{Result: e-net} below. The result for the bracketing entropy with the $L^2$-metric follows readily from that of the $L^1$-metric since the densities are uniformly bounded under the eigenvalue condition, hence the $L^2$ norm is bounded by a multiple of the $L^1$ norm. 
\end{proof}
\begin{lemma}\label{Result 1}
    The collection of pfds $\{ f(\bfw|F), F\in S_a\}$ is equicontinuous. So is $\{ f(\bfx|F), F\in S_a\}$. In fact, the result holds under the eigenvalue condition: $0<\underline{\lambda}< \overline{\lambda}< \infty$.
\end{lemma}

\begin{proof}
    The result follows from the following inequality which holds for all $\bfw$.
    \begin{eqnarray*}
        &&\left\|\frac{\partial  f(\bfw|F)}{\partial \bfw}\right\|\\
        &\le& \int_S \frac{1}{(2\pi)^{(p+1)/2}|\Sigma_{\bfw,\bfw}|^{1/2}} \|\Sigma_{\bfw, \bfw}^{-1} (\bfw-\mu_\bfw)\|\exp(-(\bfw-\mu_\bfw)^\top\Sigma_{\bfw, \bfw}^{-1} (\bfw-\mu_\bfw)/2) dF(\mu_\bfW, \Sigma_{\bfW, \bfW}) \\
        &\le & \frac{1}{(2\pi)^{(p+1)/2}\underline{\lambda}^{(p+2)/2}}\times C_1
    \end{eqnarray*}
where the constant $C_1=\max_{x\ge 0} x\exp(-x^2/2)=\sqrt{2}\exp(-1)$. The claim regarding the collection $\{ f(\bfx|F): F\in S_a\}$ follows similarly. 
    
\end{proof}

The proofs of the following two lemmas are straightforward and hence omitted.  
\begin{lemma}\label{Result 2}
    For any $F \in S_a$ and compact set $\tilde{K}$, it holds that there exists an $\eta>0$ such that 
\begin{equation}
    \inf_{x\in \tilde{K}} f(x|F) \ge \eta 
\end{equation}
where $\eta$ can be chosen as $\frac{1}{(2\overline{\lambda}\pi)^{p/2}}\exp(-\overline{\lambda}\max(\|\bfx\|^2: \bfx \in \tilde{K})/2)$.
Moreover, 
\begin{equation}
\sup_{\bfx\in \mathbb{R}^p} f(\bfx|F)\le   \frac{1}{(2\underline{\lambda}\pi)^{p/2}}, \qquad 
\sup_{\bfw\in \mathbb{R}^{p+1}} f(\bfw|F)\le   \frac{1}{(2\underline{\lambda}\pi)^{(p+1)/2}}.
\end{equation}
\end{lemma} 

\begin{lemma}\label{Result 3}
 For any $F, F' \in S_a$ and for $M=\max\left(2a, \sqrt{8} \,\overline{\lambda} \left( \log \frac{1}{\epsilon}\right)^{1/2}\right)$
\begin{eqnarray*}
    \sup_{|\bfw|\ge M} |f(\bfw| F)-f(\bfw| F')| &\le & \frac{2}{(2\underline{\lambda}\pi)^{(p+1)/2}}\exp(-M^2/(8\overline{\lambda}^2))\\
    &\le& \frac{2}{(2\underline{\lambda}\pi)^{(p+1)/2}} \epsilon.
\end{eqnarray*}
Similarly,
\begin{eqnarray*}
    \sup_{|\bfx|\ge M} |f(\bfx| F)-f(\bfx| F')| &\le & \frac{2}{(2\underline{\lambda}\pi)^{p/2}} \epsilon.
 \end{eqnarray*}   
\end{lemma}

\begin{lemma}\label{Result 4}
    Let $0 < \epsilon < \frac{1}{2}$ and suppose $F\in S_a$,  where $a \leq L (\log \tfrac{1}{\epsilon})^\gamma$ for constants $L > 0$ and $\gamma \geq \tfrac{1}{2}$. Then there exists a discrete probability measure $F'$ with  at most
\[
N \leq (\log \tfrac{1}{\epsilon})^{2\gamma(p+1)}
\]
atoms in $S_a$ such that
\[
\sup_{\bfw} \left\| f(\bfw|F) - f(\bfw |F') \right\|_\infty \lesssim \epsilon.
\]
\end{lemma}
\begin{proof}

We outline the proof below as it is similar to that of \citet[Theorem 3.1]{ghosal2001entropies}.
We approximate $\phi(\bfw|\mu, \Sigma)$, where we write $\mu, \Sigma$ for $\mu_\bfW, \Sigma_{\bfW, \bfW}$ using Taylor expansion. For $u < 0$,
\begin{equation}
    \left| e^u - \sum_{j=0}^{k-1} \frac{u^j}{j!} \right| \leq \frac{(e|u|)^k}{k^k}.
    \label{Eq: Taylor expansion error bound}
\end{equation}
Set $u^2 = (\bfw-\mu)^\top \Sigma^{-1} (\bfw-\mu)/2$ to bound the truncation error:
\begin{equation}
    \left| \phi(\bfw|\mu,\Sigma)-\sum_{j=0}^{k-1} \frac{(-1)^j u^{2j}}{2^j(2\pi)^{(p+1)/2}j!|\Sigma|^{1/2}}\right| \le \underline{\lambda}^{-(p+1)/2} \frac{(e^{1/2}  u)^{2k}}{(2\pi)^{(p+1)/2} k^k}.
\end{equation}
Write $\bfw=(w_1,\cdots, w_{p+1})^\top, \mathbf{d}=(d_1,\cdots, d_{p+1})^\top$ where the $d$'s are non-negative integers. Define $\bfw^\mathbf{d}=\prod_{i=1}^{p+1} w_i^{d_i}$ and $|\mathbf{d}|=\sum_{i=1}^{p+1} d_i$.  Let $P_{2k-2}(\bfw) = \sum_{|\mathbf{d}|\le 2k-2} c_{\mathbf{d}}(\mu, \Sigma) \bfw^\mathbf{d}$ be the Taylor polynomial of up to degree $2k-2$ for $\frac{1}  {2^{(p+1)/2}|\Sigma|^{1/2}}\exp(-u/2)$. It can be readily checked that the $c$'s are   continuous functions of $(\mu, \Sigma)\in S_a$. Let $M$ be as in Lemma~\ref{Result 3}. Now
\begin{eqnarray*}
    &&\sup_{\|\bfw\| \le M} |f(\bfw | F) - f(\bfw | F')|\\
    &\leq& \left| \int_{S_a} P_{2k-2}(\bfw) \, d(F - F')(\mu, \Sigma) \right| + \text{error}\\
    &=& \left| \sum_{|\mathbf{d}|\le 2k-2} \bfw^\mathbf{d} \int_{S_a}  c_{\mathbf{d}}(\mu, \Sigma) \, d(F - F')(\mu, \Sigma) \right| + \text{error}.
\end{eqnarray*}

To eliminate the first term, we match the expectation of  $c_{\mathbf{d}}(\mu, \Sigma)$ under $F'$ to that under the true mixing distribution, for all $0<|\mathbf{d}|\le 2k-2$. The number of matched ``moments" is
\[
N_{\text{moments}} = \binom{2k - 2 + (p+1)}{p+1} - 1.
\]
To control the error from the Taylor expansion, note that $$|u| = (2\underline{\lambda})^{-1/2}(\|\bfw\|+\|\mu\|)\le \frac{3M}{\sqrt{8\underline{\lambda}}}\le (2\underline{\lambda})^{-1/2}\max(3L, \sqrt{18} \, \overline{\lambda})\left(\log\frac{1}{\epsilon}\right)^\gamma$$
hence  the remainder term in the approximation satisfies
\[
\text{error} \lesssim \frac{(c (\log \tfrac{1}{\epsilon})^\gamma)^{2k}}{k^k},
\]
where $c=e^{1/2} 2^{-1/2}\underline{\lambda}^{-1/2}\max(3L, \sqrt{18} \, \overline{\lambda})$. If $k$ is chosen to be the smallest integer exceeding $(1+c^2)(\log \frac{1}{\epsilon})^{2\gamma}$, then the 
$ \text{error} \lesssim \epsilon $.


Using \citet[Lemma A.1]{ghosal2001entropies}, $F'$ can be chosen to be a discrete distribution with the number of atoms at most equal to 
\[
N = \binom{2k + p - 1}{p+1} \lesssim k^{p+1} \sim (\log \tfrac{1}{\epsilon})^{2\gamma(p+1)}.
\]
Combining the tail bound in Lemma~\ref{Result 4} and the polynomial approximation over the central region completes the proof.
\end{proof}

\begin{lemma} \label{Result: e-net1}
   Let $\epsilon$ be  a positive constant.  Uniformly for any pair $(\mu_i, \Sigma_i)\in S_a, \, i=1,2$ such that $\max(\|\mu_1-\mu_2\|, \|\Sigma_1-\Sigma_2\|)\le \epsilon,$ it holds that $\sup_{\bfw \in \mathbb{R}^{p+1}} |\phi(\bfw|\mu_1, \Sigma_1)-\phi(\bfw|\mu_2, \sigma_2)|\lesssim \epsilon.$
\end{lemma}
\begin{proof}
    Consider
    \begin{eqnarray*}
|\phi(\bfw|\mu_1, \Sigma_1)-\phi(\bfw|\mu_2, \sigma_2)| &\le& A+B+C        
    \end{eqnarray*}
    where 
    \begin{eqnarray*}
A &=&  \frac{1}{(2\pi)^{(p+1)/2}}\left||\Sigma_{1}|^{-1/2} -|\Sigma_{2}|^{-1/2}\right|\exp(-(\bfw-\mu_1)^\top\Sigma_{1}^{-1} (\bfw-\mu_1)/2)  \\
B &=& \frac{1}{(2\pi)^{(p+1)/2}|\Sigma_2|^{1/2}}\left| 
\exp(-(\bfw-\mu_1)^\top\Sigma_{1}^{-1} (\bfw-\mu_1)/2) -\exp(-(\bfw-\mu_1)^\top\Sigma_{2}^{-1} (\bfw-\mu_1)/2)
\right| \\
C &=& \frac{1}{(2\pi)^{(p+1)/2}|\Sigma_2|^{1/2}}\left| 
\exp(-(\bfw-\mu_1)^\top\Sigma_{2}^{-1} (\bfw-\mu_1)/2) -\exp(-(\bfw-\mu_2)^\top\Sigma_{2}^{-1} (\bfw-\mu_2)/2)
\right|
    \end{eqnarray*}
each of which is uniformly $\lesssim \epsilon$.  To see this, first note that  $$\left||\Sigma_{1}|^{-1/2} -|\Sigma_{2}|^{-1/2}\right|=|\Sigma_2^{-1}(\Sigma_2-\Sigma_1)\Sigma_1^{-1}|/ (|\Sigma_1|^{-1/2}+|\Sigma_2|^{-1/2}).$$ The claim that $A\lesssim \epsilon$ follows from (i) the definition of $S_a$, (ii) the  determinant inequality \citep[Eqn (1)]{gil2021perturbation} that for any two $q\times q$ matrices $M_1, M_2$, $||M_1|-|M_2||\le q \max(\|M_1\|, \|M_2\|)^{q-1} \|M_1-M_2\|$ where $\|\cdot\|$ denotes the spectral norm of an enclosed square matrix  and (iii) the bound $\|M\|\le \sqrt{q} \|M\|_F$. Using the inequality that for any non-negative numbers $y_1, y_2 $
$$ |\exp(-y_1)-\exp(-y_2)|\le \max(\exp(-y_1), \exp(-y_2))|y_1-y_2|$$ 
and, after some algebra, we have 
\begin{eqnarray*}
  B &\le &  \frac{1}{(2\pi)^{(p+1)/2}\underline{\lambda}^{(p+1)/2}}\sqrt{p+1}\,\epsilon\, \frac{\|\bfw-\mu_1\|^2 }{2\underline{\lambda}^2} \exp\left(  -\frac{-\|\bfw -\mu_1\|^2}{2\overline{\lambda}}\right) \\
  &\le& \frac{1}{(2\pi)^{(p+1)/2}\underline{\lambda}^{(p+1)/2}}\sqrt{p+1}\,\epsilon\, \frac{\overline{\lambda}e^{-1} }{2\underline{\lambda}^2}\\
  &\lesssim & \epsilon, 
\end{eqnarray*}
since $\max_{x\ge 0} x\exp(-x)=e^{-1}$. Similarly,
\begin{eqnarray*}
    C &\le & \frac{1}{(2\pi)^{(p+1)/2}\underline{\lambda}^{(p+1)/2}}\sqrt{p+1}\,\epsilon\, \frac{\|\bfw-\mu_1\|+2a }{\underline{\lambda}} \exp\left(  -\frac{-\|\bfw -\mu_1\|^2}{2\overline{\lambda}}\right) \\
    &\le & \frac{1}{(2\pi)^{(p+1)/2}\underline{\lambda}^{(p+1)/2}}\sqrt{p+1}\,\epsilon\, \left(\frac{2a }{\underline{\lambda}}+ \frac{\overline{\lambda}^{1/2} e^{1/2}}{\underline{\lambda}}  \right)\lesssim \epsilon,
\end{eqnarray*}
because  $\max_{x\ge 0} x\exp(-x^2)=(2e)^{-1/2}$.
\end{proof}

\begin{lemma} \label{Result: e-net}
    For any pair $(\mu_i, \Sigma_i)\in S_a, \, i=1,2$,
\begin{eqnarray*}
&&\|f_\bfW(\cdot|\mu_1,\Sigma_1)-f_\bfW(\cdot|\mu_2,\Sigma_2)\|_1
\\
&\lesssim& \|f_\bfW(\cdot|\mu_1,\Sigma_1)-f_\bfW(\cdot|\mu_2,\Sigma_2)\|_\infty \times \\
&&\left( \max\left\{ 
\sqrt{\log_+\left(\frac{1}{\|f_\bfW(\cdot|\mu_1,\Sigma_1)-f_\bfW(\cdot|\mu_2,\Sigma_2)\|_\infty}\right)}, a,2
\right\}\right)^{(p+1)/2},
\end{eqnarray*}    
where $f_\bfW(\cdot|\mu,\Sigma)$ is the joint normal pdf of $\bfW$ given the mean $\mu$ and covariance matrix $\Sigma$.    
\end{lemma}

\begin{proof}
 Because the $L^1$-norm between two densities is bounded by 2, the inequality holds trivially if the $L^\infty$ norm between the two densities is greater than 1. Thus, with no loss of generality, consider the case that  $\|f_\bfW(\cdot|\mu_1,\Sigma_1)-f_\bfW(\cdot|\mu_2,\Sigma_2)\|_\infty<1.$ For any $T\ge 2 a$ and $\|\bfw\|>T$,
 $$
 f(\bfw|\mu,\Sigma)\le 
 \frac{1}{(2\pi \underline{\lambda})^{(p+1)/2}} \exp\left(-\frac{(\|\bfw\|-a)^2}{2\overline{\lambda}}\right)
 \le  \frac{1}{(2\pi \underline{\lambda})^{(p+1)/2}} \exp\left(-\frac{\|\bfw\|^2}{8\overline{\lambda}}\right).
 $$
 Hence, $\|f_\bfW(\cdot|\mu_1,\Sigma_1)-f_\bfW(\cdot|\mu_2,\Sigma_2)\|_1$ is bounded by 
 \begin{eqnarray*}
     && 2\int_{\|\bfw\|>T} \frac{1}{(2\pi \underline{\lambda})^{(p+1)/2}} \exp\left(-\frac{\|\bfw\|^2}{8\overline{\lambda}}\right) d\bfw + T^{p+1} \nu_{p+1}\|f_\bfW(\cdot|\mu_1,\Sigma_1)-f_\bfW(\cdot|\mu_2,\Sigma_2)\|_\infty\\
     &\lesssim & T^{p-1}\exp\left(-\frac{T^2}{8\overline{\lambda}} \right)+ T^{p+1} \|f_\bfW(\cdot|\mu_1,\Sigma_1)-f_\bfW(\cdot|\mu_2,\Sigma_2)\|_\infty,
 \end{eqnarray*}
 where $\nu_{q}$ is the volume of the unit ball in $\mathbb{R}^q$. The result follows by choosing $T= \max\left(\sqrt{ 8 \overline{\lambda} \log\left( \frac{1}{\|f_\bfW(\cdot|\mu_1,\Sigma_1)-f_\bfW(\cdot|\mu_2,\Sigma_2)\|_\infty}\right)}, 2a\right)$.
\end{proof}

\begin{lemma}\label{lemma: K and V}
    Suppose that $F(S_B)>1/2$ for some constant $B>0$ and $F^*$ is a probability measure satisfying $F^*(S_t^c)\lesssim \exp(-b't^2)$ for some constant $b'>0$, then for all $\epsilon=d(f_{\bfW}(\cdot|F^*), f_{\bfW}(\cdot| F))<1/2$,
    \begin{eqnarray}
        K(f_{\bfW}(\cdot|F^*), f_{\bfW}(\cdot| F) )&\lesssim & \epsilon^2\log\frac{1}{\epsilon} \label{eq: K-inequality} \\
        V(f_{\bfW}(\cdot|F^*), f_{\bfW}(\cdot| F) ) &\lesssim & \epsilon^2\left(\log\frac{1}{\epsilon}\right)^2, 
        \label{eq: V-inequality}
    \end{eqnarray}
where for any density $p_0$ and corresponding probability measure $P_0$,  $K(p_0, p)=\int \log(p_0/p)dP_0$ and $V(p_0,p)=\int (\log(p_0/p))^2 dP_0$.
\end{lemma}
\begin{proof}
    The proof is similar to that of \citet[Lemma 4.1]{ghosal2001entropies}, with a few modifications. For completeness, we present the proof. For simplicity, we drop the subscript $\bfW$. Note that
    $$
    f(\bfw|F^*)\le \frac{1}{(2\pi \underline{\lambda})^{(p+1)/2}}.
    $$
But a tighter upper bound can be obtained as follows. 
\begin{eqnarray}
   f(\bfw|F^*) &=& \int_{S_{\|\bfw\|/2}} \phi(\bfw|\mu, \Sigma) dF^* +  \int_{S_{\|\bfw\|/2}^c} \phi(\bfw|\mu, \Sigma) dF^* \nonumber \\
   &\le& \frac{1}{(2\pi \underline{\lambda})^{(p+1)/2}}\left\{\exp(-\|w\|^2/(8\overline{\lambda}))+ F^*\left(S_{\|\bfw\|/2}^c\right)\right\} \lesssim \exp(-b^* \|w\|^2) \label{Eq: tighter bound}
\end{eqnarray}
where $b^*= \min(1/(8\overline{\lambda}), b'/4)>0$.    
    If $F(S_B)>1/2$,
    $$
    f(\bfw|F)\ge \frac{1}{(2\pi \overline{\lambda})^{(p+1)/2}}\int_{S_B} \exp\{-\|\bfw -\mu\|^2/(2\underline{\lambda})\}dF\ge \frac{1}{2(2\pi \overline{\lambda})^{(p+1)/2}}\exp\{-(\|\bfw\|^2+B^2)/\underline{\lambda}\}.
    $$
    Therefore for some $c$ (depending on $B$),
    $$
    \frac{f(\bfw|F^*)}{f(\bfw|F)}\lesssim \exp(c \|\bfw\|^2),
    $$
    hence, in view of \eqref{Eq: tighter bound}, for some $\delta>0$,
    $$
    \int \left(   \frac{f(\bfw|F^*)}{f(\bfw|F)}\right)^\delta
    f(\bfw|F^*) d\bfw <\infty.
$$
    The result then follows from Theorem 5 of \citet{wong1995probability}.
\end{proof}

 The following Theorem generalizes Theorem 5 of \cite{wong1995probability}.
\begin{theorem}\label{Thm: generalization of Thm 5 of Wong and Shen}
    Let $p, q$ be two pdfs such that $\int (p^{1/2} - q^{1/2})^2 \le \epsilon^2$. Suppose that $M_\delta=\int_{\{p/q\ge e^{1/\delta}\}} p (p/q)^\delta < \infty$ for some $\delta\in (0,1]$. Then for all  $\epsilon^2 \le e^{-\gamma^2/\delta}$ 
    and  $\gamma \ge 2$, we have 
$$
\int p \left|\log\left(\frac{p}{q}\right)\right|^\gamma \le K\epsilon^2 (\log(1/\epsilon))^\gamma
$$
for some  positive constant $K$ that depends on $\gamma$ and $M_\delta$ only.  
\end{theorem}
\begin{proof}
The integral can be decomposed into a sum of three terms:
$$
\int p \left|\log\left(\frac{p}{q}\right)\right|^\gamma= A+B+C
$$
where 
\begin{eqnarray*}
    A&=& \int_{\{p/q\le e^{-\gamma}\}} p \left|\log\left(\frac{p}{q}\right)\right|^\gamma \\
    B&=& \int_{\{e^{-\gamma}<p/q\le K\}} p \left|\log\left(\frac{p}{q}\right)\right|^\gamma  \\ 
    C&=& \int_{\{K<p/q\}} p \left|\log\left(\frac{p}{q}\right)\right|^\gamma 
\end{eqnarray*}
where $K>e^{\gamma/\delta}$ is to be determined. $A$ can be bounded as follows:
\begin{eqnarray*}
  A&=& \int_{\{p/q\le e^{-\gamma}\}} q \left|\log\left(\frac{q}{p}\right)/ \left(\frac{q}{p}\right)^{1/\gamma}\right|^\gamma \\ 
  &\le& \left(\frac{\gamma}{e}\right)^\gamma \int_{\{p/q\le e^{-\gamma}\}} q  \\
  &\le& \left(\frac{\gamma}{e}\right)^\gamma (1-e^{-\gamma/2})^{-2} \epsilon^2
\end{eqnarray*}
where the first inequality holds because the $\log(x)/x^{1/\gamma}$ is a decreasing function for $x\ge e^{\gamma}$ and the second inequality follows from Lemma 2 of \citet{wong1995probability}. To bound $B$, let $y=\sqrt{p/q}-1$. Hence, $\log(p/q)=\log\{(1+y)^2\}=2\log(1+y)$ and $(p^{1/2}-q^{1/2})^2=qy^2$. The inequality $y/(1+y)\le \log(1+y)\le y$ for $y>-1$ will be applied below. Note that  the lower bound can be set to be 0 for  $y\ge 0$. Let $c=e^{\gamma/\delta}$. Now,
\begin{eqnarray*}
     B&=& \int_{\{e^{-\gamma}<p/q\le K\}} p \left|\log\left(\frac{p}{q}\right)\right|^\gamma  \\ 
     &\le& \int_{\{e^{-\gamma/2}-1 <y\le K^{1/2}-1\}} q p/q  \left|\log\left(\frac{p}{q}\right)\right|^\gamma\\
      &\le& \int_{\{e^{-\gamma/2}-1 <y\le K^{1/2}-1\}} q (1+y)^2 2^\gamma |\log(1+y)|^\gamma 
      \\ &\le& 2^\gamma\int_{\{e^{-\gamma/2}-1 <y\le c  \}} (1+y)^2 |y|^{\gamma-2}\left(\left(\frac{1}{1+y}\right)^\gamma  I(y<0) +  I(y\ge 0)\right) q y^2\\
      && \quad  +2^\gamma\int_{\{c <y\le   K^{1/2}-1\}} (1+y^{-1})^2  |\log(1+y)|^\gamma  q y^2\\
      &\le& 
     \epsilon^2\times( 2^\gamma\max(e^{-\gamma}e^{\gamma^2/2}(1-e^{-\gamma/2})^{\gamma-2}, (c+1)^2c^{\gamma-2} )  + (1+c^{-1})^2(\log(K))^\gamma).
\end{eqnarray*}
$C$ is  bounded as follows:
\begin{eqnarray*}
    C&=& \int_{\{K<p/q\}} p \left|\log\left(\frac{p}{q}\right)\right|^\gamma \\
    &=& \int_{\{K<p/q\}} p (p/q)^\delta\left|\log\left(\frac{p}{q}\right)/(p/q)^{\delta/\gamma}\right|^\gamma \\
    &\le& M_\delta(\log(K))^\delta/K^\gamma 
\end{eqnarray*}
for $K>e^{\gamma/\delta}$, since $\log(x)/x^{\delta/\gamma}$ is a decreasing function for $x>e^{\gamma/\delta}$. Suppose 
$e^{-\gamma^2/\delta}>\epsilon^2>0$.  The desired bound then follows  by choosing $K=\epsilon^{-2/\gamma}$.
\end{proof}

\subsection{Joint Mixture Normals implies Conditional are Mixture Normal}

For simplicity, we denote the density of a multivariate Normal distribution evaluated at $\bfy$ with mean $\vect\mu$ and variance $\vect\Sigma$ as $\mathcal{N}(\bfy \mid \vect{\mu}, \vect{\Sigma})$. Recall that when the joint density is Normal i.e., not a mixture, both the marginal and conditional densities are also Normal. 

\begin{proof}
Let $\bfX_1$ and $\bfX_2$ represent a partition of $\bfX$, where $f(\bfx) = \sum_{i=1}^k \pi_i\mathcal{N}_i(\bfx \mid\vect{\mu}_i, \vect{\Sigma}_i)$, where $\sum_{i=1}^k \pi_i = 1$

First, writing out the joint density we have,
\[
f(\bfx_1, \bfx_2) = f(\bfx) =  \sum_{i=1}^k \pi_i\mathcal{N}_i(\bfx \mid \vect{\mu}_i, \vect{\Sigma}_i).
\]

Next, looking at the conditional density,
\[
f(\bfx_1 \mid \bfx_2) = \frac{f(\bfx_1, \bfx_2)}{f(\bfx_2)}.
\]

Expanding the denominator first we have,
\[
f(\bfx_2) = \int f(\bfx_1, \bfx_2)d\bfx_1 = \sum_{j=1}^k \pi_j \int \mathcal{N}_j(\bfx_1, \bfx_2 \mid \vect{\mu}_j, \vect{\Sigma}_j)d\bfx_1 = \sum_{j=1}^k \pi_j \mathcal{N}_j(\bfx_2\mid \vect{\mu_{2, j}}, \vect{\Sigma}_{2, j}),
\]
where the last equality follows from the fact that if the joint density is Normal, the marginal will also be Normal.

Now, looking at the numerator of the conditional density, we factor the joint density using Bayes' Theorem (dropping the conditioning on parameters for simplicity),
\[
f(\bfx_1, \bfx_2) =  \sum_{i=1}^k \pi_i\mathcal{N}_i(\bfx_1, \bfx_2 ) = \sum \pi_i \mathcal{N}_i(\bfx_1 \mid\bfx_2) \mathcal{N}_i(\bfx_2),
\]
where the last inequality follows from factoring the joint into a conditional times a marginal, and uses the fact  that a joint Normal density leads to marginal and conditional Normal densities.

If we combine these two results back into the form of the conditional distribution,
\[
f(\bfx_1 \mid \bfx_2) = \frac{f(\bfx_1, \bfx_2)}{f(\bfx_2)} = \frac{ \sum \pi_i \mathcal{N}_i(\bfx_1 \mid\bfx_2) \mathcal{N}_i(\bfx_2) }{ \sum_{j=1}^k \pi_j \mathcal{N}_j(\bfx_2) } =\sum_{i=1}^k  \Bigg\{\frac{\pi_i \mathcal{N}_i(\bfx_2)}{ \sum_{j=1}^k \pi_j \mathcal{N}_j(\bfx_2) }\Bigg\} \mathcal{N}_i(\bfx_1 \mid\bfx_2).
\]

The term in $\{ \}$ can be seen as a weight, which is a function of only $\bfx_2$. It is also easy to see that it sums to be 1. Therefore, we can write the conditional distribution as,
\[
f(\bfx_1 \mid \bfx_2) = \sum_{i=1}^k\pi_i(\bfx_2) \mathcal{N}_i(\bfx_1 \mid\bfx_2).
\]
\end{proof}

\section{ACF \& PACF Plots}

\begin{figure}[ht]
    \centering
\includegraphics[scale = 0.35]{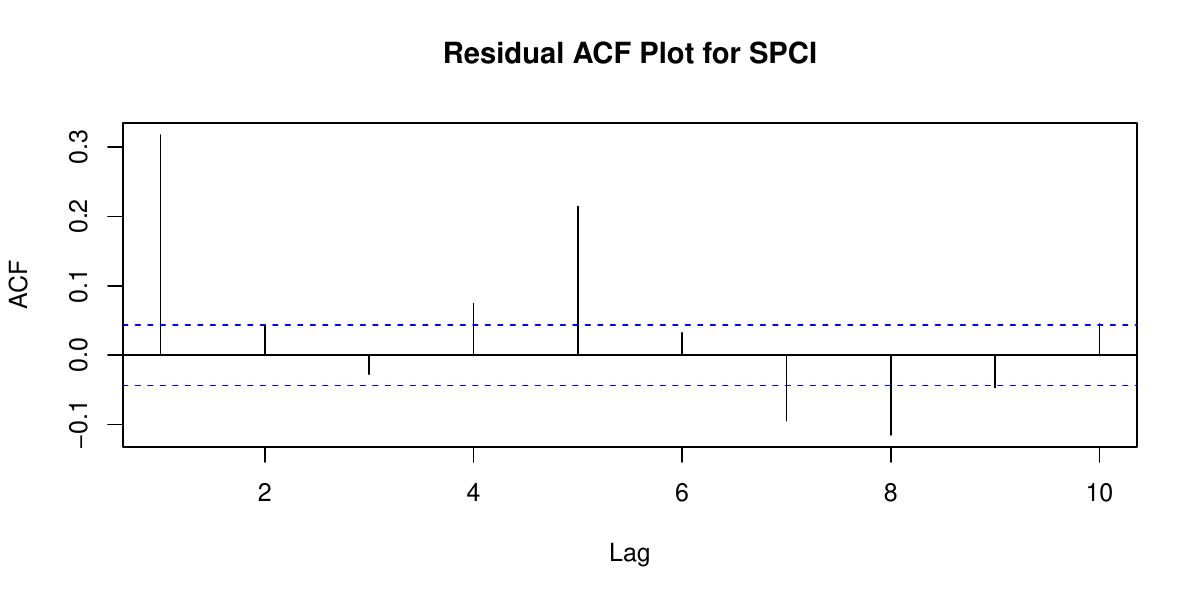}
\includegraphics[scale = 0.35]{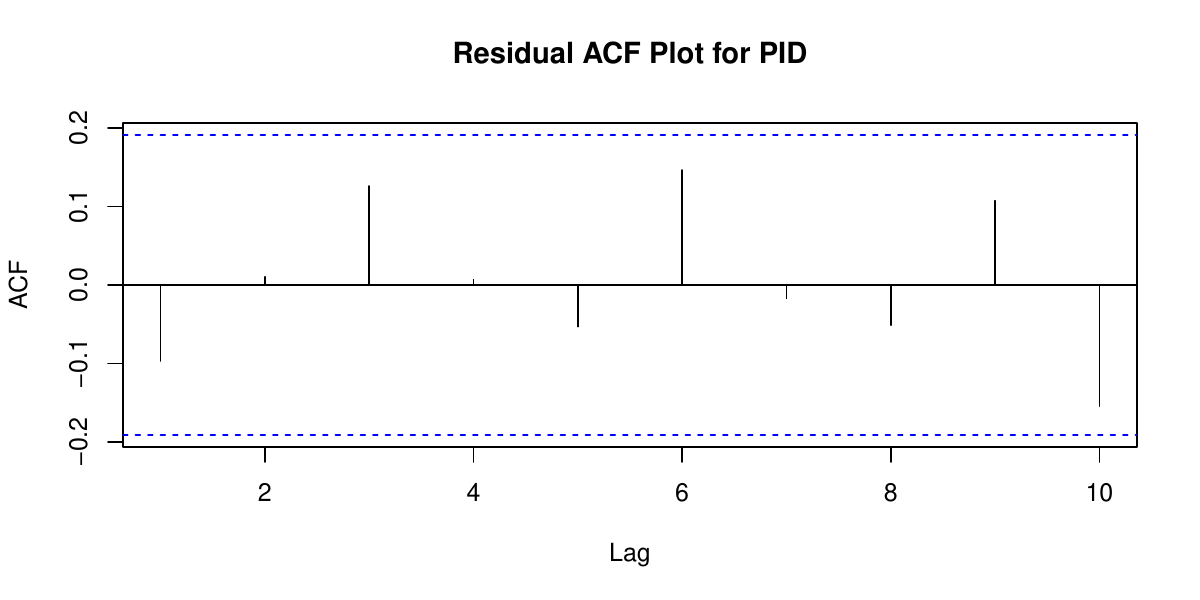}
\includegraphics[scale = 0.35]{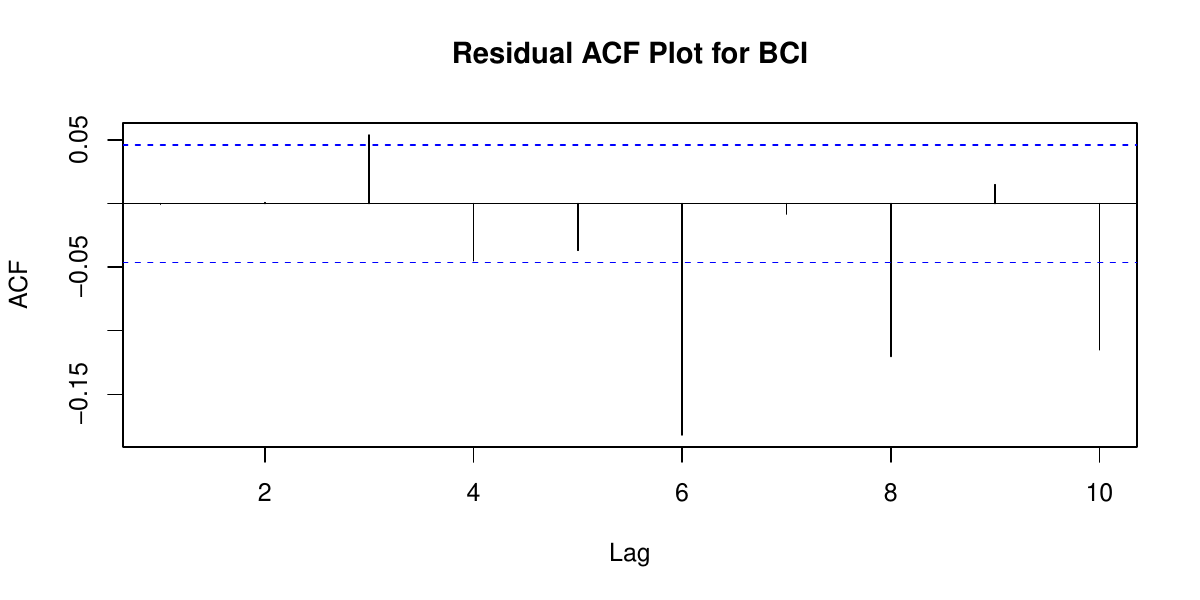}
\includegraphics[scale = 0.35]{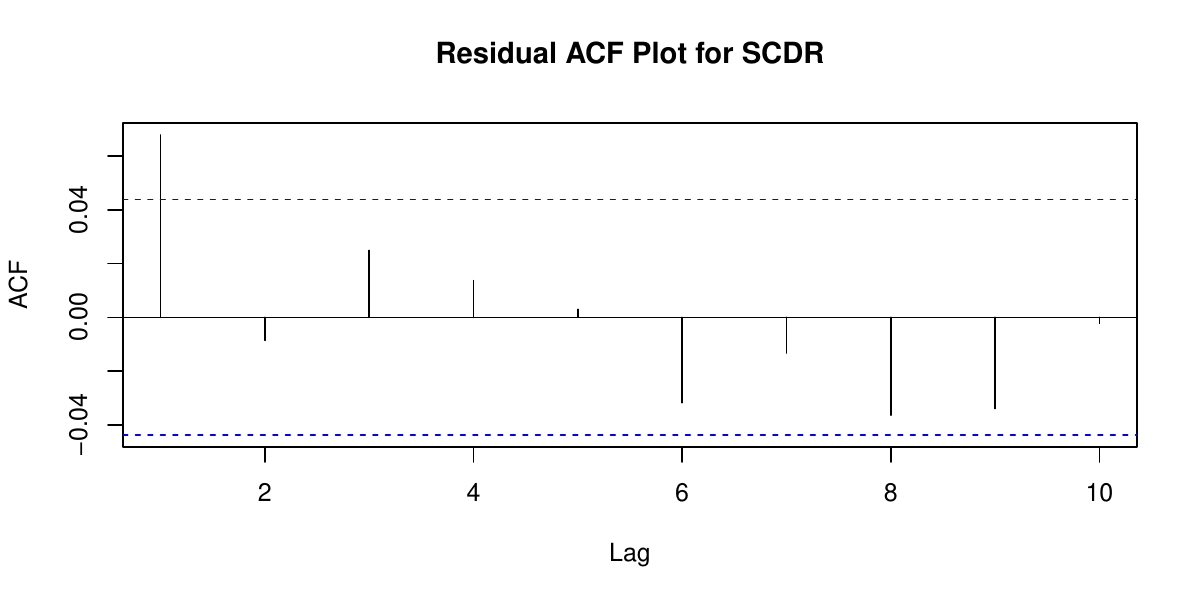}

\caption{ACF Plots for the 2,000 Observed Data Points, Bootstrap Approach Simulation. From top left to right: SPCI, PID, BCI, SCDR}\label{fig:acf_non_stat_boot_TT2000}
\end{figure}

\begin{figure}[ht]
    \centering
\includegraphics[scale = 0.35]{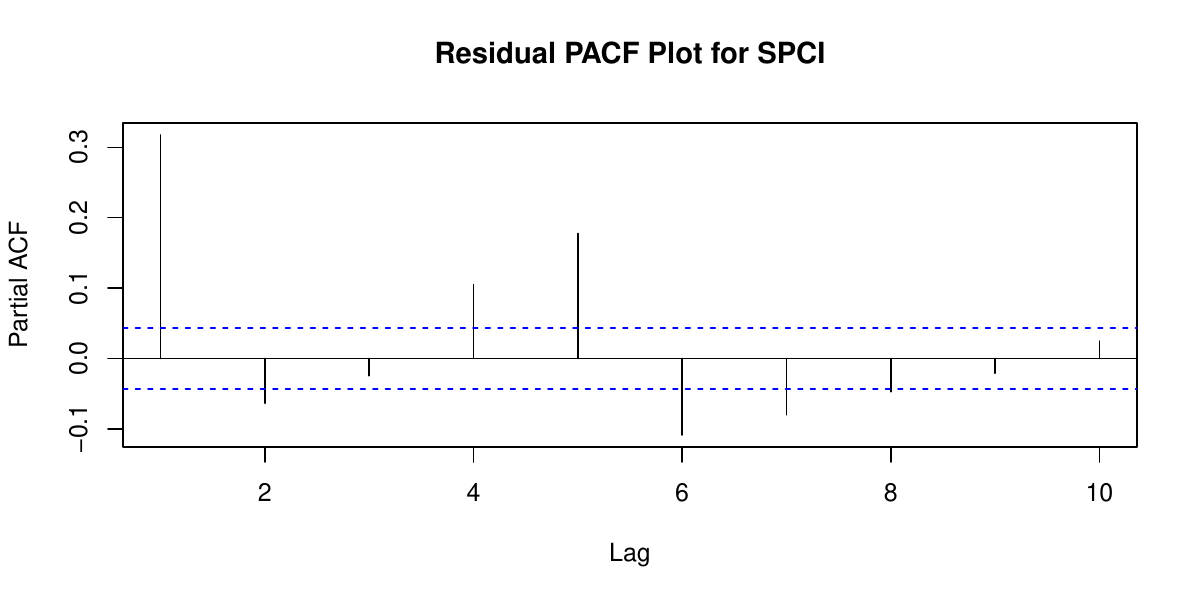}
\includegraphics[scale = 0.35]{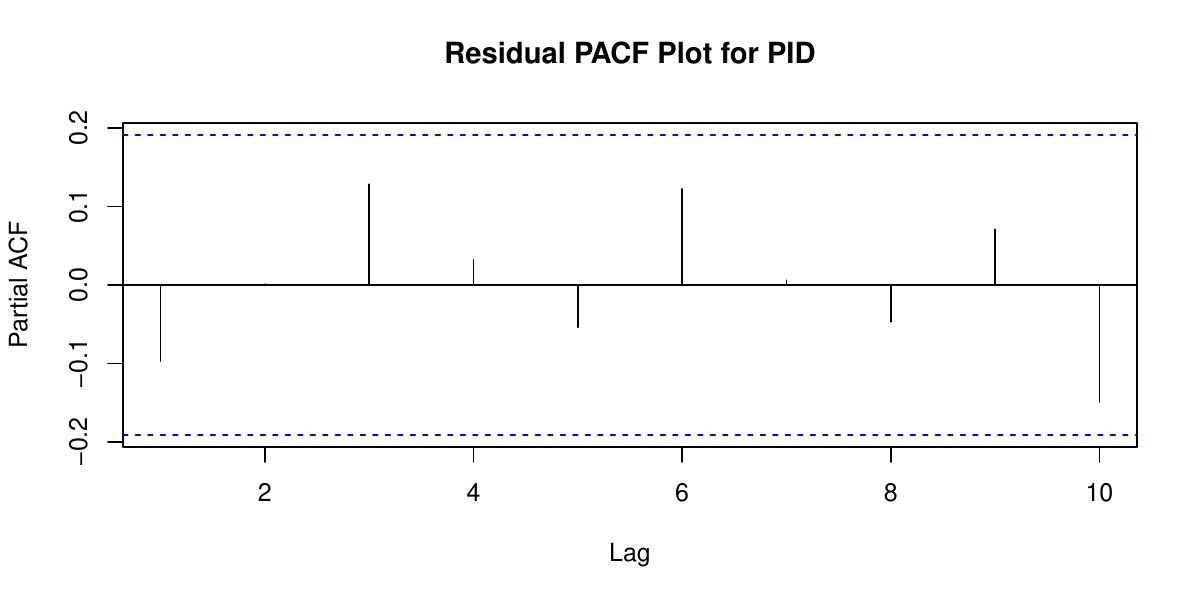}
\includegraphics[scale = 0.35]{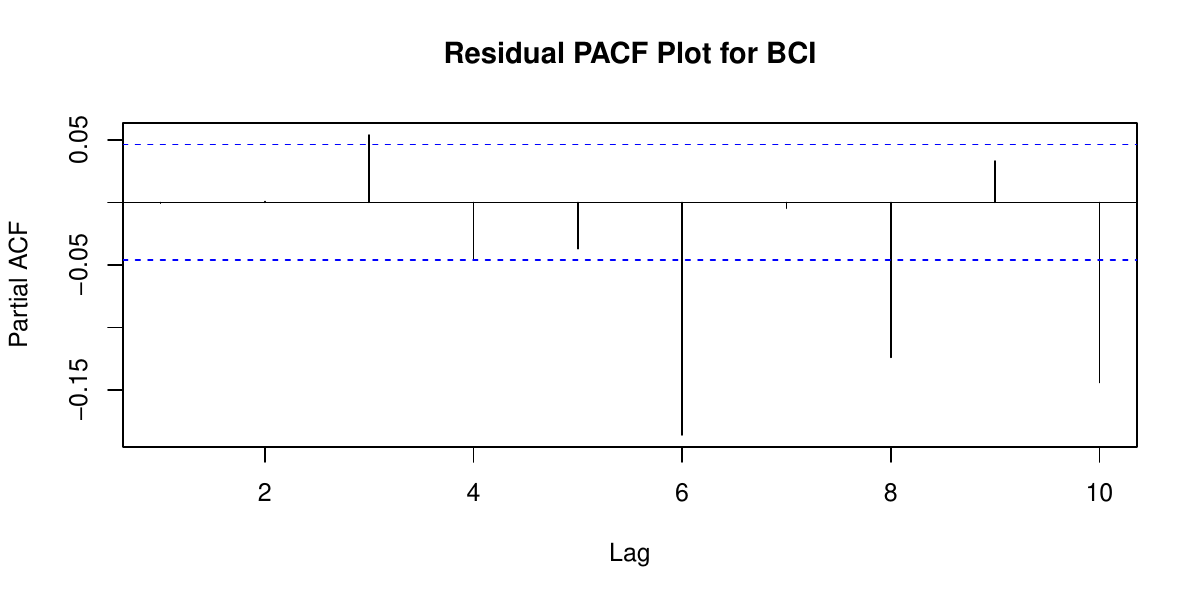}
\includegraphics[scale = 0.35]{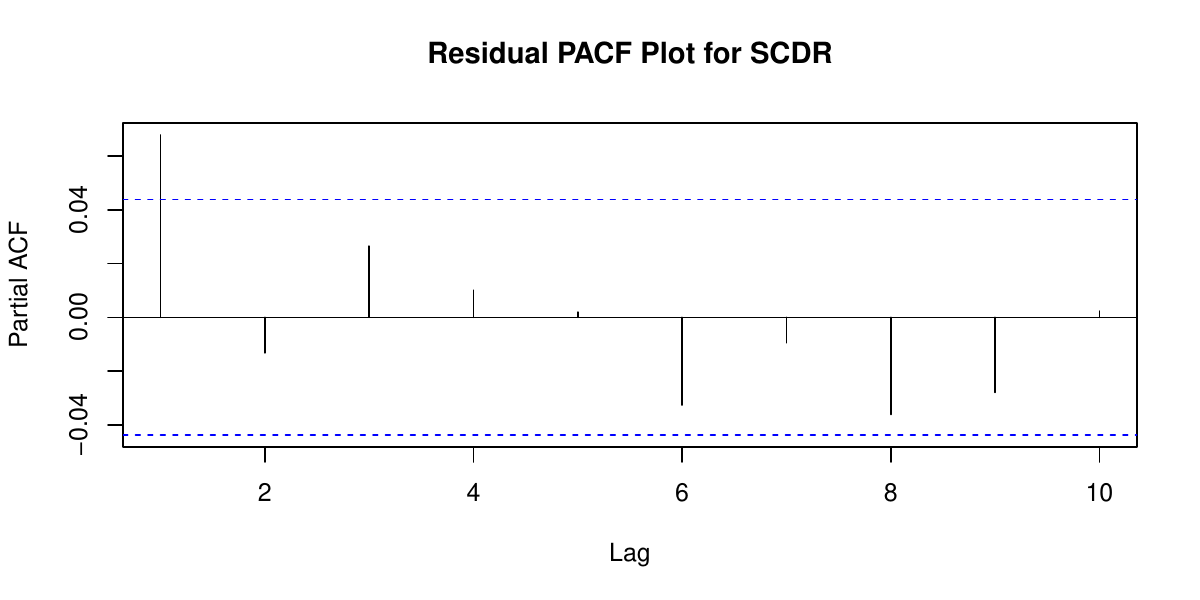}

\caption{Partial ACF Plots for the 2,000 Observed Data Points, Bootstrap Approach Simulation. From top left to right: SPCI, PID, BCI, SCDR }\label{fig:pacf_non_stat_boot_TT2000}
\end{figure}

\begin{figure}[ht]
    \centering
\includegraphics[scale = 0.35]{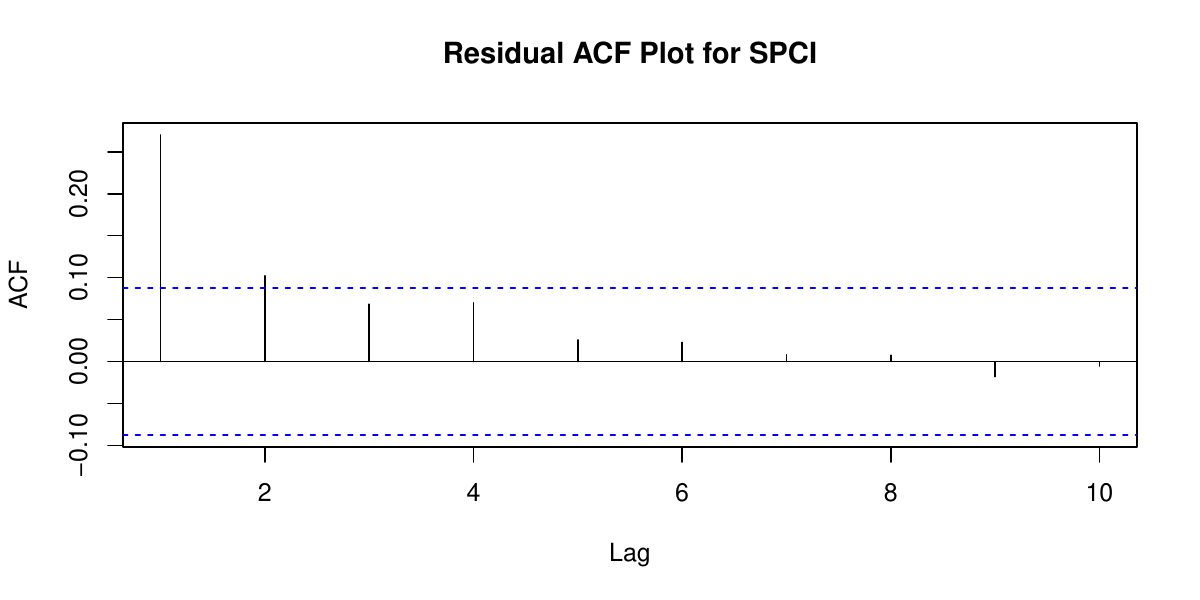}
\includegraphics[scale = 0.35]{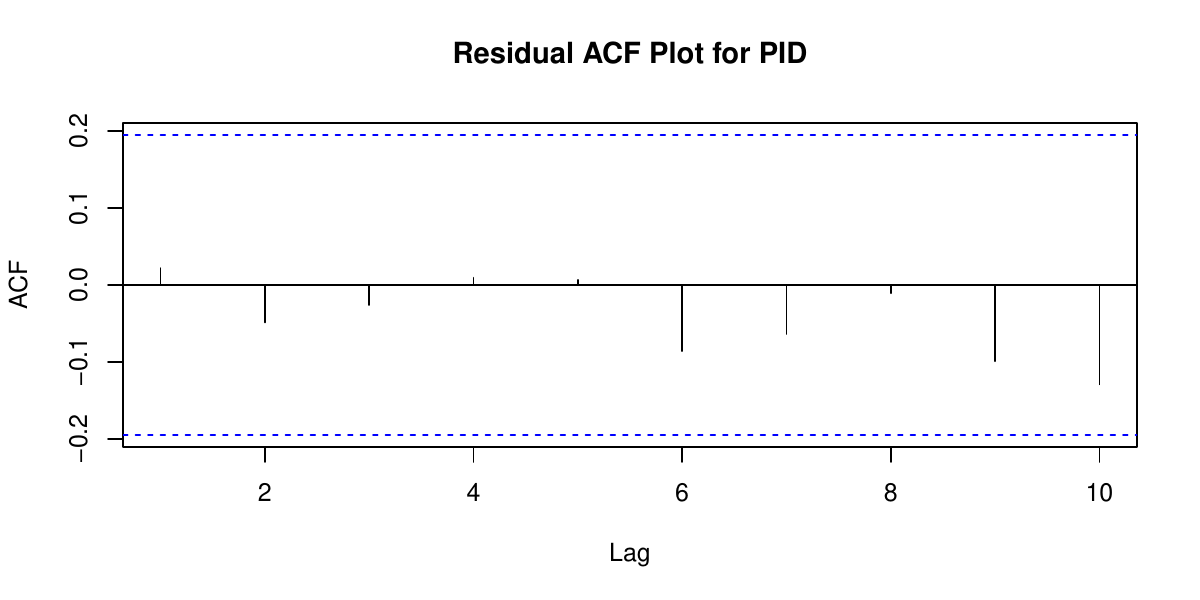}
\includegraphics[scale = 0.35]{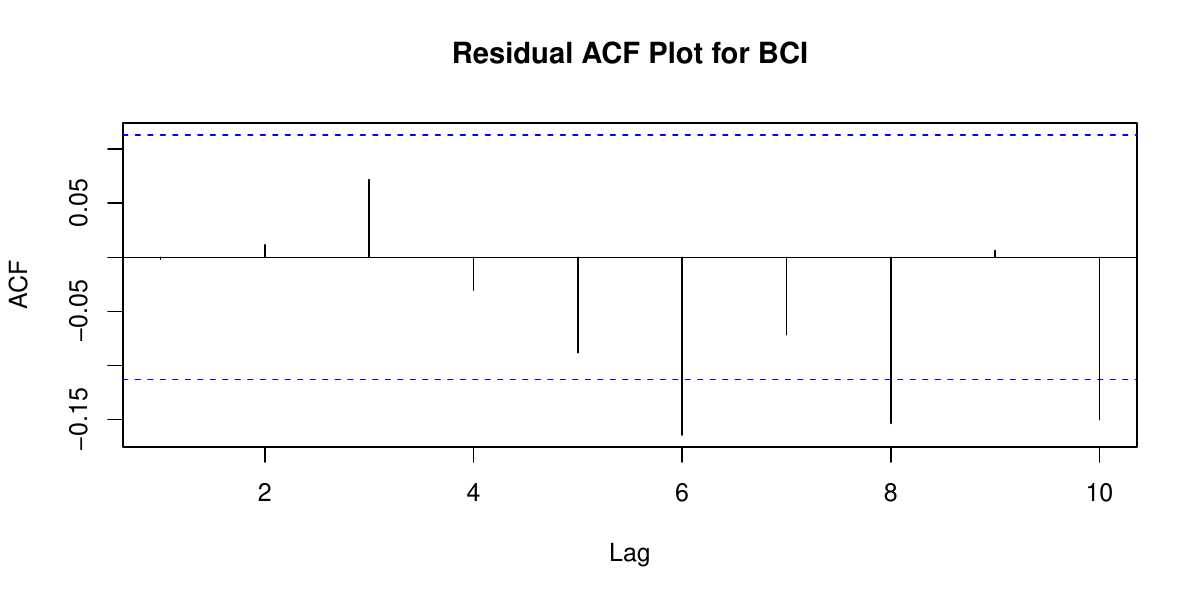}
\includegraphics[scale = 0.35]{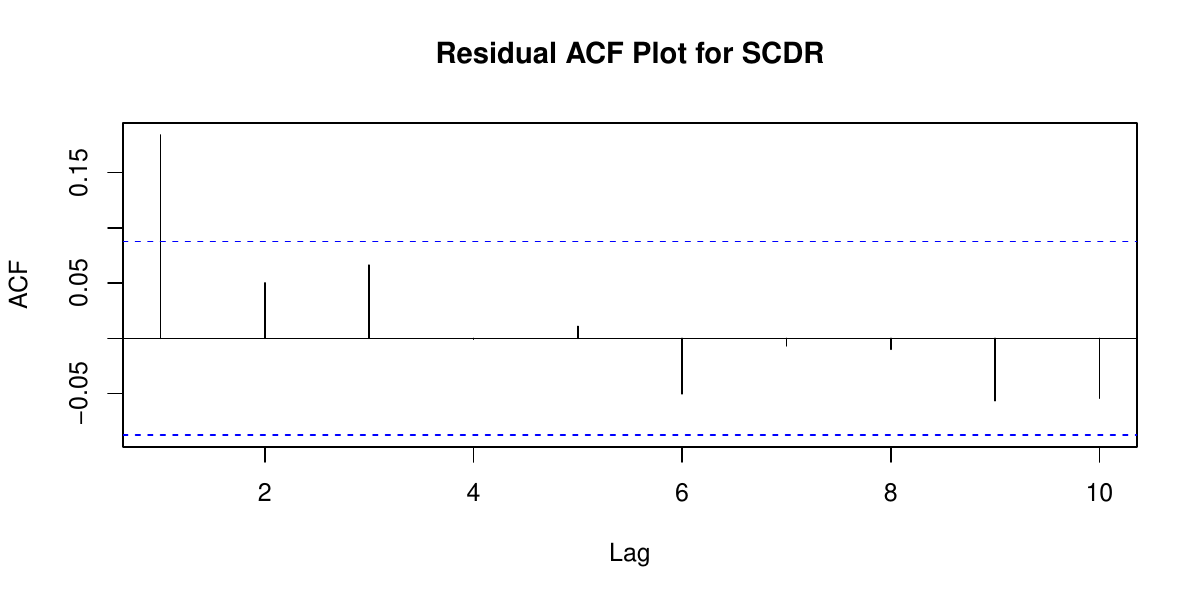}

\caption{ACF Plots for the 500 Observed Data Points, Leave One Out Approach Simulation. From top left to right: SPCI, PID, BCI, SCDR}\label{fig:acf_non_stat_loo_TT500}
\end{figure}

\begin{figure}[ht]
    \centering
\includegraphics[scale = 0.35]{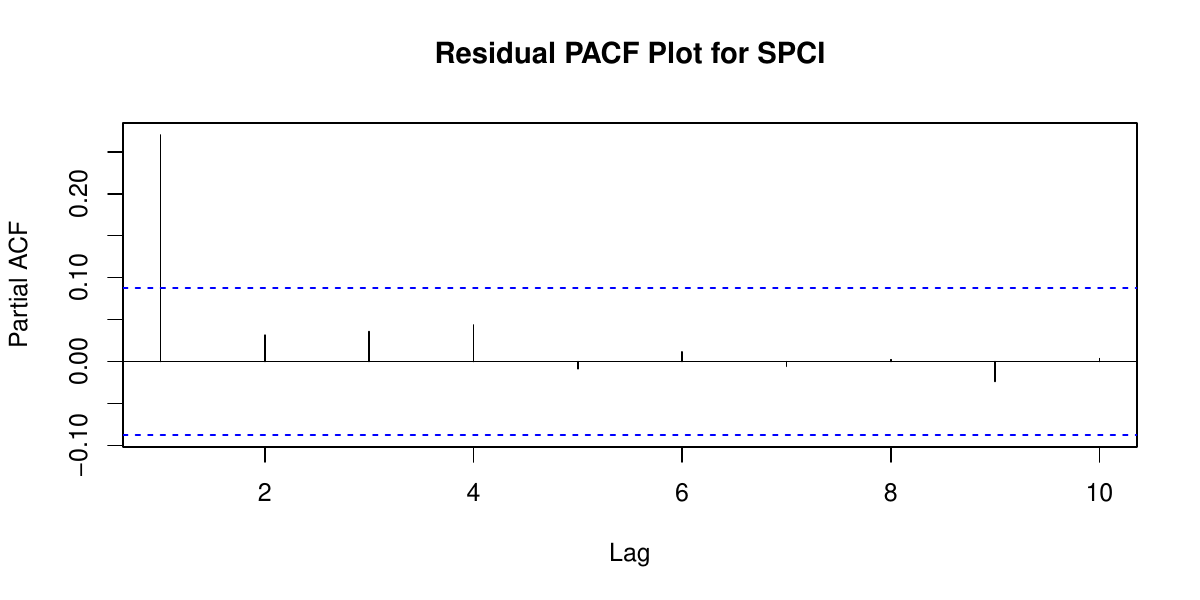}
\includegraphics[scale = 0.35]{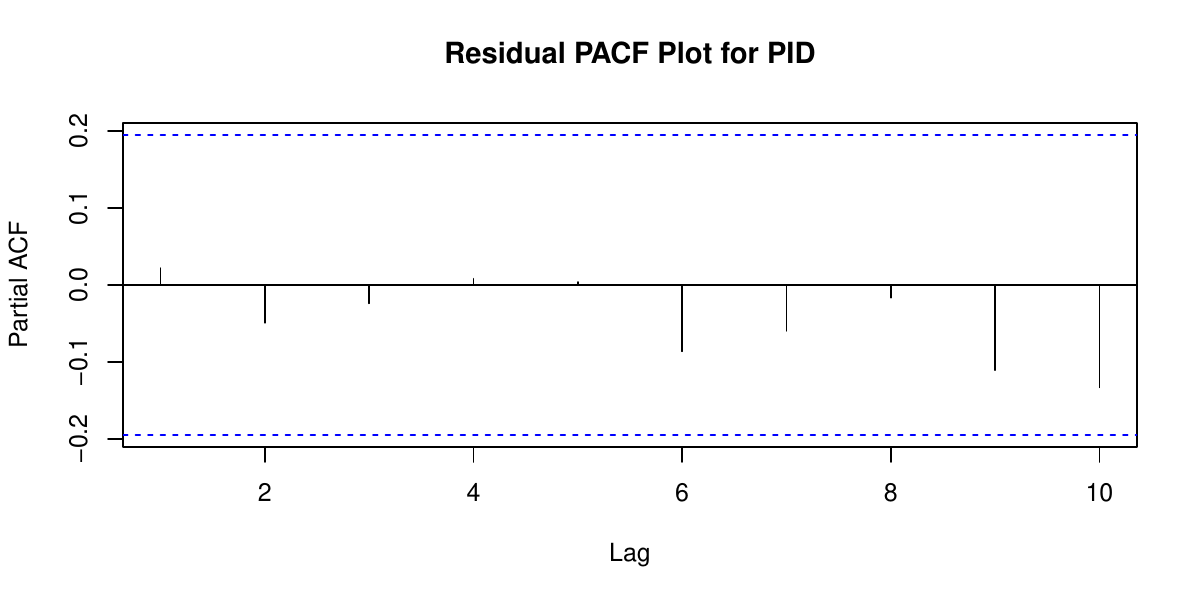}
\includegraphics[scale = 0.35]{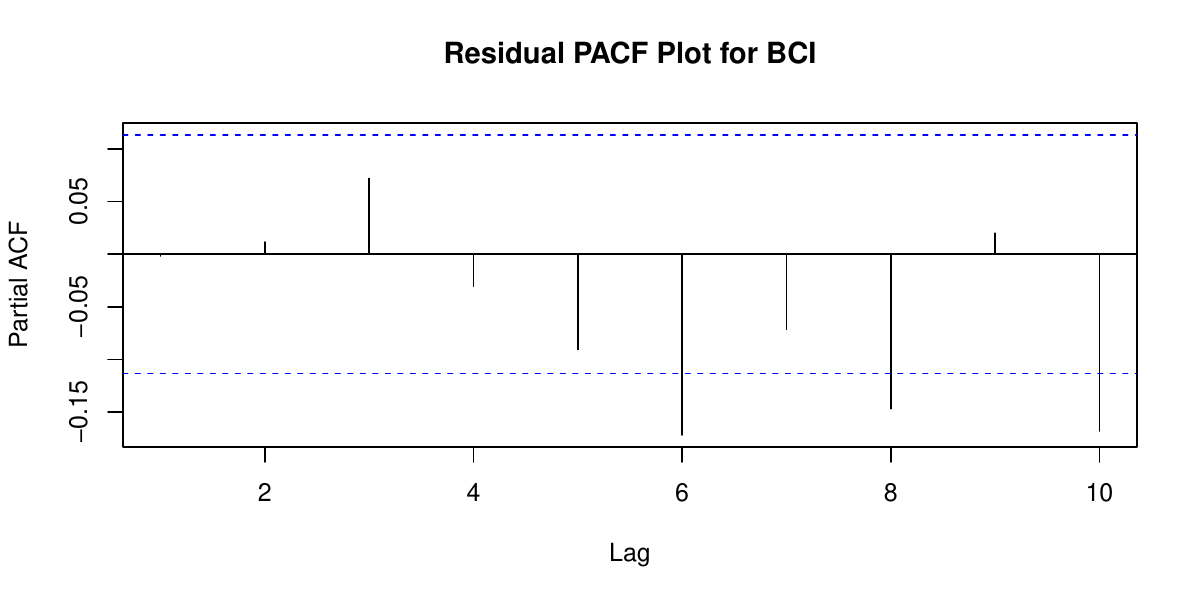}
\includegraphics[scale = 0.35]{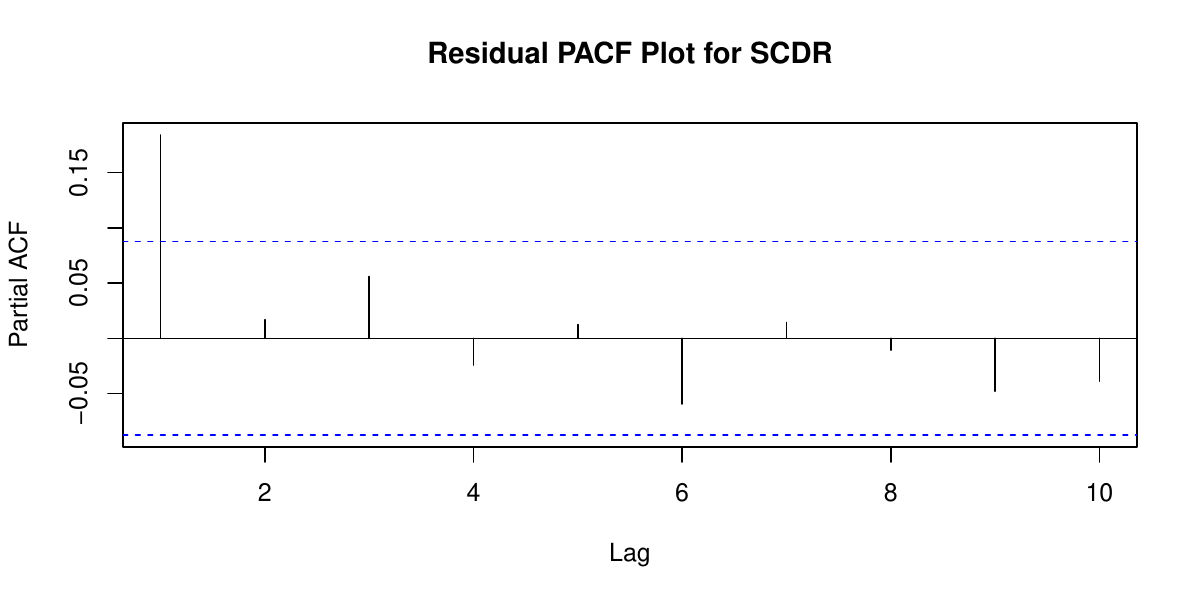}

\caption{Partial ACF Plots for the 500 Observed Data Points, Leave One Out Approach Simulation. From top left to right: SPCI, PID, BCI, SCDR}\label{fig:pacf_non_stat_loo_TT500}
\end{figure}

\begin{figure}[ht]
    \centering
\includegraphics[scale = 0.35]{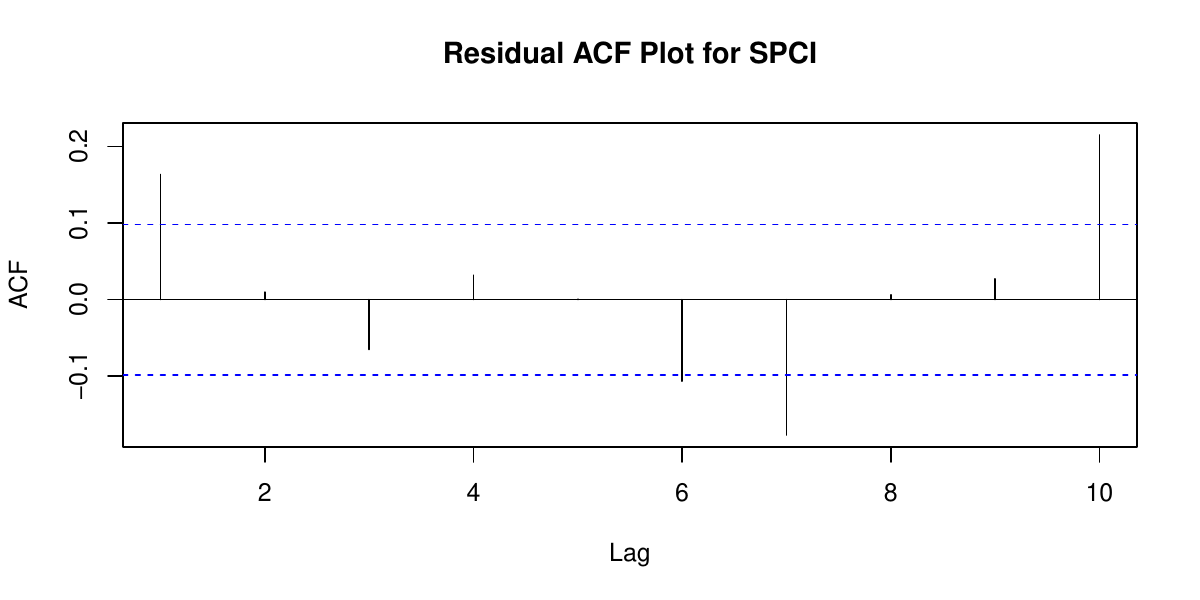}
\includegraphics[scale = 0.35]{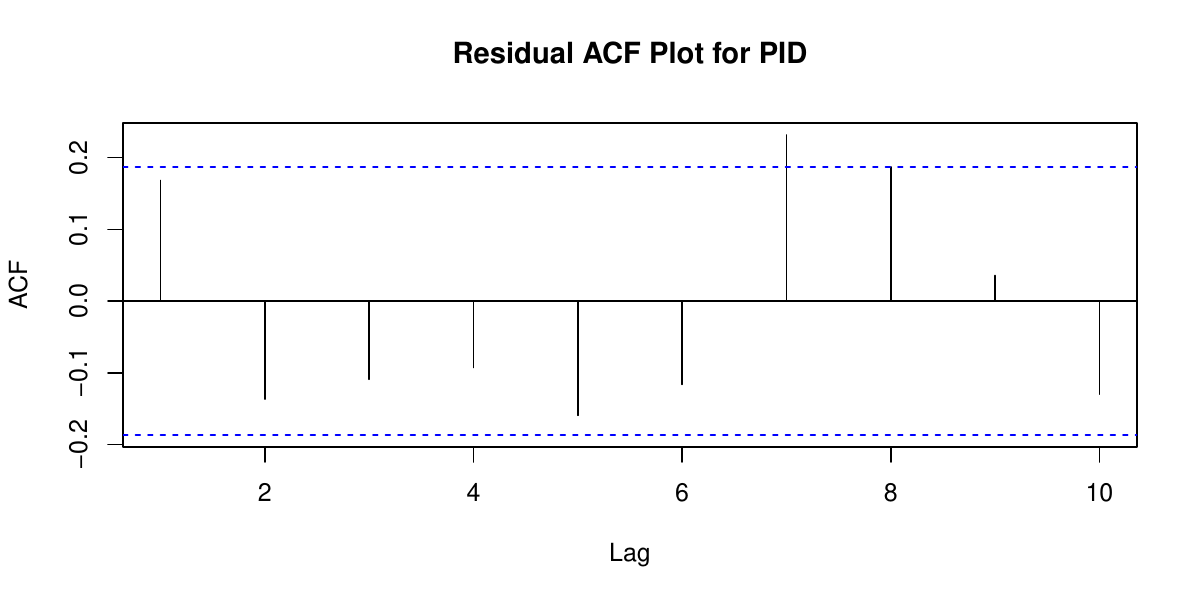}
\includegraphics[scale = 0.35]{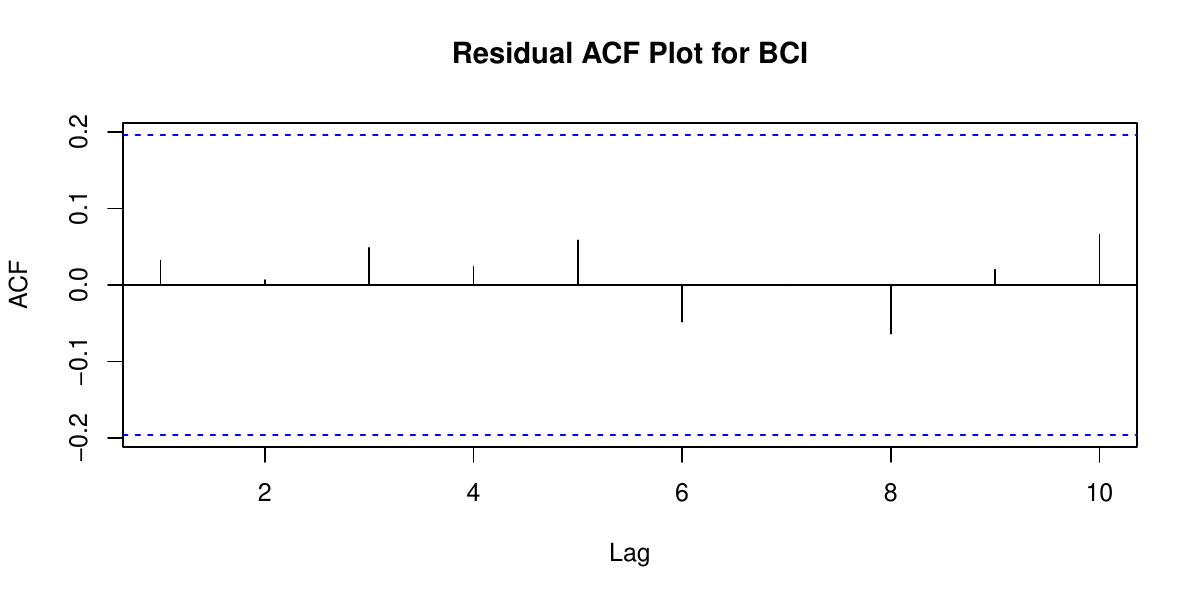}
\includegraphics[scale = 0.35]{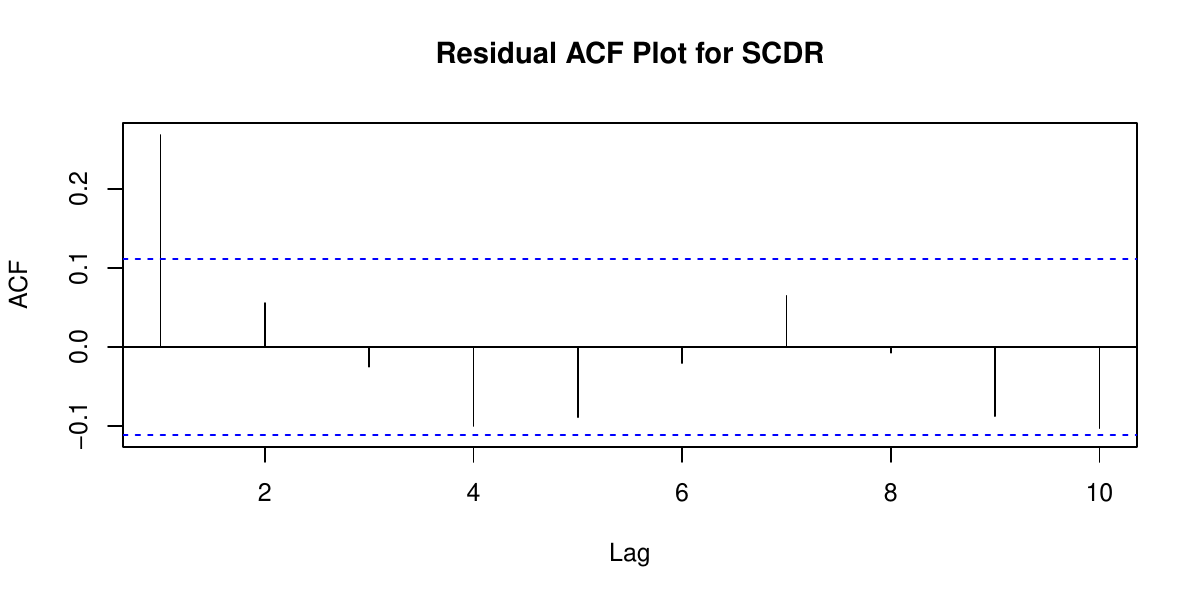}

\caption{ACF Plots for the AR(2) Simulation. From top left to right: SPCI, PID, BCI, SCDR}\label{fig:acf_AR2}
\end{figure}

\begin{figure}[ht]
    \centering
\includegraphics[scale = 0.35]{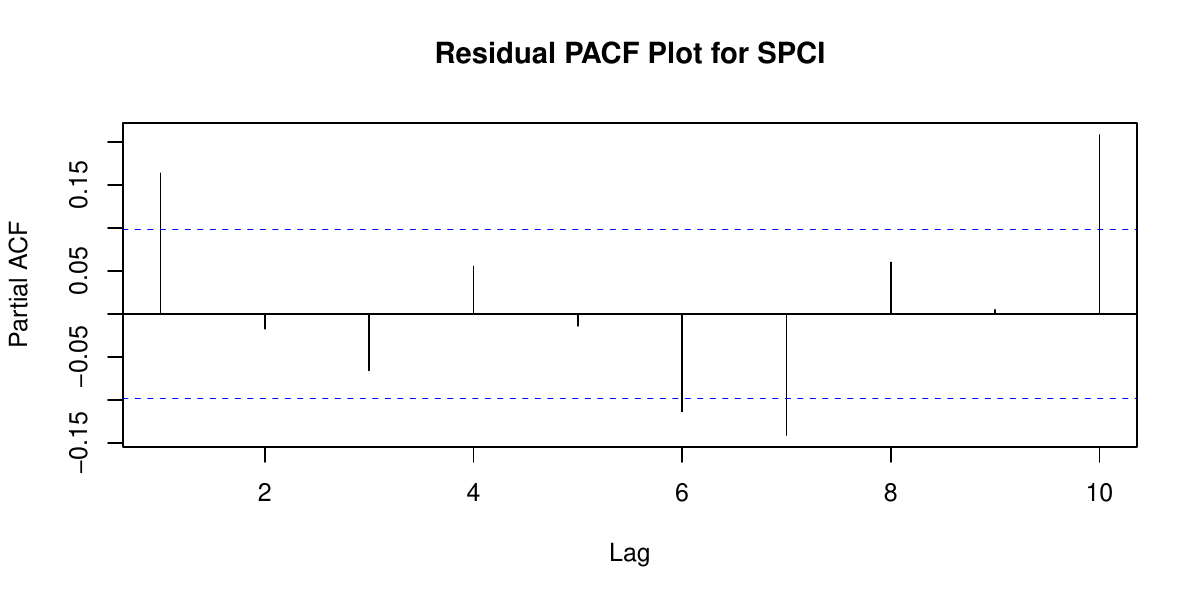}
\includegraphics[scale = 0.35]{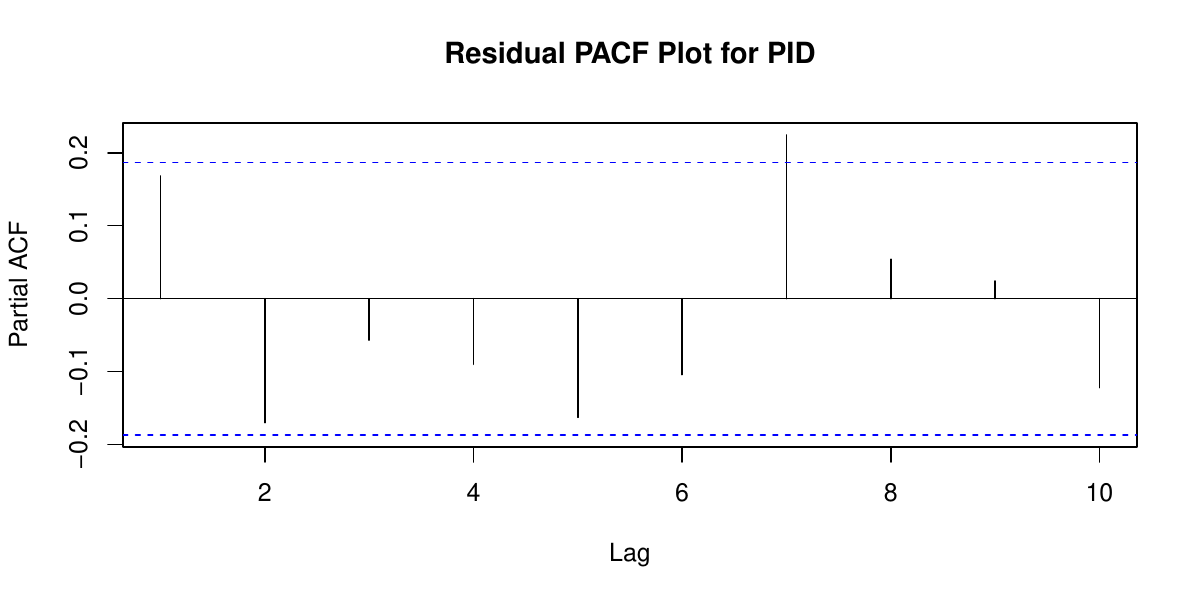}
\includegraphics[scale = 0.35]{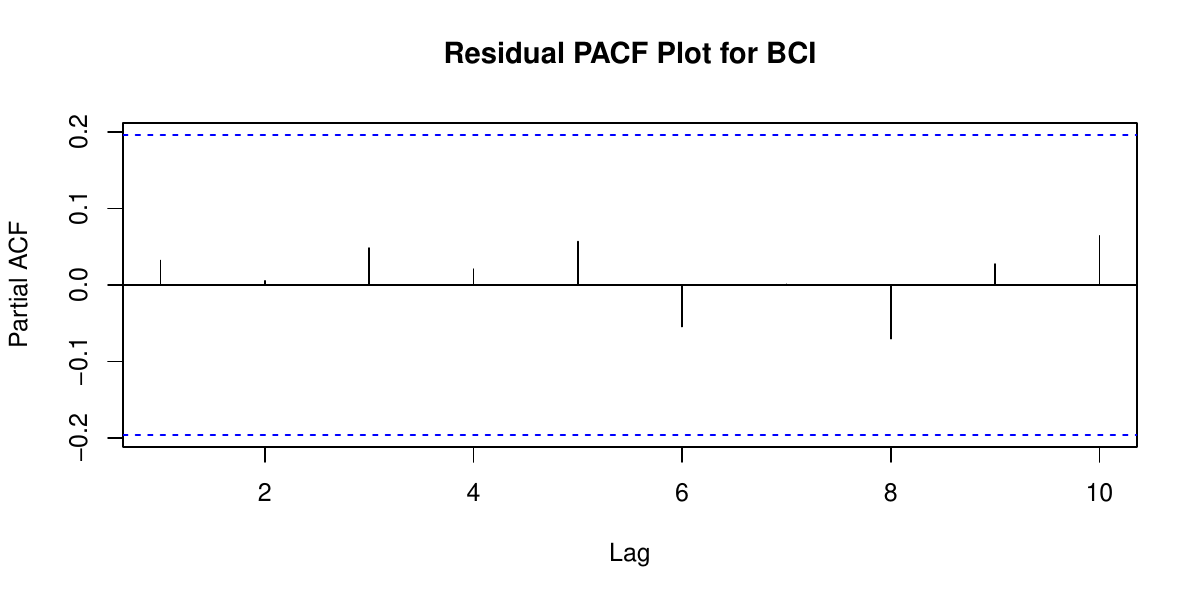}
\includegraphics[scale = 0.35]{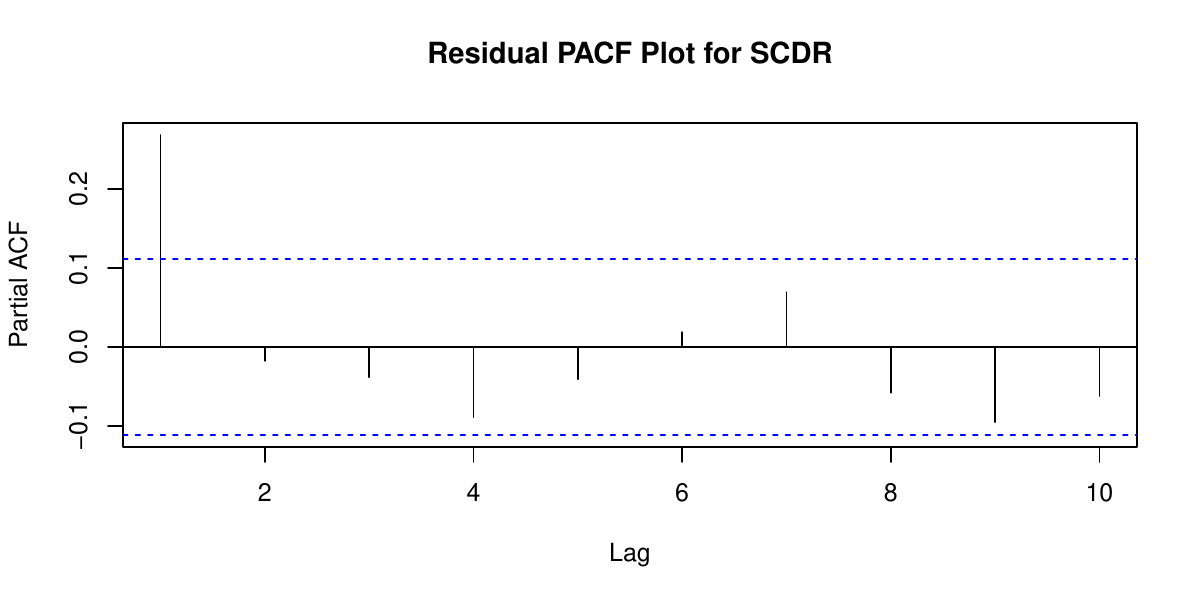}

\caption{Partial ACF Plots for the AR(2) Simulation. From top left to right: SPCI, PID, BCI, SCDR}\label{fig:pacf_AR2}
\end{figure}

\begin{figure}[ht]
    \centering
\includegraphics[scale = 0.35]{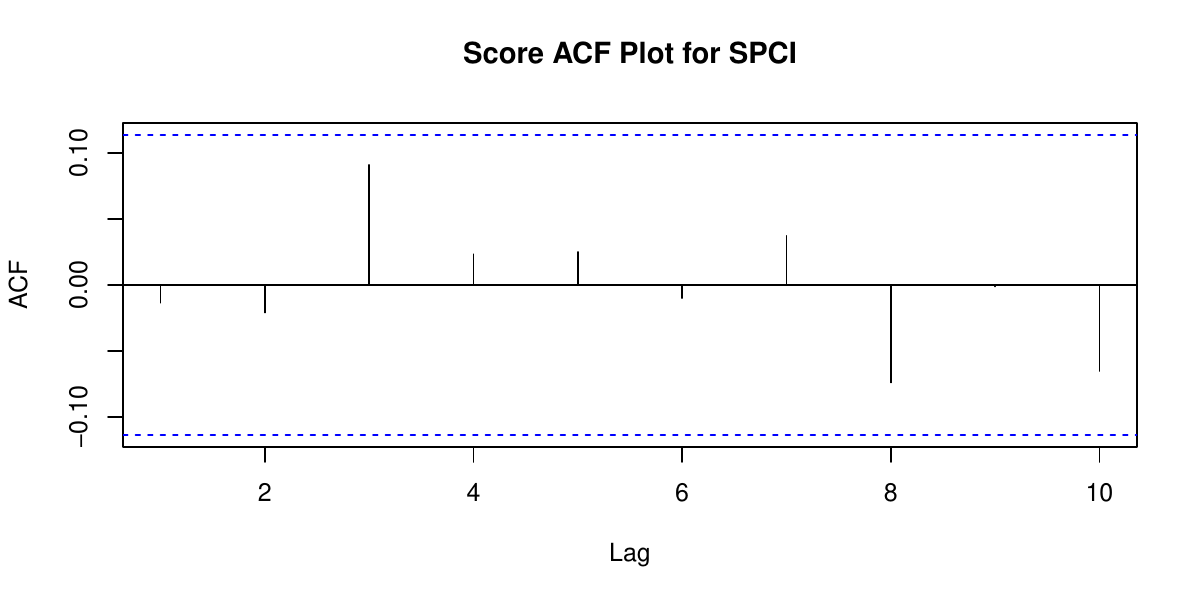}
\includegraphics[scale = 0.35]{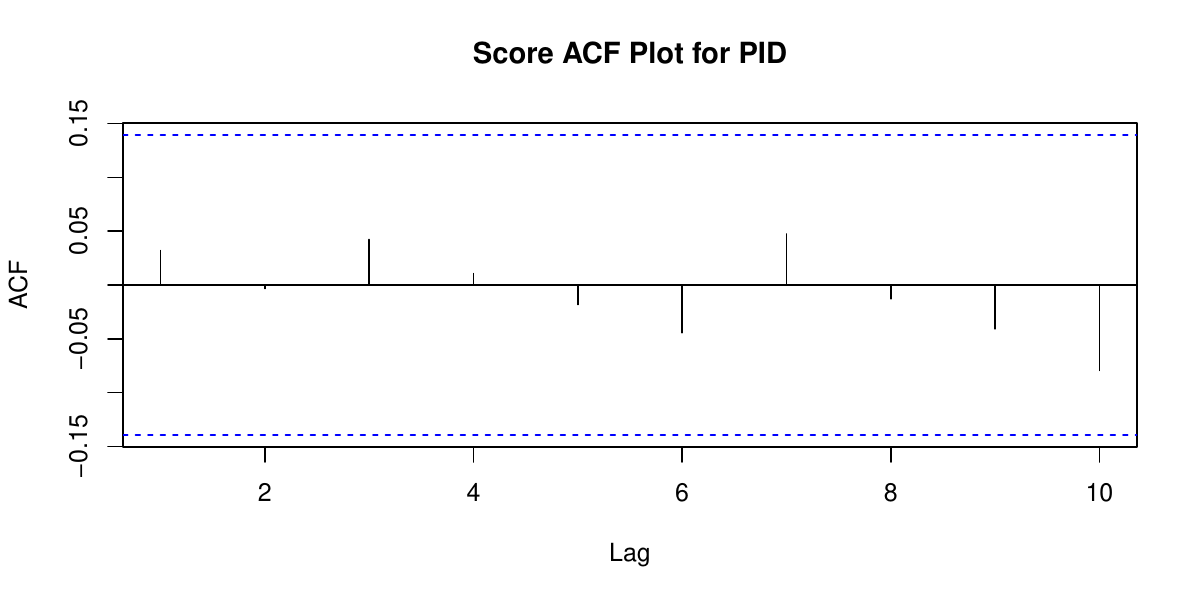}
\includegraphics[scale = 0.35]{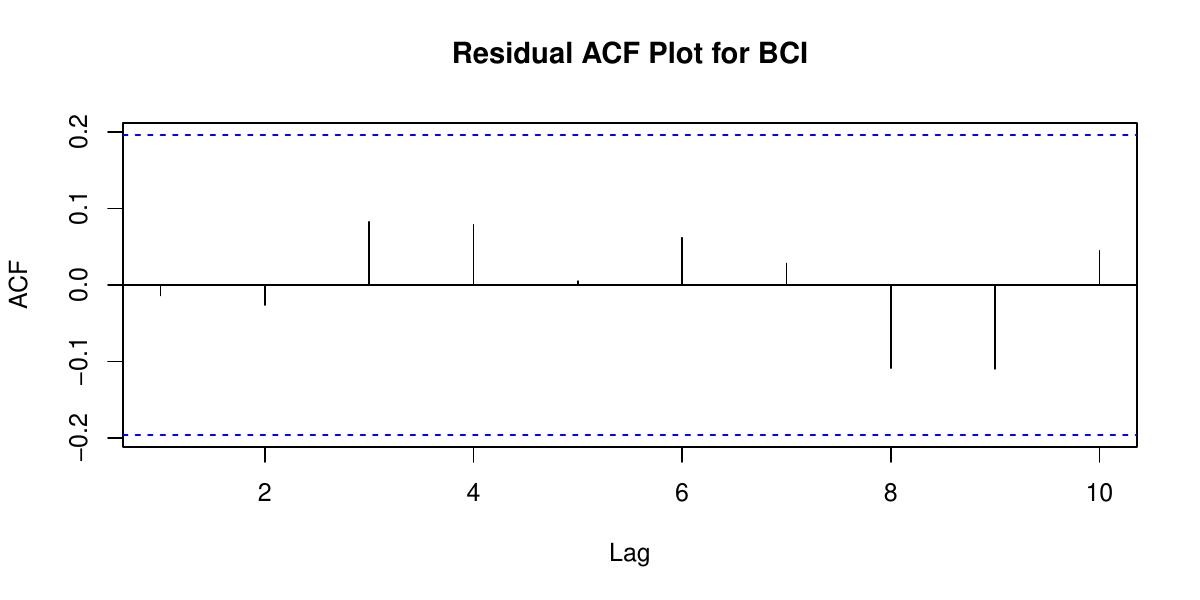}
\includegraphics[scale = 0.35]{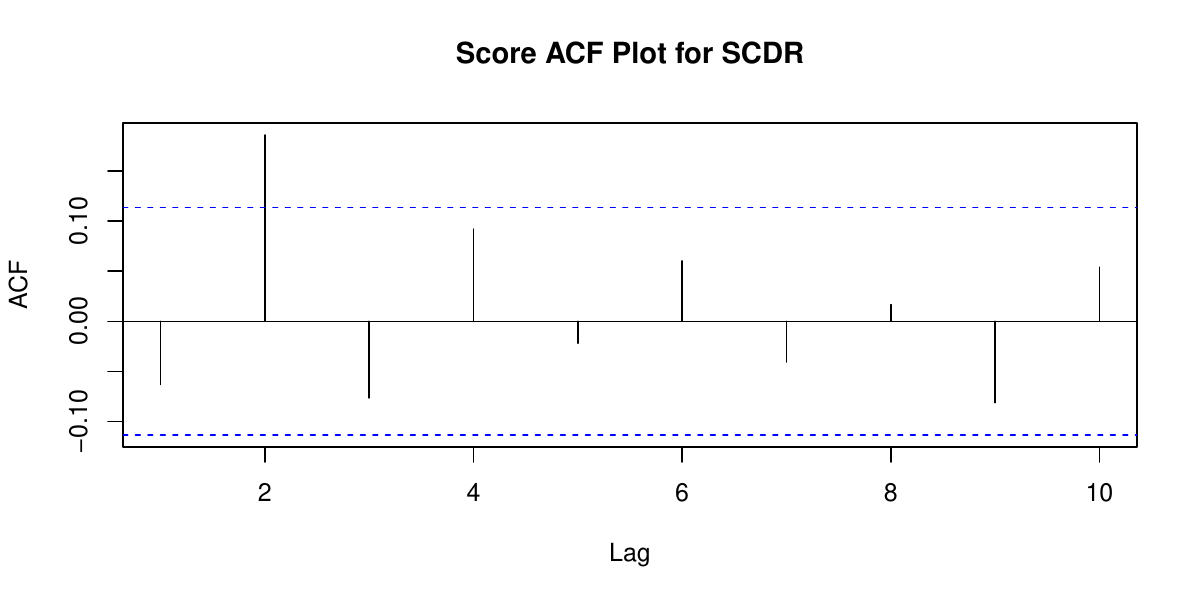}

\caption{ACF Plots for the Old Faithful Eruption Scores. From top left to right: SPCI, PID, BCI, SCDR}\label{fig:acf_geyser}
\end{figure}

\begin{figure}[ht]
    \centering
\includegraphics[scale = 0.35]{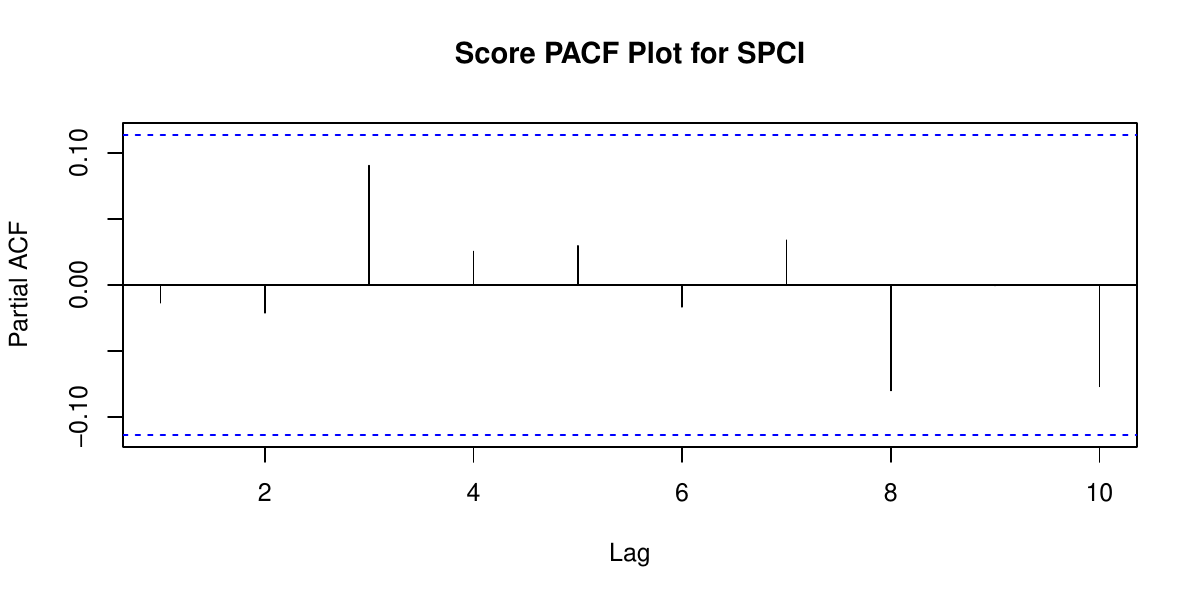}
\includegraphics[scale = 0.35]{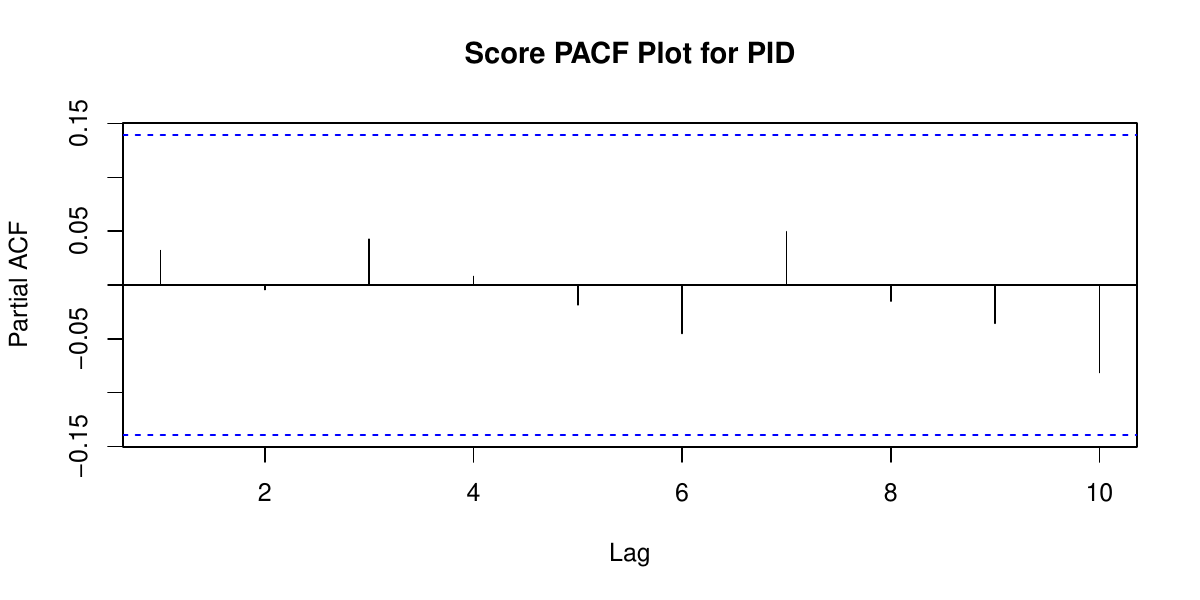}
\includegraphics[scale = 0.35]{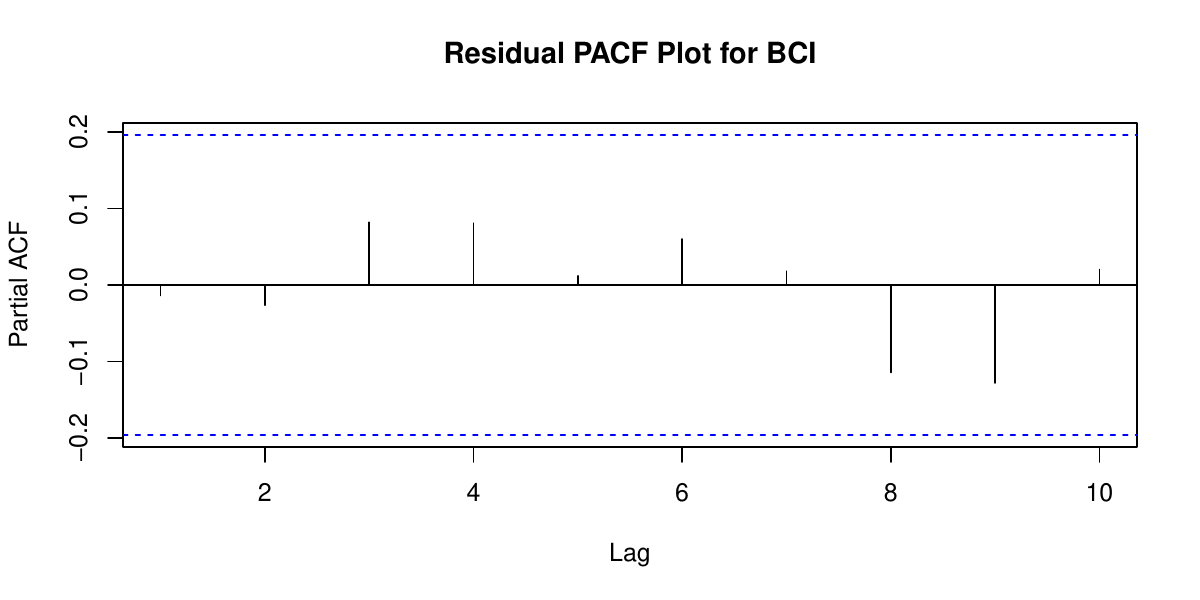}
\includegraphics[scale = 0.35]{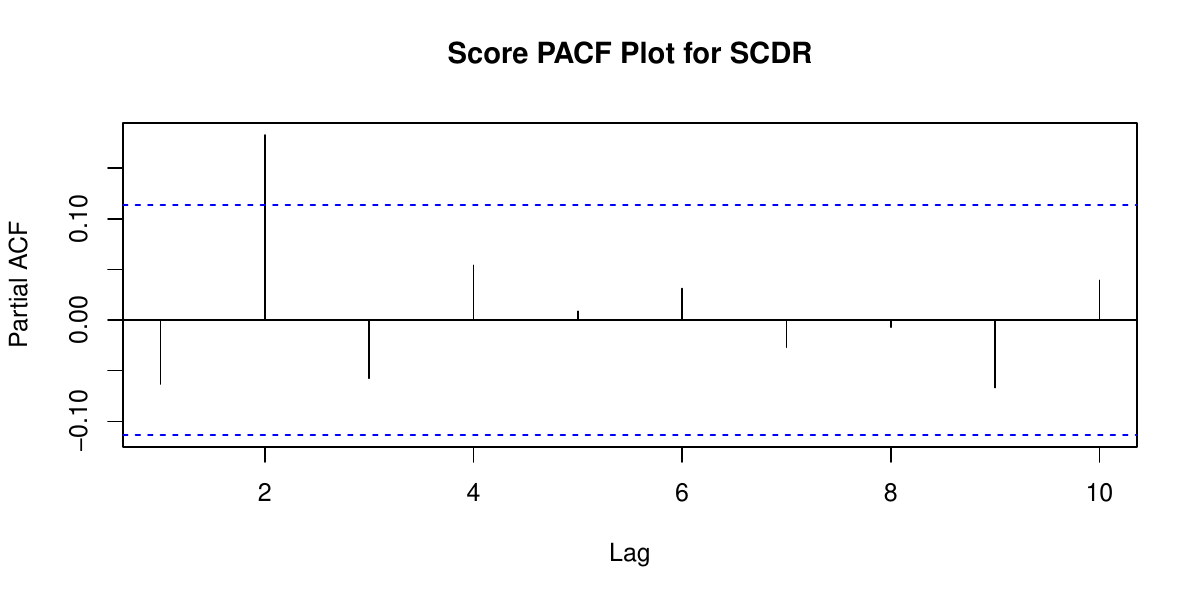}

\caption{PACF Plots for the Old Faithful Eruption Scores. From top left to right: SPCI, PID, BCI, SCDR}\label{fig:pacf_geyser}
\end{figure}

\begin{figure}[ht]
    \centering
\includegraphics[scale = 0.35]{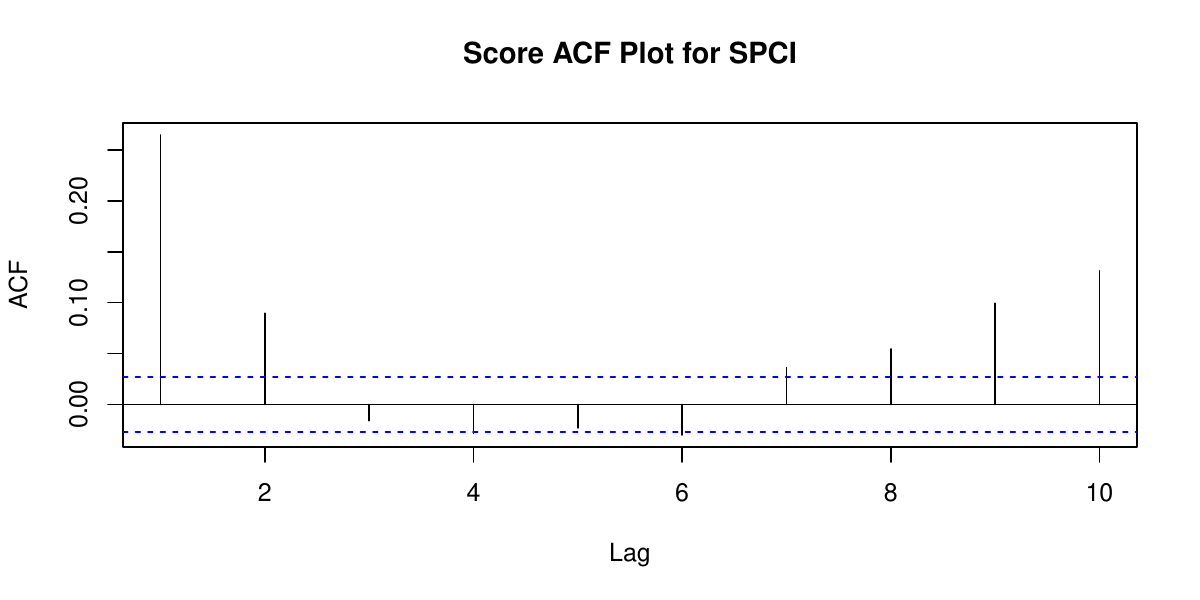}
\includegraphics[scale = 0.35]{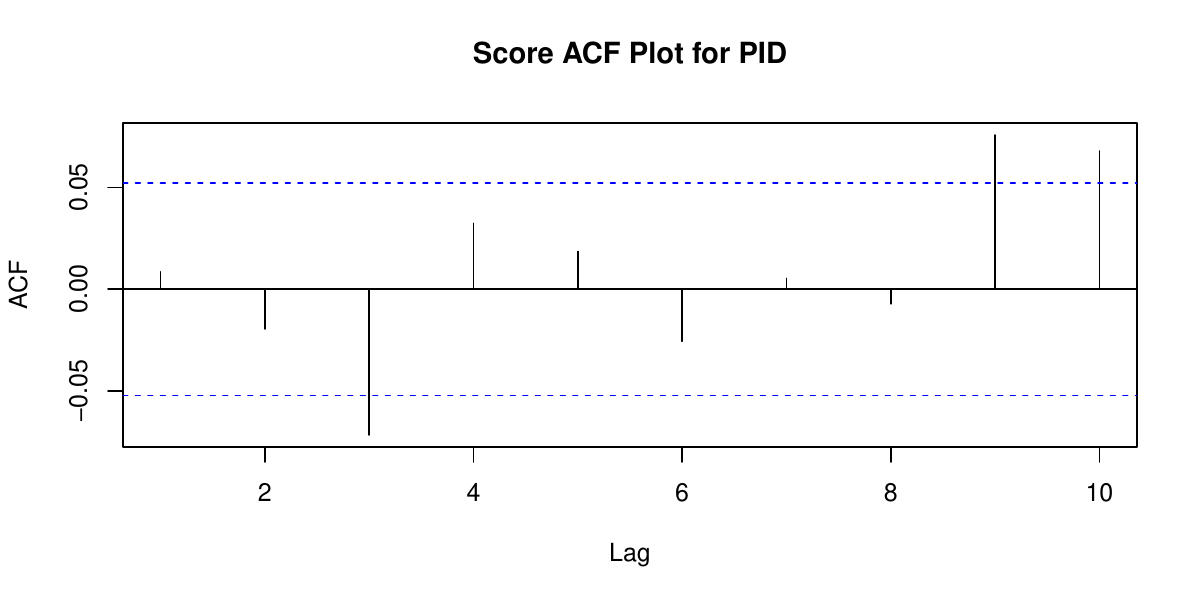}
\includegraphics[scale = 0.35]{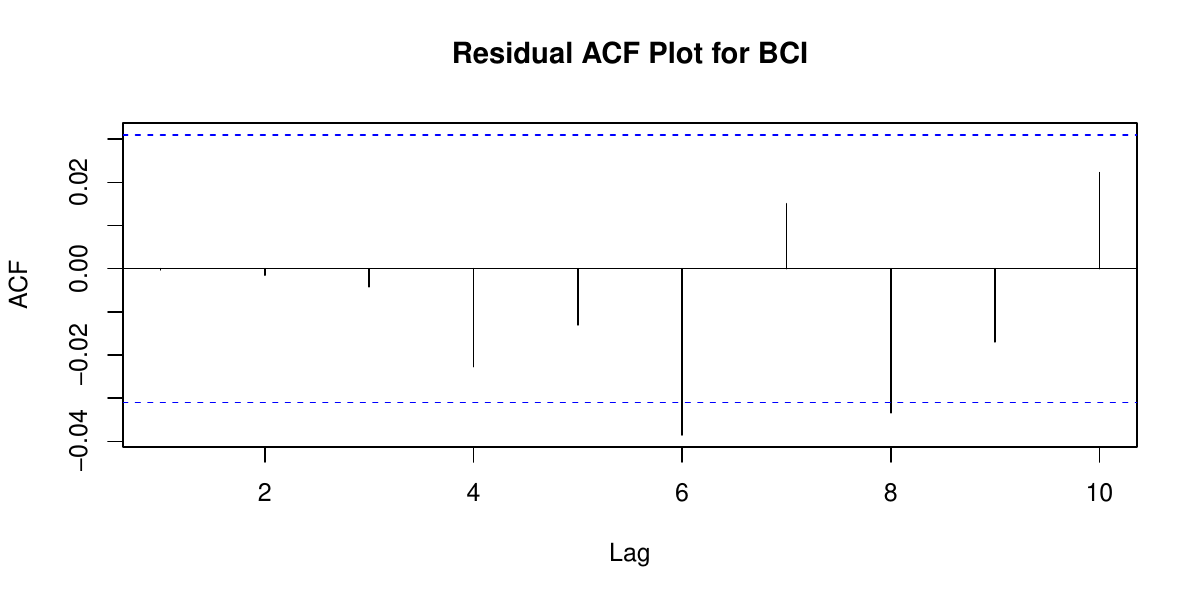}
\includegraphics[scale = 0.35]{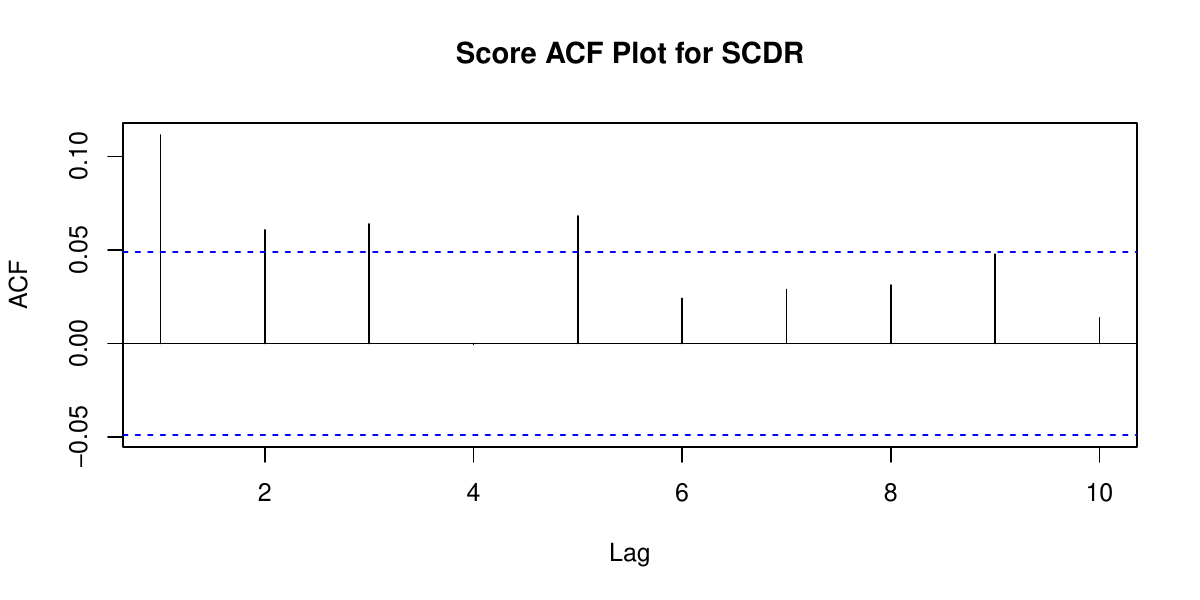}

\caption{ACF Plots for the Electricity Scores. From top left to right: SPCI, PID, BCI, SCDR}\label{fig:acf_elec}
\end{figure}

\begin{figure}[ht]
    \centering
\includegraphics[scale = 0.35]{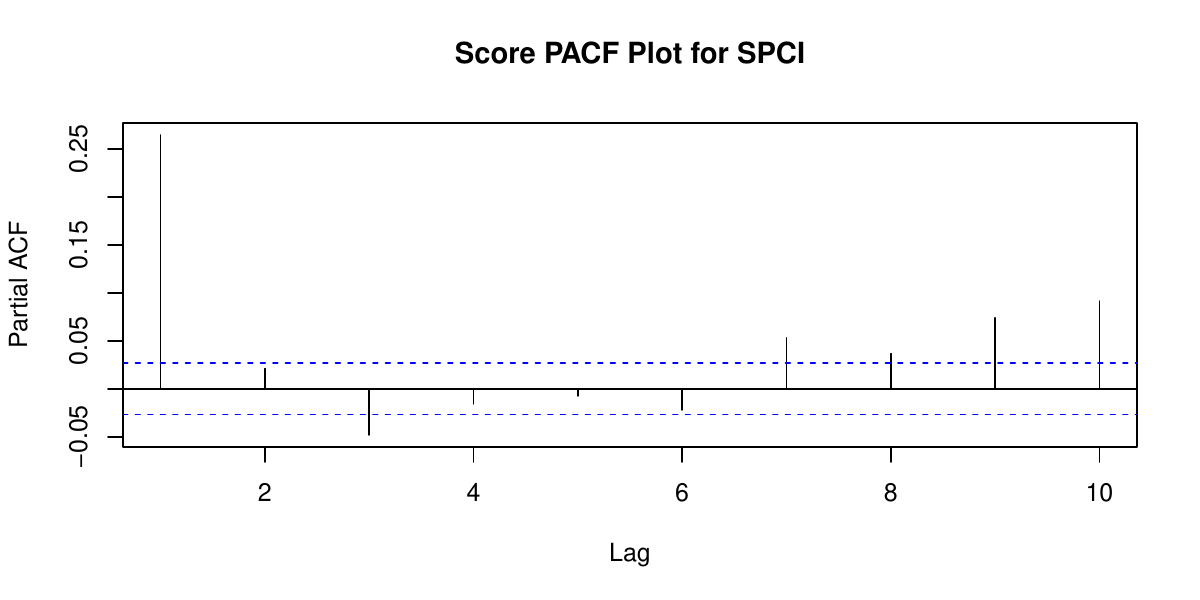}
\includegraphics[scale = 0.35]{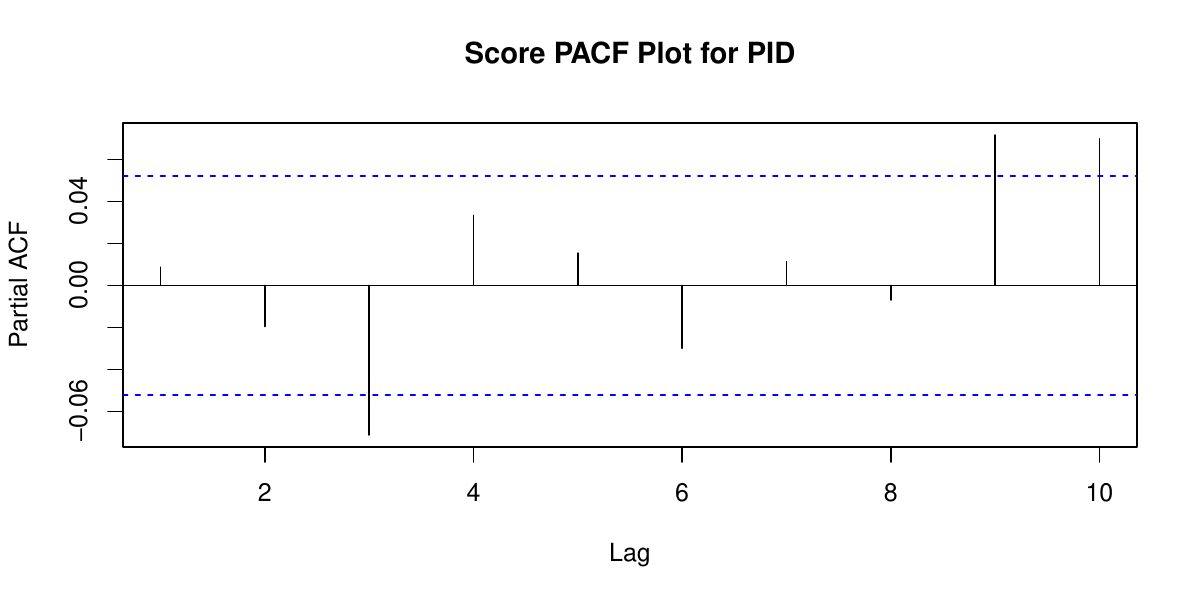}
\includegraphics[scale = 0.35]{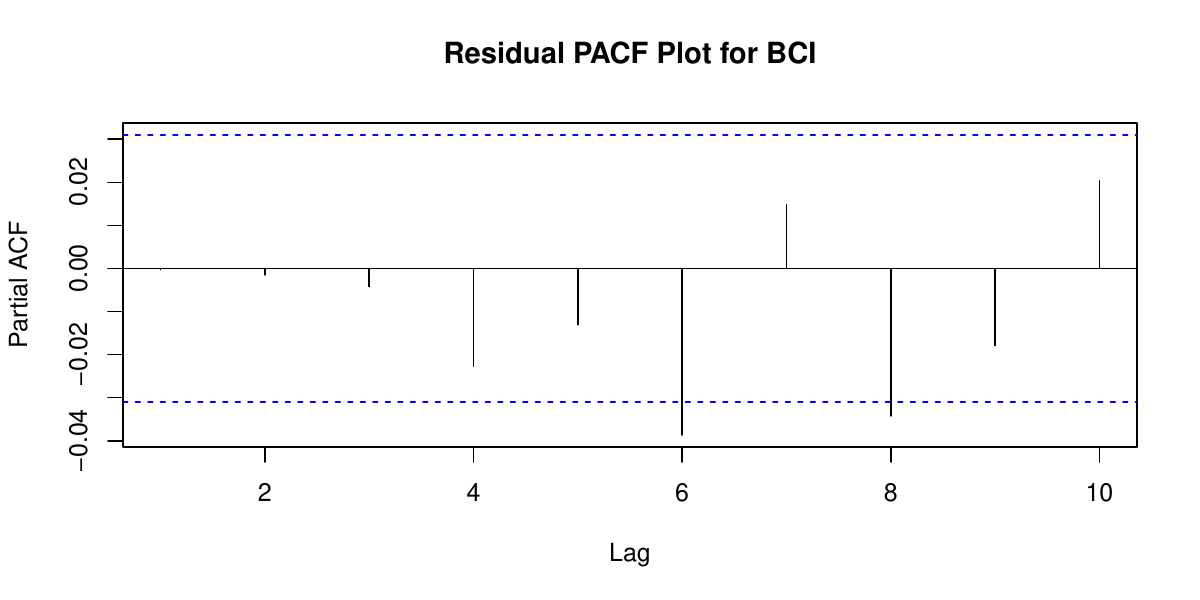}
\includegraphics[scale = 0.35]{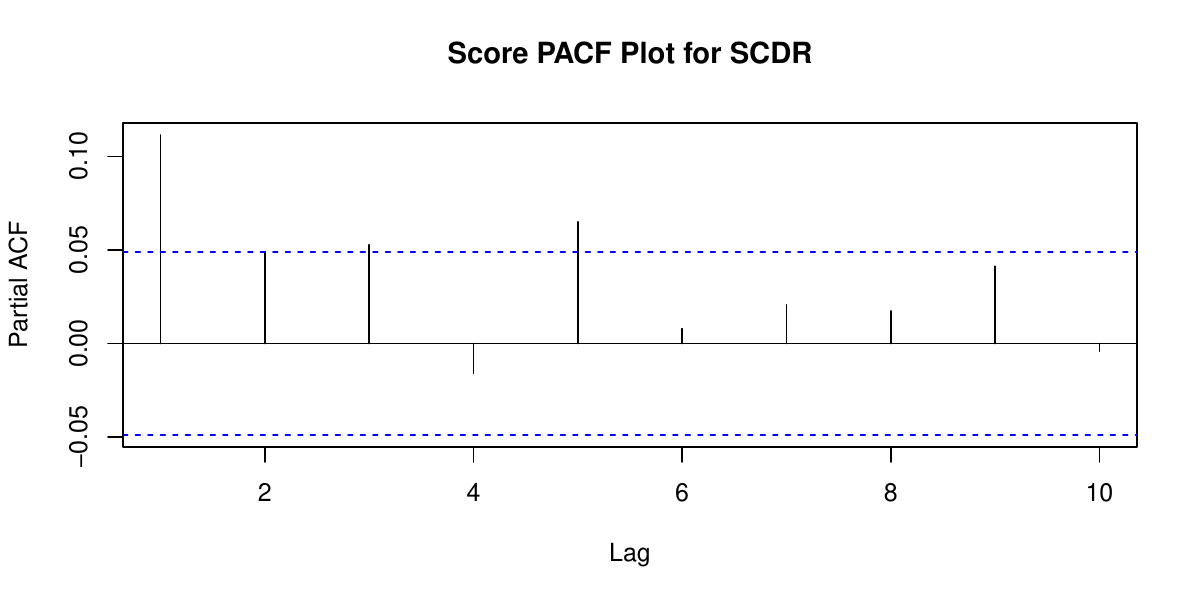}

\caption{PACF Plots for the Electricity Scores. From top left to right: SPCI, PID, BCIm SCDR}\label{fig:pacf_elec}
\end{figure}

\section{References}
\printbibliography[heading=none]